\documentclass[review,longtitle]{elsarticle}
\usepackage{fixltx2e}
\usepackage[table]{xcolor}
\usepackage{multirow}
\usepackage{appendix}
\usepackage{graphicx}
\usepackage{txfonts}
\usepackage{tabularx}
\usepackage{longtable}
\usepackage{caption}
\usepackage{subcaption}
\usepackage{lscape}
\usepackage{footnote}
\usepackage{hyperref}
\usepackage{url}

\journal{Astroparticle Physics}

\newcommand{\flux}{\mbox{${\rm \, cm^{-2} \, sr^{-1} \, s^{-1} \, MeV^{-1}}$}} 

\newcommand{\codename}{PICARD }

\begin{document}
\renewcommand{\UrlBreaks}{\do\/\do\a\do\b\do\c\do\d\do\e\do\f\do\g\do\h\do\i\do\j\do\k\do\l\do\m\do\n\do\o\do\p\do\q\do\r\do\s\do\t\do\u\do\v\do\w\do\x\do\y\do\z\do\A\do\B\do\C\do\D\do\E\do\F\do\G\do\H\do\I\do\J\do\K\do\L\do\M\do\N\do\O\do\P\do\Q\do\R\do\S\do\T\do\U\do\V\do\W\do\X\do\Y\do\Z}

\begin{frontmatter}

\title{Spiral arms as cosmic ray source distributions}

\author[uibk]{M. ~Werner \corref{cor}\fnref{fn1}}
\ead{michael.werner@uibk.ac.at}
\author[uibk]{R. ~Kissmann}
\author[mpe]{A. W. ~Strong}
\author[uibk]{O. ~Reimer}

\cortext[cor]{Corresponding author}
\fntext[fn]{\textit{Tel:} +43 51250752073}

\address[uibk]{Institut f\"ur Astro- und Teilchenphysik and Institut f\"ur Theoretische Physik, Leopold-Franzens-Universit\"at Innsbruck, A-6020 Innsbruck, Austria}
\address[mpe]{Max Planck Institut f\"ur extraterrestrische Physik, Postfach 1312, D-85741 Garching, Germany}

\begin{abstract}
The Milky Way is a spiral galaxy with (or without) a bar-like central structure. There is evidence that the distribution of suspected cosmic ray sources, such as supernova remnants, are associated with the spiral arm structure of galaxies. It is yet not clearly understood what effect such a cosmic ray source distribution has on the particle transport in our Galaxy. We investigate and measure how the propagation of Galactic cosmic rays is affected by a cosmic ray source distribution associated with spiral arm structures.

We use the \codename code to perform high-resolution 3D simulations of electrons and protons in galactic propagation scenarios that include four-arm and two-arm logarithmic spiral cosmic ray source distributions with and without a central bar structure as well as the spiral arm configuration of the NE2001 model for the distribution of free electrons in the Milky Way. Results of these simulation are compared to an axisymmetric radial source distribution. Also, effects on the cosmic ray flux and spectra due to different positions of the Earth relative to the spiral structure are studied.

We find that high energy electrons are strongly confined to their sources and the obtained spectra largely depend on the Earth's position relative to the spiral arms. Similar finding have been obtained for low energy protons and electrons albeit at smaller magnitude. We find that even fractional contributions of a spiral arm component to the total cosmic ray source distribution influences the spectra on the Earth. This is apparent when compared to an axisymmetric radial source distribution as well as with respect to the Earth's position relative to the spiral arm structure. We demonstrate that the presence of a Galactic bar manifests itself as an overall excess of low energy electrons at the Earth.

Using a spiral arm geometry as a cosmic ray source distributions offers a genuine new quality of modelling and is used to explain features in cosmic ray spectra at the Earth that are else-wise attributed to other propagation effects. We show that realistic cosmic ray propagation scenarios have to acknowledge non-axisymmetric source distributions.
\end{abstract}

\end{frontmatter}



\section{Introduction}
\label{intro}
Galactic cosmic rays (CRs) are particles that are accelerated to relativistic energies by astrophysical objects such as supernovae remnants \cite{Ackermann2013}. They are charged particles (e.g. protons, electrons and heavier nuclei) which interact with radiation fields and matter as they propagate from their sources through our Galaxy. Understanding the propagation of CRs in the Milky Way is essential for the interpretation of data from experiments that measure CRs and their secondaries at the Earth. A comprehensive overview of CR propagation and interaction is given in \cite{Schlickeiser2002,Strong2007}.

Unfortunately, we can only sample the distribution of Galactic CRs in our Galaxy at the Earth. Therefore, it is necessary to devise models of the propagation of galactic CRs to gain insight into the propagation physics. Formulating a realistic and accordingly complex physics model of CR propagation in the Milky Way as well as designing a numerical scheme that can solve the underlying transport equations accurately and efficiently are two of the main challenges in modelling CR transport. They are met by making several more or less justified simplifications in the underlying CR propagation physics and/or the input parameter space (i.e. Galactic magnetic field, gas distribution, radiation fields, spatial diffusion). Consequently, available transport codes feature CR propagation models of varying complexity \cite{Strong2007}, from single species fluid models to those that account for a nuclear reaction network of multiple CR species. Current propagation codes include DRAGON \cite{Evoli2008}, GALPROP \footnote{Available at \texttt{http://galprop.stanford.edu/} and \texttt{http://sourceforge.net/projects/galprop/}} \cite{GALPROP} and \codename \cite{Kissmann2013}, although other more specialized codes such as \textit{Usine} \cite{Putze2011} exist. CR propagation models can be tested against measurements of the local Galactic CR flux and measurements of the global distribution of galactic CRs via secondary particles, such as the Galactic diffuse $\gamma$-ray background \cite{Ackermann2012}, produced by interaction with the matter and radiation fields in our Galaxy.

In the past, modelling of CR propagation in the Milky Way has been done essentially in two dimensions by assuming azimuthal symmetry, thus treating the Galaxy as a disk with only radial ($r$) and height ($z$) dependencies in cylindrical coordinates. However, two-dimensional propagation scenarios have been rendered obsolete by tighter constraint on the input parameters as provided by new experimental results, i.e. matter distributions, magnetic field models, radiation fields, CR source distributions, and perhaps more importantly by the increase in computational power that allows a more realistic 3-dimensional non-isotropic treatment of galactic CR propagation physics. Presently, a paradigm shift towards full-fledged three-dimensional simulations that does not involve any geometrical simplifications is in progress \cite{Effenberger2012,Gaggero2013a,Werner2013}. Three-dimensional simulations allow for the treatment of propagation scenarios that were impossible to implement in two dimensional simulations and thus allow a more realistic modelling of CR propagation in the Milky Way.

In the past, due to limitations intrinsically inherent to two dimensional models, propagation scenarios featured CR source distributions that exhibit a radial and z-dependency only. These CR source distributions are based either on studies of the radial distribution of pulsars \cite{Yusifov2004,Lorimer2006} or supernova remnants \cite{Case1998} in our Galaxy. There is a widespread consensus that the Milky Way is a spiral galaxy with or without a central bar-like structure, e.g. \cite{Churchwell2009}. The number of spiral arms, their geometry as well as their position relative to the Earth, however, is the topic of ongoing research and depends on the tracers used to identify them as well as on the interpretation of the corresponding analysis results \cite{Vallee2008,SteimanCameron2010,Dame2011}. There is evidence that OB associations \cite{Higdon2013} and star forming regions \cite{SteimanCameron2010}, both of which are proposed to be connected to CR sources such as pulsars and supernova remnants, trace spiral arms. Therefore, it is reasonable to postulate that some fraction of the Galactic CR sources are associated with the spiral arms of the Milky Way. This hypothesis also effects the propagation of Galactic CRs in the Milky Way and was first postulated by \cite{Kniffen1973} (SAS-2) and \cite{Bignami1975} (COS-B) in order to interpret measurements of the Galactic diffuse $\gamma$-ray emission. A study by \cite{Rogers1988} which investigated the $\gamma$-ray emission originating from directions believed to be associated with the Orion spiral arm, claims to find a flatter $\gamma$-ray spectrum in on-arm than in inter-arm regions. Now, due to the advent of three-dimensional CR simulations can this hypothesis be tested and the resulting implications investigated \cite{Effenberger2012,Gaggero2013b,Benyamin2013, Werner2013}.

It also has been postulated that a periodic variation of the total CR flux incident on the Earth is convoluted with the observed periodicity in fossil records \cite{Rohde2005} and long-term climate cycles \cite{Shaviv2002,Shaviv2003}. The variation of the CR flux could arise from the changes in the Earth's position relative to the spiral arms in the Milky Way as the Sun orbits the Galactic Center. Three-dimensional simulations of the CR propagation can characterize the variation of the CR flux \cite{Benyamin2013,Effenberger2012} and test the plausibility of this hypothesis and allow us to quantify the resulting variations in CR flux.

Using the \codename code \cite{Kissmann2013} we study several propagation scenarios for protons and electrons with different spiral CR source distributions. We implement source distributions based on the logarithmic spiral arm structures given by \cite{Churchwell2009}, \cite{SteimanCameron2010}, \cite{Dame2011} and the spiral arm geometry used in \cite{Cordes2002}. We introduce our simulation set-up and detail the propagation scenarios in Section \ref{setup}. In Section \ref{results} we present the results of our simulations. We discuss the implications when compared to an axisymmetric CR source distribution for which we use one based on the radial distribution of pulsar in the Milky Way \cite{Yusifov2004} in Section \ref{discussion} and provide our conclusion in in Section \ref{conc}.

\section{Simulation Set-up}
\label{setup}
We use the recently developed \codename code \cite{Kissmann2013} to perform the CR propagation modelling presented herein. \codename uses a Gauss-Seidel multi-grid method to calculate the steady state solution of the CR transport equation. \codename is fast, accurate, and able to use parallel computing architectures efficiently. \codename utilizes the \textit{HDF5} \cite{hdf5} data model to store the simulation output in order to allow parallel and fast handling of very large data files. We refer to \cite{Kissmann2013} for a complete description of the underlying numerics and a thorough discussion of the capabilities and limitations of \codename. Because of computational constraints we consider CR protons and electrons, only. This is done because we want to isolate the effects on the propagation of CR without entangling them with effects attributable to the nuclear reaction network. It also reduced the computational demand considerably.
\subsection{CR Propagation Parameters}
\label{crProp}
All modelling was performed on a Cartesian spatial grid with 257 $\times$ 257 $\times$ 65 grid points for the x-,y- and z-axis, respectively. The spatial grid represents a simulation volume centred on the Galactic Center with an extension of $\pm 20$ kpc in the x- and y-dimension and $\pm 4$ kpc in the z-dimension. This corresponds to a spatial resolution of $0.15625$ kpc for the x- and y-dimension and $0.125$ kpc for the z-dimension. The Earth's nominal position is set to $x=8.5$ $\mbox{kpc, }  y = 0$ $\mbox{kpc, } z = 0$ $\mbox{kpc}$. We simulate protons and electrons with kinetic energies $E_{kin}$ from 100 MeV to 1 PeV. The momentum grid uses 129 logarithmically equidistant grid points. For consistency and comparison with published propagation scenarios all propagation parameters, except those of describing the CR source distribution, are identical to those given in Table 1 of \cite{Strong2010} (model ``z04LMPDS''), i.e. the model does not include diffusive reacceleration. For quick reference a summary of relevant propagation parameters is given in Table \ref{table:PropParams}. This model has been tuned to reproduce to CR and $\gamma$-ray data. We demonstrate similar agreement between \codename simulation results and CR data by showing the B/C-ratio in \ref{dataFit}. 

We use the same source distribution for electrons as for protons. The main aim of this work is to determine what genuine qualities can be obtained by using a CR source distribution following spiral arms instead of an axisymmetric CR source distribution. Therefore, we study only the effects induced by a change in the CR source distribution while keeping all other parameters constant \footnote{Parameter files are available upon request}. We do not yet aim to fit any data and investigate only the difference between different CR source distribution models. Determining what source distributions fit the data best does not serve any purpose in our attempt to isolate the impact of the different CR source distributions. The CR propagation problem is highly degenerate. Different sets of propagation parameters may lead to the same CR spectrum at the Earth. For example, effects on the resulting CR spectrum due to changes in one propagation parameter may be compensated by changes in another one.

\begin{table}
\caption{Propagation parameters used for all propagation scenarios detailed herein. Adapted from \cite{Strong2010}, see text for details.}
\label{table:PropParams}
\begin{minipage}{1.0\textwidth}
\centering
\renewcommand{\arraystretch}{1.5}
\begin{tabular}{l | c | c}
Parameter & Value  & unit \\
\hline
Halo height													& 4 		& $\left[\mbox{kpc}\right]$ \\
Diffusion coefficient $D_{xx}$ 								& 3.4  	& $10^{28} \, \left[ \mbox{cm}^{2} \, \mbox{s}^{-1} \right]$ \\
Reference rigidity for $D_{xx}$ 								& 	4 	& $\left[\mbox{MV}\right]$\\
Break energy (protons)										& 	9	& $\left[\mbox{GeV}\right]$\\
Index below break (protons)	 								&	1.8 &\\
Index above break (protons)									&	2.25&\\
Break energy (electrons)										&	4	& $\left[\mbox{GeV}\right]$\\
Index below break (electrons)	 							&	1.8 &\\
Index above break (electrons)								&	2.25&\\
\end{tabular}
\end{minipage}
\end{table}

As the representative for an axisymmetric CR source distribution we chose a parametrization for the source distribution that is determined by the radial $r$ dependence of the Milky Way’s pulsar surface density $\rho\left(r\right)$ based on the one given in \cite{Yusifov2004}
\begin{equation}
\rho\left(r\right) =  \left( \frac{r}{R_{\odot}} \right)^{\alpha} \exp \left[ - \beta \left( \frac{r - R_{\odot} }{R_{\odot}} \right) \right]
\label{eq:yusifov}
\end{equation}

\noindent wherein $\alpha$, $\beta$  are free parameters and $R_{\odot}$ is the galactocentric distance of the sun. We use $R_{\odot} = 8.5 \, \mbox{ kpc}$. For consistency with \cite{Strong2010}, we modified Equation \ref{eq:yusifov} by setting $r = r_{const} = 10 \, \mbox{kpc}$ for $ 10 \, \mbox{kpc} \, \leq \, r \, \leq  \, 15 \, \mbox{kpc}$ and $\rho\left(r\right) = 0 $ for $r \, > \, 15 \, \mbox{kpc}$. We use values for $\alpha = 0.475063$ and $\beta = 2.16570$ that were obtained by fitting \textit{Fermi}-LAT observation of the diffuse Galactic $\gamma$-ray emission \cite{Strong2011,Orlando2013}. Henceforth that model will be referred to as the \textit{Reference}-Model.

\codename normalizes the obtained CR density distribution by setting the CR flux at the nominal position of the Earth to $N^{p} = 5 \times 10^{-9}$ \flux for protons with  a $E_{kin} = 100$ GeV and $ N^{e} = 3.2 \times 10^{-10}$ \flux for electrons with a $E_{kin} = 34.5$ GeV. These values are based on direct measurements of protons and electrons. This normalization conforms to common practice and is adapted from \cite{Strong2010}. All spectra shown herein are normalized to these fluxes at the given energies. This means that in our simulations the total amount of energy carried by all CRs in our Galaxy is determined by the applied normalization. 

\subsection{Spiral Arm Models}
\label{spirals}
Even though the Milky Way's spiral structure has been studied over the past five decades discordances regarding the number of spiral arms, their geometry, their position within the Galaxy and the existence of a bar remain. This ambiguity arises due to the fact that observations of suitable tracers indicative for the distribution of matter in spiral arms are limited to one viewpoint of the Milky Way, i.e. the Earth, and that observations of different tracers (stellar populations, molecular clouds, dust, star-forming regions etc.), subsequent analysis and interpretation lead to different spiral arm models. However, logarithmic spirals seem to be favoured by the majority of spiral arm models with a dichotomy between four-arm and two-arm models. For a summary and discussion of current research into the Milky Way's spiral structure, we refer to \cite{Vallee2008} and \cite{SteimanCameron2010}.

One defining feature of spiral arms is the presence of star-forming regions \cite{SteimanCameron2010}. Star-forming regions give birth to the progenitors of CR sources such as supernova remnants. Therefore it is a viable assumption that at least a fraction of all CR sources in our Galaxy is also associated with the spiral arm structure. We implemented four different logarithmic spiral arm models as cosmic ray source distributions which we compare with the \textit{Reference}-Model 

Representative for a four-arm logarithmic spiral model we chose the one given in  \cite{SteimanCameron2010}. It is derived using a statistical analysis of data from observations of FIR cooling lines, $\left[ CII \right]$ and $\left[ NII \right]$, of the interstellar medium (ISM) using the \textit{Far Infrared Absolute Spectrophotometer} of the \textit{COsmic Background Explorer} (COBE). $\left[ CII \right]$ and $\left[ NII \right]$ trace increased density, UV radiation fields and are also indicative for the existence of star-forming regions and consequently the progenitors of supernova remnants. For each of the four arms we use the parametrization (in cylindrical coordinates $r$,$\phi$ and $z$) as given in \cite{SteimanCameron2010} for the $i$-th spiral arm CR source distribution
\begin{equation}
\mbox{Sp}_{i} = \exp \left[ - \frac{\left| r - R_{3} \right|}{\sigma_{r}} \right] \exp \left[- \frac{\left(\phi - \phi_{i} \right)^{2}} {\sigma_{\phi}^{2}} \right] \exp \left[- \frac{z^{2}}{2 \sigma_{z}^{2} } \right]
\end{equation}
wherein $\sigma_{r}$, $\sigma_{\phi}$, $\sigma_{z}$ are factors representing the characteristic scale-length of the spiral arm in the $r$-, $\phi$- and $z$-dimension respectively. $R_{3}$ is a parameter introduced to allow for a cusp in the inner Galaxy and $\phi_{i}$ being the center line  definition of the $i$-th spiral arm
\begin{equation}
\phi_{i} = \frac{\ln\left( \frac{r}{a_{i}}\right)}{\alpha_{i}}
\end{equation}
wherein $a_{i}$ determines the orientations of the spiral arm and $\alpha_{i}$ the pitch angle $\theta = \arctan\left(\alpha_{i} \right)$. We use the values given in \cite{SteimanCameron2010} for all parameters. A comparison of the four-arm spiral CR source distribution to the one that is used in the \textit{Reference}-Model is shown in Figure \ref{fig:rDepYusifov}. This four-arm spiral model is henceforth called \textit{Steiman}-Model. 

CR sources in the Milky Way may not be associated exclusively with the spiral arm structure. This can be due to the presence of CR sources such as supernova remnants outside the spiral arms and the fact that the dynamics of the spiral arms cause a smearing of the CR source distribution on time-scales relevant for CR propagation. Accordingly, we constructed models that combine the characteristics of the \textit{Steiman}-Model and the \textit{Reference}-Model in order to study the effects of different fractional contribution of CR sources in spiral arms to the total CR source distribution systematically. We refer to these models as the \textit{Mix}-Models and use the nomenclature given in Table \ref{table:mix} to denote the percentage of the spiral arm component and the axisymmetric component.
\begin{table}
\caption{Nomenclature of the \textit{Mix}-Models and corresponding contribution of spiral arm and the \textit{Reference}-Model component. Here the given relative contributions are computed from the integral source strength, referring to the integral of the given component of the source distribution over the entire numerical domain.}
\label{table:mix}
\centering
\renewcommand{\arraystretch}{1.5}
\begin{tabular}{l| c | c}
\hline
 & \textit{Reference}-Model contribution   & \textit{Steiman}-Model contribution\\
\hline
\textit{Mix-Model-10}	&	90.9\%	& 9.1 \%  \\
\textit{Mix-Model-50}	&	66.7\%	& 33.3\%\\
\textit{Mix-Model-100}	&	50\%		& 50\%\\
\textit{Mix-Model-200}	& 	33.3\%  & 66.7\%\\
\textit{Mix-Model-1000}	& 	9.1\%	& 90.9\%  \\
\hline
\end{tabular}
\end{table}

The \textit{Scutum-Crux} and \textit{Perseus} arm of the \textit{Steiman}-Model is qualitatively consistent with the arms of a barred two-arm model presented in \cite{Churchwell2009}. \cite{Dame2011} found evidence of a molecular spiral arm in the far outer Galaxy. This newly discovered spiral arm is consistent with an extension of the \textit{Scutum-Centaurus} arm as given in \cite{Churchwell2009} (see \cite{Dame2011}). Owing to the strong similarities between both models we construct a two-arm model that uses the parametrization of the Steiman-Model but includes only the Scutum-Crux and the Perseus arms and add a bar component with a half-length of $3.2$ kpc and the following parametrization (here in Cartesian coordinates) for the CR source distribution $B_{cr}$  
\begin{equation}
B_{cr}= \exp \left[-\frac{ \left( z^{2} + \left(y - x \sin \left(\varphi_{bar} \right)\right)^{2}\right) }{\delta_{bar}^{2}} \right]
\end{equation}
wherein $\varphi_{bar}$ is the bar rotation angle relative to the Solar-Galactic Center line and $\delta_{bar}$ represents a scale factor for the thickness of the Galactic bar. We adopt a value of $\varphi_{bar} = 20^{\circ}$ \cite{Churchwell2009} and $\delta_{bar} = 0.31$ kpc. The Galactic bar and the spiral arms contribute equally to the total integrated CR source distribution. This model will be referred to as the \textit{Dame-Model}. To investigate the effect of a Galactic bar on CR propagation we constructed several models that have different fractional contributions of the spiral component and a Galactic bar to the total CR source distribution. The model designations and the corresponding contributions of the Galactic bar are given in Table \ref{table:bar}.
\begin{table}
\caption{Nomenclature of two arm models and corresponding Galactic bar contribution to the CR source distribution relative to the spiral arm component. Here the given relative contributions are computed from the integral source strength, referring to the integral of the given component of the source distribution over the entire numerical domain.}
\label{table:bar}
\centering
\renewcommand{\arraystretch}{1.5}
\begin{tabular}{l| c | c}
\hline
 &  Galactic bar contribution & Spiral arm contribution.\\
\hline
\textit{Dame-No-Bar}-Model	&	0\%		& 100\%\\
\textit{Dame-Bar-10}-Model	&	9.1\%	& 90.9\%\\
\textit{Dame-Bar-50}-Model	&	33.3\%	& 66.7\%\\
\textit{Dame}-Model			& 	50\%		& 50\%\\
\textit{Dame-Bar-200}-Model	& 	33.3\%	& 66.7\%\\
\textit{Dame-Bar-1000}-Model	& 	90.9\%	& 9.1\%\\
\hline
\end{tabular}
\end{table}

Finally, we implement the spiral model that is used by \cite{Cordes2002} in the NE2001 model of the free electron density in the Milky Way. We take the spiral arm configuration, which is based on the work of \cite{Wainscoat1992}, directly from the publicly available code provided on the project homepage\footnote{Available here \texttt{http://www.astro.cornell.edu/\~cordes/NE2001/}}. Henceforth, this model will be called \textit{NE2001}-Model. Compared to the \textit{Steiman}-Model (0.1 kpc scale) this model features much broader (1 kpc scale) spiral arms, including a nearby (Orion-Cygnus) spiral arm segment, but does not have an exponential decline in the $\phi$-direction.

A visualisation of Earth's location relative to the nearest spiral arms in the different spiral arm models can be found in Figure \ref{FigCRFluxNearEarth}. The three spiral arm models can also be characterized by their CR source densities in Earth's vicinity. Due to the nearby (Orion-Cygnus) spiral arm segment, which can be considered as a local CR source, the \textit{NE2001}-Model has the highest CR source density near the Earth. In contrast, the \textit{Dame}-Model features only a comparably low CR source density in the Earth's immediate neighbourhood, while the CR source density in the \textit{Steiman}-Model lies somewhere between theses two extremes.

\section{Results}
\label{results}
In the following we present results obtained using the different source models. We show spatial distributions of Galactic CR protons and electrons in Section \ref{distri}. In Section \ref{spectra} we compare the spectra and the relative deviations of each CR source model to the \textit{Reference}-Model. In Section \ref{frac} we investigate the proton and electron spectra at the Earth when using the \textit{Mix}-Models. In Section \ref{bars} we study the effects of a Galactic bar on the propagation of electrons and protons. We compare and quantify the variations of the integral CR flux in Section \ref{spectrogram}, where we also investigate the effects on the CR electron and proton spectrum at the Earth as a function of the Earth's position relative to the spiral arms are investigated.
\subsection{CR Density Distributions}
\label{distri}
We calculated the CR density distribution function at every spatial and momentum grid point using \codename. Figure \ref{Hdistributions} shows examples of x,y-slices at $z=0$ and x,z-slices at $y=0$ of the density distributions of protons with kinetic energies of $E_{kin} = 1.1$ GeV and $E_{kin} = 1.1$ TeV for the different source models while Figure \ref{Edistributions} shows the same for electrons. The imprint of the underlying CR source model is visible in the resulting CR proton and electron distributions. However, for protons and low energy electrons the structure of the CR source distributions is smeared out due to the relatively large mean energy loss length when compared to the scale of the spiral arm structure. Note that the spatial proton density distribution has a very weak energy dependence i.e. only the total amplitude changes with energy while the spatial distribution remain unchanged. Inverse Compton (IC) scattering causes severe energy losses for high energy electrons. Therefore, as can be seen in Figure \ref{Edistributions}, high energy electrons are closely confined to the underlying source distribution. Due to the broad spiral arms of the \textit{NE2001}-Model the imprint of the CR source distribution is no longer readily apparent in the resulting distribution of protons and electrons with kinetic energies below 1 TeV. Note, that the bright CR hotspot near the Earth's nominal position (see Figure \ref{FigCRFluxNearEarth}) in the \textit{NE2001}-Model coincides with the presence of a nearby (Orion-Cygnus) spiral arm segment unique to that model \cite{Cordes2002}.
\subsection{CR Spectra at the Earth}
\label{spectra}
The proton and electron spectra at the Earth's nominal position are shown in Figure \ref{HESpectra}. A comparison of our simulations using the \textit{Steiman}-Model with proton data is discussed in \ref{dataFit}. For protons all source models yield spectra that are seemingly indistinguishable in the chosen representation. To quantify the remaining differences Figure \ref{relHESpectra} shows the ratio between the proton spectrum obtained with the spiral arm CR source distribution models and the proton spectrum obtained using the \textit{Reference}-Model. Using the \textit{Steiman}-Model we find a deficit of protons with $E_{kin} < 1$ GeV that increases towards lower energies when compared with the \textit{Reference}-Model. The deficit reaches a maximum of approximately 10\% at the lowest kinetic energy $E_{kin} = 100$ MeV considered. In contrast, for protons with $E_{kin} > 1$ GeV an excess is observed. The excess increases with a very shallow slope and never exceeds 2\% at even the highest energy $E_{kin} = 1$ PeV. Comparable findings are obtained when examining the spectra of the \textit{Dame-Model}. However, in the \textit{Dame-Model} the deficit is more pronounced. The deficit starts at a kinetic energy $E_{kin} < 30$ GeV and it reaches a maximum of 60 \% at $E_{kin} = 100$ MeV when compared to the \textit{Reference}-Model. At $E_{kin} = 100$ MeV nearly twice as many low energy protons arrive at the Earth in the axisymmetric source model than in the \textit{Dame-Model}. For energies above $E_{kin} > 30$ GeV the behaviour mimics the one of the \textit{Steiman}-Model not only qualitatively but also quantitatively. In contrast, in the case of the \textit{NE2001}-Model there is an excess (instead of a deficit) of low energy protons with kinetic energies $E_{kin} < 14$ GeV, increasing towards lower energies with a maximum of approximately 8\% at the $E_{kin} = 100$ MeV. We also find a proton deficit for energies $E_{kin} > 14$ GeV that reaches values of approximately 1\% for the highest energies.

\begin{figure}[t]
\centering
\includegraphics[width=12cm]{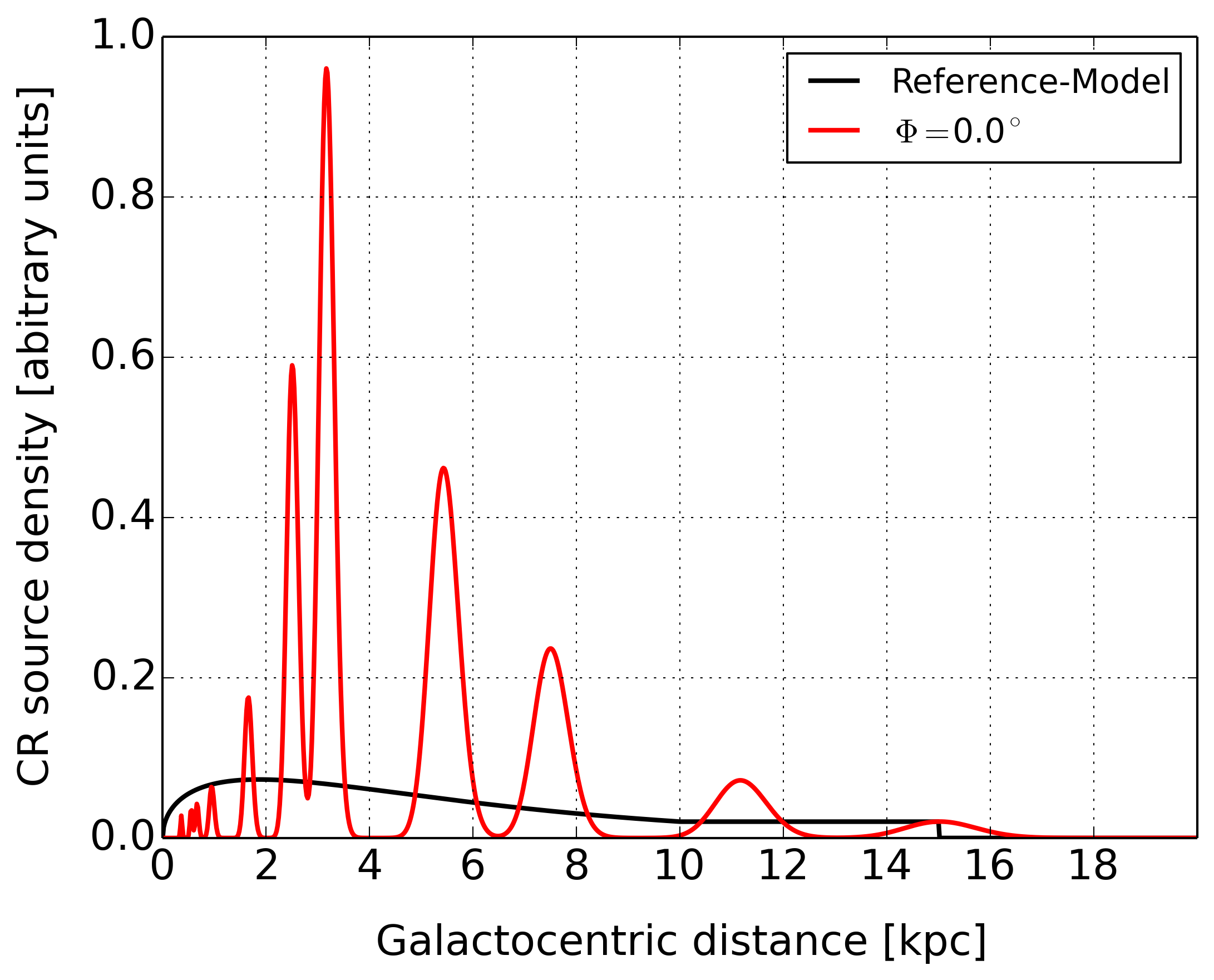}
\caption{Comparison of the axisymmetric \textit{Reference}-Model and a slice at $\phi =0^{\circ}$ of the \textit{Steiman}-Model. This slice represents a cut long the line connecting the Galactic Center and the Earth's nominal position (x = 8.5 kpc, y = 0 kpc, z = 0 kpc). The relative normalisation between these source distributions relates to equal cosmic ray flux at Earth at an energy of 100 GeV, and the CR source density is given in arbitrary units. \label{fig:rDepYusifov}}
\end{figure}

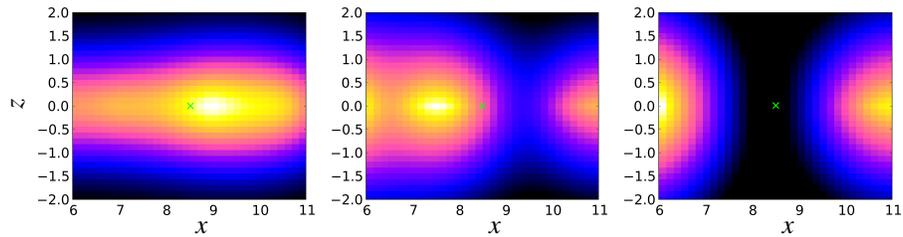
\begin{figure}[h]
    \setlength{\unitlength}{0.001\textwidth}
    \begin{subfigure}{300\unitlength}
        \begin{picture}(300,300)
        		\put(0,135){\rotatebox{90}{$z$}}
        		\put(200,0){$x$}
            \put(30,20){\includegraphics[trim=0cm 0cm 0cm 0cm, clip=true,width=300\unitlength]{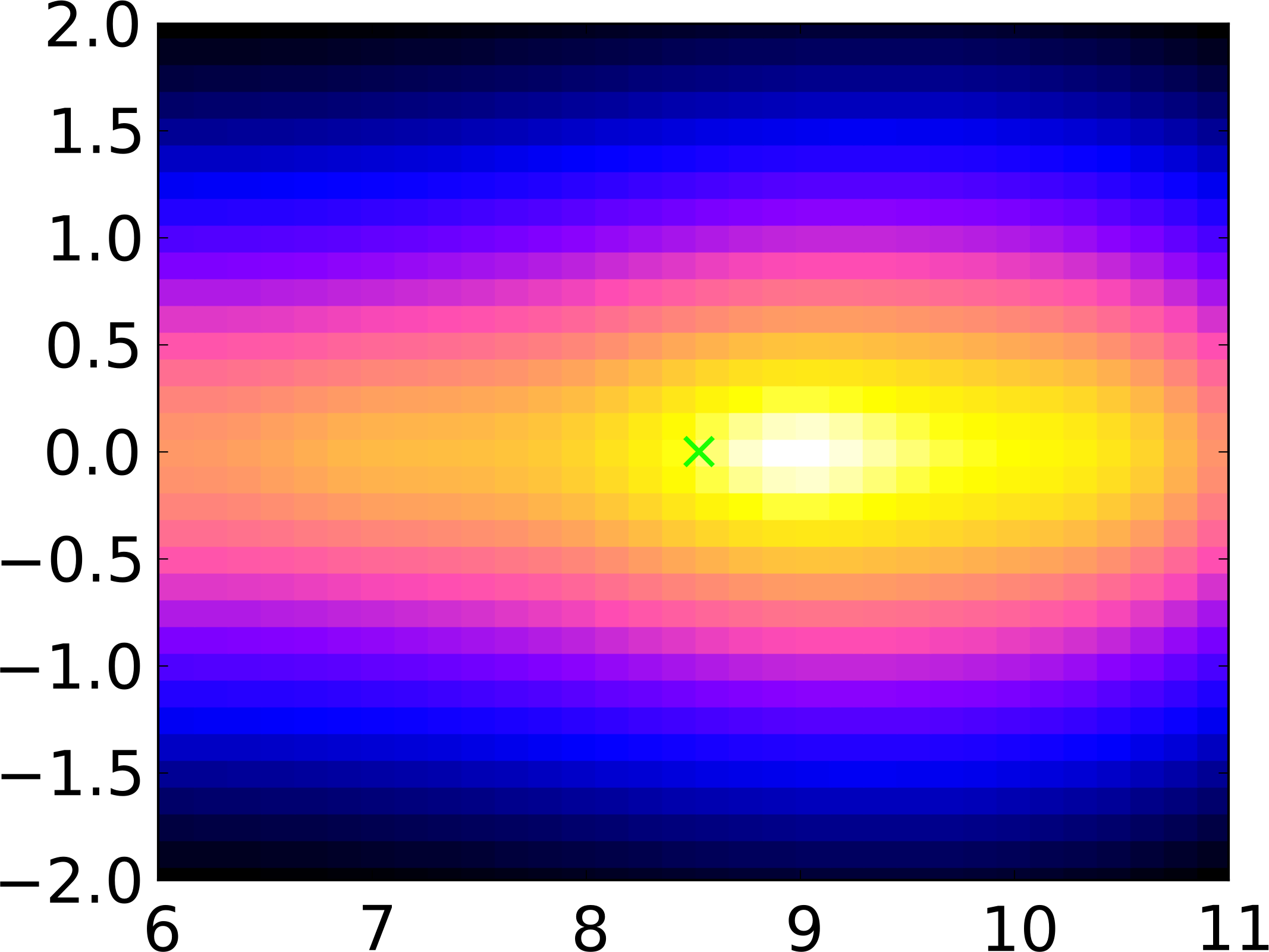}}
        \end{picture}
    \end{subfigure}
\,
     \begin{subfigure}{300\unitlength}
        \begin{picture}(300,300)
        		\put(200,0){$x$}
            \put(30,20){\includegraphics[trim=0cm 0cm 0cm 0cm, clip=true,width=300\unitlength]{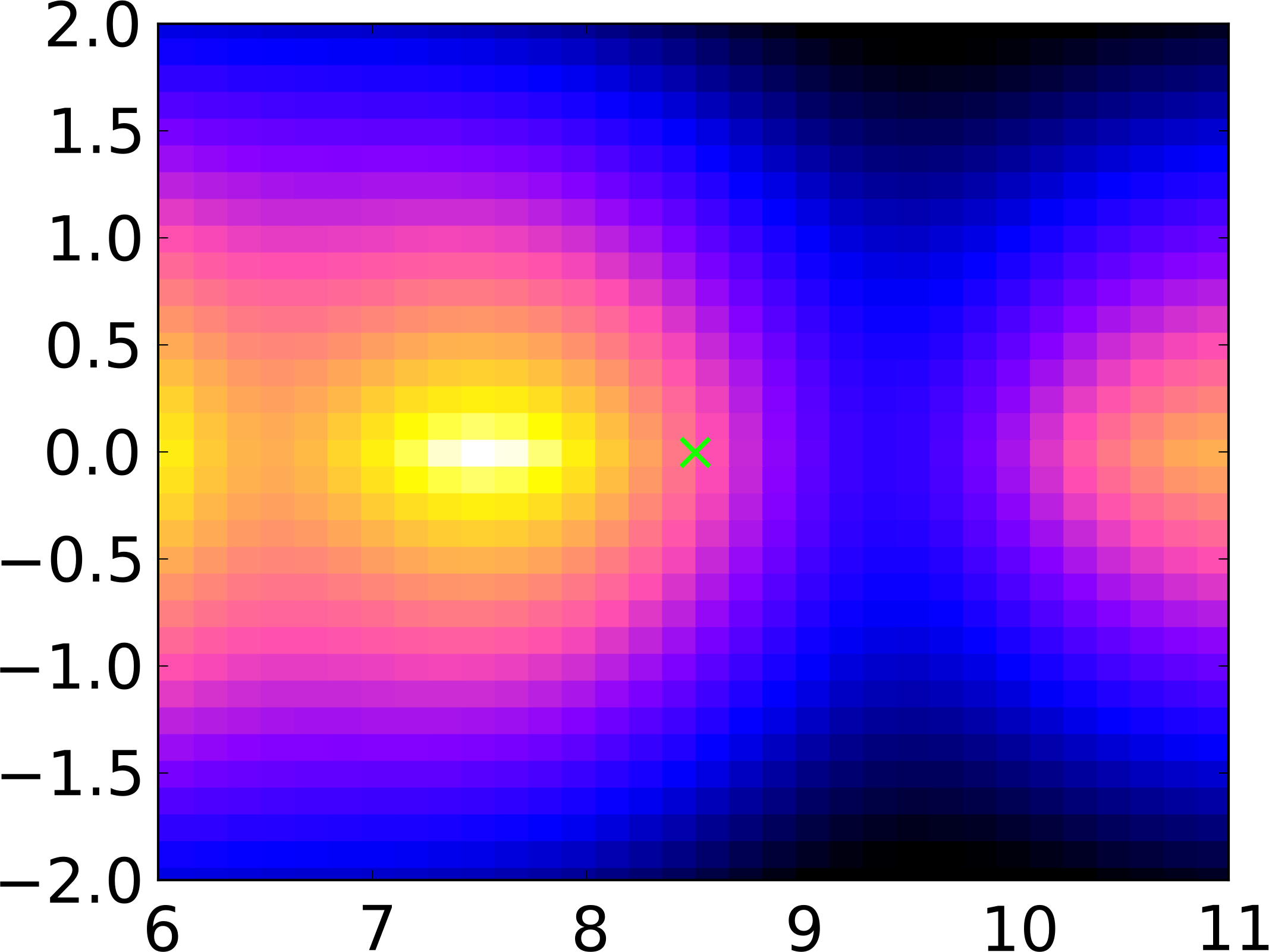}}
        \end{picture}
    \end{subfigure}
\,
        \begin{subfigure}{300\unitlength}
        \begin{picture}(300,300)
        		\put(200,0){$x$}
            \put(30,20){\includegraphics[trim=0cm 0cm 0cm 0cm, clip=true,width=300\unitlength]{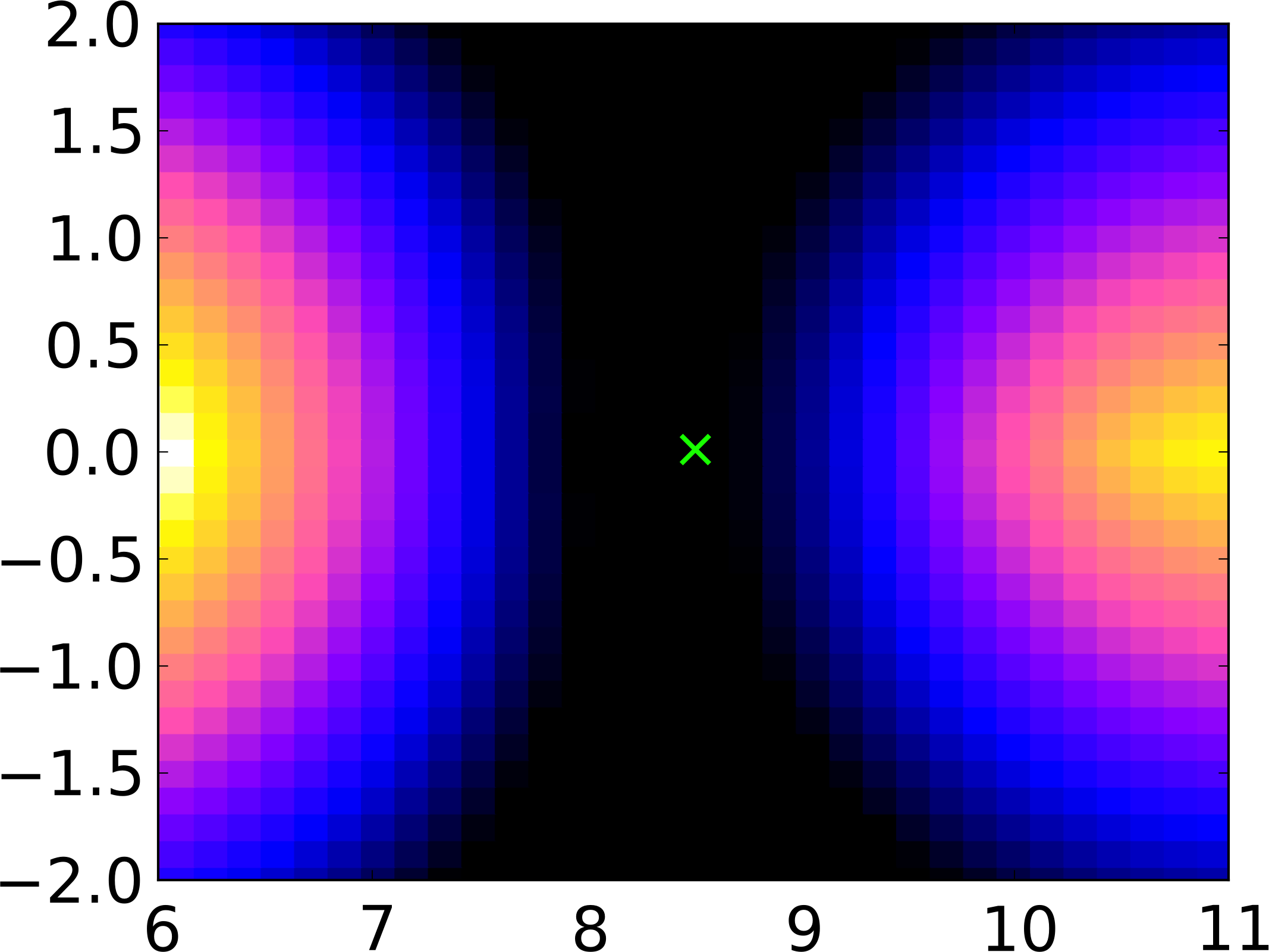}}
        \end{picture}
    \end{subfigure}
\caption{Visualisation of Earth's location relative to the nearest spiral arms indicated  through by the cosmic ray electron density at 1\,TeV. Results are shown in the $x-z$ plane for $y=0$, $x=[6,11]$ kpc and $z=[-2,2]$ kpc. The position of the Earth is marked by a green cross. Results are shown for the \textit{NE2001}-Model (\textit{left}), the four-arm \textit{Steiman}-Model (\textit{middle}) and the two-arm \textit{Dame}-Model (\textit{right}). Used color scales are arbitrary.}
\label{FigCRFluxNearEarth}
\end{figure}

This apparent dichotomy in the behaviour of the spiral arm models can be understood when we take the position of the Earth relative to the spiral arm structure into account. As we show in Figure \ref{FigCRFluxNearEarth}, in both the \textit{Steiman}-Model and the \textit{Dame-Model} the Earth is located in an inter-arm region, away from the dense CR source regions which are confined inside the spiral arms. Also, the gradients in the CR source distributions are much steeper in the \textit{Steiman}-Model and the \textit{Dame-Model} than in the \textit{Reference}-Model (see Figure \ref{fig:rDepYusifov}). As a consequence the bulk of the low energy protons fail to reach the Earth due to Coulomb and ionisation losses. These losses are the dominant energy loss processes for kinetic energies below $\approx$ 1 GeV. The excess of higher energy protons found in both spiral models can be attributed to the applied normalization which can shift the spectrum by a constant factor. The excess indicates that, above 1 GeV in the \textit{Steiman}-Model and above 30 GeV in the \textit{Dame-Model}, more protons reach the Earth. In contrast to the two other spiral arm models, in the \textit{NE2001}-Model the Earth is located in close proximity (see Figure \ref{FigCRFluxNearEarth}) of the local (Orion-Cygnus) spiral arm segment. In addition the \textit{NE2001}-Model has much broader spiral arms that extend further towards the Earth than the spiral arms in the \textit{Steiman}-Model and the \textit{Dame-Model}. When compared to the \textit{Reference}-Model the excess of low energy protons can be understood by considering the close proximity of the Earth to nearby CR sources: More low energy protons can reach the Earth before their energy is dissipated. The observed deficit for protons with energies $E_{kin} > 14$ GeV can be attributed to the applied normalization that shifts the entire spectrum downwards. 

The electron spectra at the Earth's nominal position obtained for the different CR source models are shown in Figure \ref{HESpectra}. In contrast to the proton spectra the electron spectra vary significantly and allow to readily distinguish the different source models. The spectrum obtained using the \textit{Dame}-Model exhibits the largest deviations from the axisymmetric \textit{Reference}-Model. Below the normalization energy the electron flux increases relative to the \textit{Reference}-Model. At the highest energies the electron flux is orders of magnitude below the \textit{Reference}-Model. The same is true for the \textit{Steiman}-Model albeit to a much lesser extend. To quantify these deviations we plot ratios between the electron spectra obtained using the spiral arm models and the spectrum from our \textit{Reference}-Model in Figure \ref{relHESpectra}. Both the \textit{Steiman}-Model and the \textit{Dame-Model} show an increase in the low energy electron flux for kinetic energies below some 35 GeV. This increase is most pronounced in the \textit{Dame-Model}. We find a distinct maximum of the enhancement at a kinetic energy of approximately 450 MeV with a value of approximately 4 times the electron flux of that in the \textit{Reference}-Model. In case of the \textit{Steiman}-Model the enhancement is not as pronounced as in the aforementioned case but and peaks at 500 MeV. In the \textit{Steiman}-Model the enhancement of the electron flux is approximately 125\% of the electron flux of the \textit{Reference}-Model. There is no indication of an enhancement of low energy electrons in case of the \textit{NE2001}-Model. Rather, we find a minor deficit that is about 95\% of the electron flux obtained using the axisymmetric \textit{Reference}-Model at a kinetic energy of approximately 500 MeV. In all three spiral arm models we find a deficit of electrons above a certain energy ($E_{kin} > 35$ GeV in case of the \textit{Dame-Model} and the \textit{NE2001}-Model and $E_{kin} > 100$ GeV for the \textit{Steiman}-Model). Both in the \textit{NE2001}-Model and the \textit{Steiman}-Model this deficit reaches its highest value of some 40\% of the electron flux \textit{Reference}-Model at the highest energies.

The excess of low energy electrons and the deficit of high energy electrons that is apparent in both the \textit{Steiman}-Model and the \textit{Dame-Model} is due to the Earth's inter-arm position which means that most high energy electrons dissipate their energy before reaching the Earth. The \textit{Dame-Model} has even lower source density in the vicinity of the Earth and therefore the deficit is much more pronounced than in the \textit{Steiman}-Model. The enhancement of low energy electron is the consequence of the steeper spectrum and the applied normalization. In order to explain the spectra of the \textit{NE2001}-Model we have to consider the broadness of the spiral arms as well as the nearby spiral arm segment (see Figure \ref{FigCRFluxNearEarth}). Because of these, the observed deficit is not as prominent as in the \textit{Steiman}-Model or the \textit{Dame-Model}. More high energy electrons reach the Earth. We note that the bend in the spectrum at the high energy end of the distinct peak in all three model at $E_{kin} \approx 4$ GeV is due to a break in the electron injection spectrum that has been adapted from \cite{Strong2010}.

\subsection{Models with Fractional Spiral Arm Component}
\label{frac}
The ratio of the CR spectra at the Earth obtained using the \textit{Mix}-Models and the spectrum obtained using the \textit{Reference}-Model is shown in Figure \ref{relYusifov}. The enhancement of the electron flux due to the spiral arms is evident in all of the \textit{Mix}-Models. Even for a spiral arm contribution of about 10 percent the spectrum of electrons deviates as much as $\sim$ 28\% at the highest energies from the one obtained by using the \textit{Reference}-Model. The excess of electrons with kinetic energies below 100 GeV is at or below the twenty-percent level in all the \textit{Mix}-Models. For a larger spiral arm contribution to the overall CR source distribution the electron spectrum quickly approximates the one obtained using just the \textit{Steiman}-Model. The same is true for protons and a protons deficit is also seen in all of the \textit{Mix}-Models at low energies. Likewise the observed excess of protons with $E_{kin} > 1$ GeV increases as the contribution of the spiral arm component to the total CR source distribution increases.

\subsection{Galactic Bar Models}
\label{bars}
Here we investigate the effects of a Galactic bar on the electron and proton spectra at the Earth. We compare the spectrum obtained using the \textit{Dame-No-Bar}-Model with spectra resulting from the \textit{Dame-Bar}-Models with their ratios being shown in Figure \ref{relBar}. The presence of a Galactic bar causes an excess of electrons with kinetic energies below 37.1 GeV. This excess increases with increasing contribution of the Galactic bar to the overall source distribution. In the \textit{Dame-Bar-50-Model} the excess of low energy electrons amounts to 80\% at lowest energies. But even in the extreme case of the \textit{Dame-Bar-1000-Model} where the majority of CR sources in our Galaxy is contained within the Galactic bar the enhancement does not exceed a factor of 5. For kinetic energies above 37.1 GeV we observe a small decrease in the electron flux. But even for the extreme case of the \textit{Dame-Bar-1000-Model} this deficit is below 30 percent. For protons the situation is reversed. As can be seen in Figure \ref{relBar} the larger the contribution of the Galactic bar to the total source distribution the larger the deficit in proton flux at kinetic energies below some 20 GeV. The deficit increases with decreasing energy and reaches its highest value at the lowest energy considered. For the \textit{Dame-Bar-1000-Model} where most sources are confined within the Galactic bar the proton flux is decreases by 25\% compared to the \textit{Dame-Model}. Due to the applied normalization the presence of a Galactic bar also leads to a small increase of the proton flux at energies above 16 GeV. 

The Galactic bar is located at the Galactic Center far from the Earth. High energy electrons accelerated in the Galactic Bar can not reach the Earth before there energy is dissipated. The normalization and the steepening of the spectrum leads to the enhancement of low energy electrons. Due to Coulomb and ionisation losses low energy protons are confined inside the Galactic bar and do not reach the Earth.

\subsection{Spatial-Spectral Variations of CRs in the Galaxy}
\label{spectrogram}
We investigated the spectra for different positions along the Sun's Galactic orbit, i.e. we quantify the difference between on-arm and inter-arm positions. To accomplish this we determine the variations of the total proton flux as well as the changes on the electron and proton spectra at different positions on a circle around the Galactic Center with radius $r = R_{\odot} = 8.5 \mbox{kpc}$. The variations of the total proton flux for each spiral arm model is shown in Figure \ref{fig:orbitModels}. We plot the ratio of the flux at the nominal position of the Earth and the total flux $F_{pos}$ at each test position. For each test position we average over the nearest neighbouring grid points.

We find the maximum enhancement of the total proton flux inside the spiral arms when using the \textit{Dame-No-Bar}-Model, in which case the total proton flux is by a factor of up to 3.1 higher than the flux at the nominal inter-arm position of the Earth. This is because the \textit{Dame-No-Bar}-Model has fewer spiral arms leading to a larger extent of the inter-arm regions (see Figure \ref{FigCRFluxNearEarth}). Additionally, the distance of the Earth to the nearest spiral arm center line is greater than in the other spiral arm models resulting lower CR source densities. The confinement of low energy protons to the spiral arms is responsible for the strong increase of the integral proton flux inside the spiral arms. The presence of a Galactic bar in the \textit{Dame-Model} reduces the peak proton flux by approximately 1/3. Because the \textit{Steiman}-Model has inter-arm regions that are less extended and the Earth is located closer to the nearest spiral arm the integral proton flux increases by a factor 1.7 inside the spiral arms as compared to the flux at the Earth's nominal position.

When using the \textit{NE2001}-Model we find no localized variation in the total proton flux. Due to the broad spiral arms of this model and the presence of a local spiral arm segment CRs are much more uniformly distributed so that the energy loss processes responsible for the enhancement in the other spiral arm models are less apparent. Owing to the presence of the nearby Orion-Cygnus spiral arm segment the total proton flux is highest at the nominal position of the Earth.

In Figure \ref{fig:orbitYusifov} we investigated the changes in the total proton flux using the \textit{Mix}-Model in the same manner. Even for a spiral arm contribution of approximately 10\% we find a peak enhancement of 28\% when compared to the nominal position of the Earth. For larger spiral arm contributions to the total CR source distribution the magnitude of the flux variations approaches that of the \textit{Steiman}-Model.

To visualize the differences between the spectra at different positions on a circle around the Galactic Center we use spectrograms, i.e. contour-plots that show the deviation of the spectra $S_{\phi}$ at the considered test positions relative to the spectrum $S_{earth}$ at the Earth's nominal position. The positions constitute the x-axis and are given by the angle $\phi$ between the line connecting the nominal position $x=8.5 \mbox{kpc, }  y = 0 \mbox{kpc, } z = 0 \mbox{kpc}$ of the Earth with the Galactic Center and the line connecting the Galactic Center with the test position on the circle. The spectrum for each position is represented by a vertical line with the y-axis in the spectrograms denoting the kinetic energy while the flux is represented by a color bar. The flux at the test positions is normalized to the flux at the nominal position of the Earth.

The proton and electron spectrograms are shown in Figure \ref{Spectrograms} for all spiral models tested. For the \textit{Steiman}-Model the imprint of the four spiral arms is readily apparent. For protons with a kinetic energy above 1 GeV we find that the enhancement factor has a very weak energy-dependence. For protons with a kinetic energy below 1 GeV we find a softening of the spectrum inside the spiral arms that increases towards lower energies. For the \textit{Dame-Model} we observe the same effects as in case of the \textit{Steiman}-Model, albeit with larger enhancement factors. The imprint of the spiral arms on the spectrum is even more pronounced. Inside the spiral arms the enhancement factor for electrons of the highest energies is orders of magnitudes higher than outside. This results in a harder spectrum at on-arm than inter-arm positions. For protons we also find a softer spectrum at on-arm positions. As in the \textit{Steiman}-Model the enhancement for protons with kinetic energies below 1 GeV is most easily discernible while also exhibiting a strong energy-dependence. For protons with kinetic energies greater than 1 GeV we find a much smaller enhancement that also features a very weak energy dependence.

The spiral structure of \textit{NE2001}-Model is not visible in the proton spectrogram and there is no hardening or softening of the spectrum anywhere. As discussed in Section \ref{spirals} the highest protons flux is found at the nominal position of the Earth which is due to the presence of the Orion-Cygnus spiral arm segment in the Earth's vicinity. For electrons the situation is more complex. As in the case of protons we find the highest electron flux at the nominal position of the Earth. We do find a hardening of the electron spectra when going from on-arm to inter-arm positions, that effects electrons with a kinetic energy above 10 TeV. The observed changes in the proton and electron spectra can be understood when we consider the model specific characteristics used to explain the difference between the model spectra in Section \ref{spectra}.

\section{Discussion}
\label{discussion}
When compared to the axisymmetric \textit{Reference}-Model we find several interesting features in the CR spectra obtained using the spiral arm models. The non-uniformity in the distribution of sources in the spiral arm models (i.e. the number of spiral arms) and the spatial scales of their sub structure (i.e. the dimensions of the spiral arms) influences the propagation of CRs in these models. CRs are affected most if the time-scales and spatial scales of the propagation effects they are subjected to are smaller or of the same order as the corresponding time-scales or spatial scales of the spiral arm source distribution.

In our scenarios this becomes evident for electrons with kinetic energies above some 100 GeV and protons with kinetic energies below some 1 GeV. Both are subjected to energy losses that reduce their mean energy loss length to an order comparable to the widths of the spiral arms (~ 0.1 kpc) in cases of the \textit{Steiman} and the \textit{Dame}-Models. For high energy electrons the dominant energy loss mechanism is IC-scattering and synchrotron emission. Therefore, high energy electrons are confined to their source regions and trace the CR source distribution closely. 
Low energy protons are affected in a similar but lesser manner by ionisation and Coulomb losses and consequently do not trace the CR source distribution so well. As we have shown previously, due to the confinement the exact shape of the electron and proton spectrum is highly dependent on the location of the Earth relative to the spiral arms. In all our spiral arm models the Earth is located at an inter-arm position where few higher energy electrons can reach the Earth. Therefore, when compared with the axisymmetric \textit{Reference}-Model we observe a deficit of high energy electrons at the Earth in all three models. The magnitude of this deficit varies considerably. In the \textit{Dame}-Model the CR sources are distributed most unevenly and therefore all effects relating to the spiral substructure are particularly pronounced. Recently, \cite{Gaggero2013a} arrive at a similar conclusion using the DRAGON code. They attribute the anomalous behaviour of the positron fraction to a spiral arm source distribution and a nearby positron source. Due to the severe influence on the electron spectrum CR sources distributed as in our two arm spiral model seems unrealistic since we observe high energy electrons at the Earth. Indeed, a recent survey of massive young stars supports the concept of our Milky Way as a four arm spiral galaxy \cite{Urquhart2014}.

As we have shown, even a fractional spiral arm contribution of 10\% to a global axisymmetric CR source distribution leads to a deficit of 28\% of high energy electron at the Earth when compared to the \textit{Reference}-Model. We note that for higher spiral arm contributions the spectra quickly approximates that of a pure spiral arm CR source distribution. Our work clearly shows that the exact shape of the electron spectrum at the Earth strongly depends on the Earth's position relative to any non-axisymmetric source distribution. Together with other factors (such as the position of nearby sources, the shape of the injection spectrum, etc.) this makes conclusions about the electron source distribution difficult using direct measurements of the electron spectrum alone.
 
Precise measurements of the high energy electron spectrum at the Earth also provide an additional test of CR electron source models \cite{Mitchell2010}. Currently, measurements of the electron spectrum extend to TeV energies \cite{Aharonian2008} but have large systematic uncertainties. Future ground based Cherenkov telescopes such as the \textit{Cherenkov Telescope Array} (CTA) \cite{CTA2011} will improve upon this. Measurements by satellite based instruments of the electron spectrum extend to 1 TeV \cite{Fermi2010} and the forthcoming CALorimetric Electron Telescope (CALET) shall be able to measure electrons with energies up to 10 TeV directly \cite{CALET2012}. With such observations we can sample the electron distribution at Earth only. Observations of very high energy Galactic diffuse $\gamma$-ray emission might allow us to indirectly probe the global high energy electron population and infer properties of the Galactic electron CR source population. Unfortunately, the investigations of the very high energy component is complicated by the fact that the Galactic diffuse $\gamma$-ray emission depends on many properties of the Milky Way beside the CR source distribution such as e.g. gas distribution, stellar radiation fields, magnetic fields. Furthermore, the interpretation of Galactic diffuse $\gamma$-ray emission data relies on numerical simulations with realistic descriptions of all the governing properties. Presently, no simulation exists that meets all these requirements. Accurately modelling the Galactic diffuse $\gamma$-ray emission remains an ultimate test for the modelling of CR propagation in the Milky Way.

When compared to the \textit{Reference}-Model the deviations of low energy electrons and low energy protons are not as pronounced as those found for high energy electrons but can still reach levels exceeding 100-percent in the \textit{Dame}-Model and the \textit{Steinman}-Model. The enhancement of low energy electrons is also the main influence of the presence of a Galactic bar on the electron spectrum at Earth. Our finding for low energy protons may be of interest for models of the heliosphere and its interaction with the ISM for which the local flux of Galactic CRs is an input parameter \cite{Florinski2013}. 

We find a significant variation of the total proton flux when comparing on-arm and inter-arm locations on the $r=R_{\odot}$ Sun's Galactic orbit. This enhancement reaches a factor of 4.3 in case of the \textit{Dame}-Model and a factor of 2.3 in case of the \textit{Steiman}-Model. The values we obtain using the four-arm \textit{Steiman}-Model are compatible with \cite{Effenberger2012} (enhancement factor approx. 1.9; see Figure 5 therein). The amplitude of the CR variation also compares well with \cite{Shaviv2003} (enhancement factor approx. 4.2; see Figure 7 therein). When using the \textit{Mix}-Model even small spiral arm contributions have a significant effect on the protons flux when comparing on-arm and inter-arm positions. For the \textit{NE2001}-Model we see no local variations that are associated with the spiral arms; because of the local spiral arm segment (see Figure \ref{FigCRFluxNearEarth}) the integral flux is highest at the nominal position of the Earth and due to the broad spiral arms the protons have diffused sufficiently to homogenize any spatial variations along Sun's orbit around the Galactic Center. We have not considered any dynamics of the spiral arms that will potentially lead to a stronger homogenization of the total proton flux due to the effective smearing of the underlying CR spiral arm distribution.

The shape of the high energy proton $E_{kin} > 10$ GeV spectrum at Earth's nominal position does not show any deviations when comparing the spiral arm source models with the \textit{Reference}-Model. High energy protons have a mean energy loss length that is larger than the underlying structure of the spiral arm source models and therefore their propagation is not influenced noticeably by it. This is also apparent when comparing the proton spectra at on-arm positions to spectra at inter-arm positions in which case we only obtain a different total proton flux while the spectral shape shows no discernible changes. Accordingliy, the predicted change in the on-arm and inter-arm spectral indices as of \cite{Rogers1988} is not evident in our results.

Non-axisymmetric source distributions can, at least qualitatively, reproduce features in the CR spectrum at the Earth, such as the so called positron-anomaly \cite{Pamela2009,AMS2013,Gaggero2013b}, that were often attributed to processes such as dark matter annihilation \cite{Delahaye2008}. 

The work presented herein only discusses CR source distributions that are associated with the spiral arms. There is considerable evidence that other properties of the Milky Way that influence the propagation, such as, among many, the magnetic field, the gas distribution, the stellar radiation fields, are also connected to the spiral arm structure. Consequently, this will also influence the CR distribution in our Galaxy.

\section{Conclusion}
\label{conc}
Using the \codename code we have shown that CR source distributions that follow the spiral structure of the Milky Way leave a distinct imprint in the proton and electron spectrum at the Earth. We find that the magnitude of the imprint depends on the spiral model characteristics such as the Earth's location relative to the spiral arms, the number of arms and width of the spiral arms. Particularly electrons with energies greater than 100 GeV are of interest since they are confined inside their source region and therefore trace the CR source distribution closely. When compared to an axis-symmetric source distribution spiral arms lead to a deficit in high energy electrons at the Earth. This should be observable from the Earth by precision measurements of the electron spectrum at high energies even if the spiral arms only contain a small fraction of the CR sources in our Galaxy. We also quantify the changes of the total proton flux at the Earth along the Sun's orbit around the Galactic Center and find them consistent with previous results using a different treatment of CR propagation physics (e.g. \cite{Effenberger2012}).

We also show that measurements of CR electron and proton spectra alone are not sufficient to distinguish between different source models. Knowledge of the local electron and proton distribution does not allow us to infer the global distributions of these CR species. To overcome this, a global treatment of all CR observables is needed. This includes, but is not limited to, spectra of heavier CRs, CR secondary to primary ratios and neutral messengers, demanding an integrated approach in modelling the propagation of CR in the Milky Way. \codename is well suited for this task. However, an integrated treatment is complicated by the degeneracy of the different propagation parameters (e.g. halo height, diffusion coefficient, magnetic field)  and our lack of knowledge of their precise values in the Milky Way. The interactions of CR with the ISM contribute to the Galactic diffuse $\gamma$-ray emission. Consequently, it remains to be investigated if spiral arms leave a distinct imprint in the diffuse emission. Presently, observations of the Galactic diffuse $\gamma$-ray background constitute the only comprehensive way to probe the global Galactic electron and nucleon distributions and are the most promising avenue to test the CR propagation models. It has become clear that future CR modelling should go beyond the paradigm of an axisymmetric CR source distribution and investigate compound non-axisymmetric source distributions to see which effect arise that can be tested against data from current and future CR experiments.

\section*{Acknowledgements}
This research has made use of NASA's Astrophysics Data System. Support by the Austrian Ministry of Science, Research and Economics BMWFW as part of the UniInfrastrukturprogramm of the Research Platform Scientific Computing at the University of Innsbruck and the Austrian Science Fund (FWF) supported Doctorate School DK+ W1227-N16 is acknowledged. M. Werner would like to thank the Max Planck Institute for Extraterrestrial Physics for its hospitality while hosting him as a guest.



\section*{References}

\bibliography{References}

\clearpage
\onecolumn


\begin{figure}[h]
    \setlength{\unitlength}{0.001\textwidth}
    \begin{subfigure}{500\unitlength}
        \begin{picture}(500,600)
        		\put(220,0){x}
        		\put(0,350){y}
        		\put(0,80){z}
        		\put(27,20){\includegraphics[trim=0cm 2cm 0cm 0cm, clip=true,width=410\unitlength]{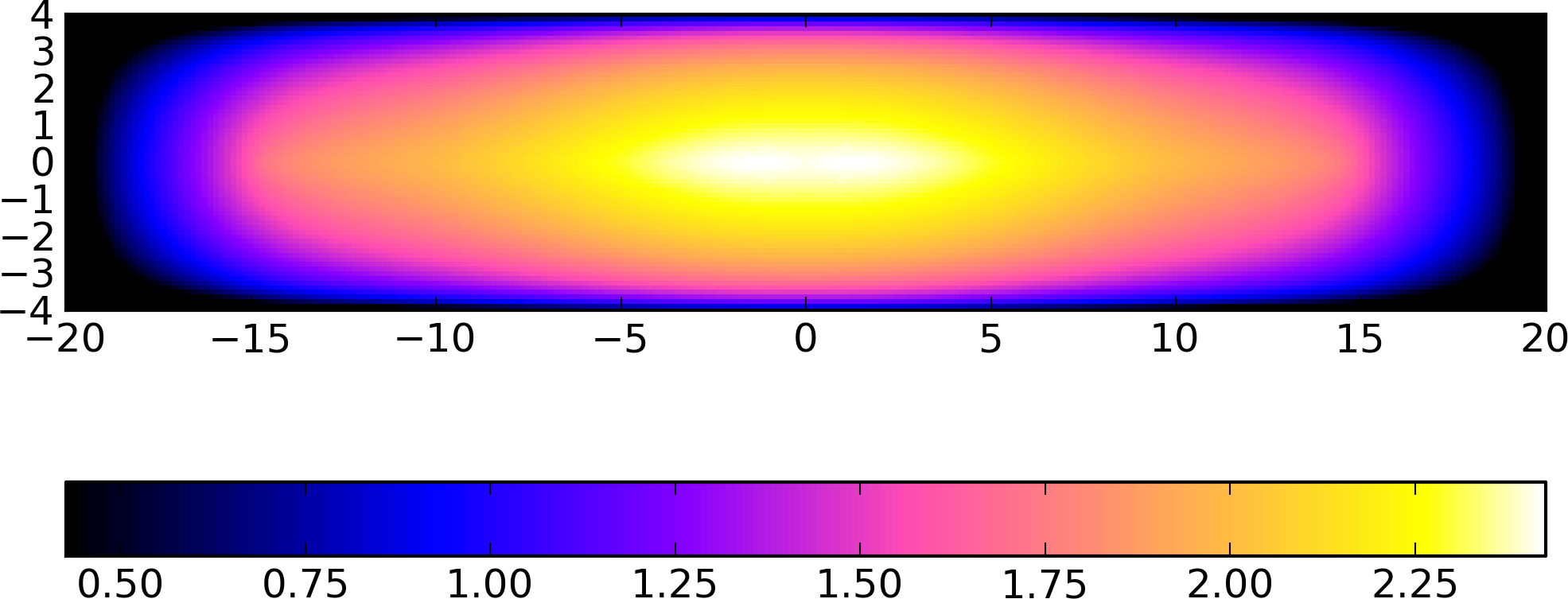}}
            \put(15,150){\includegraphics[width=500\unitlength]{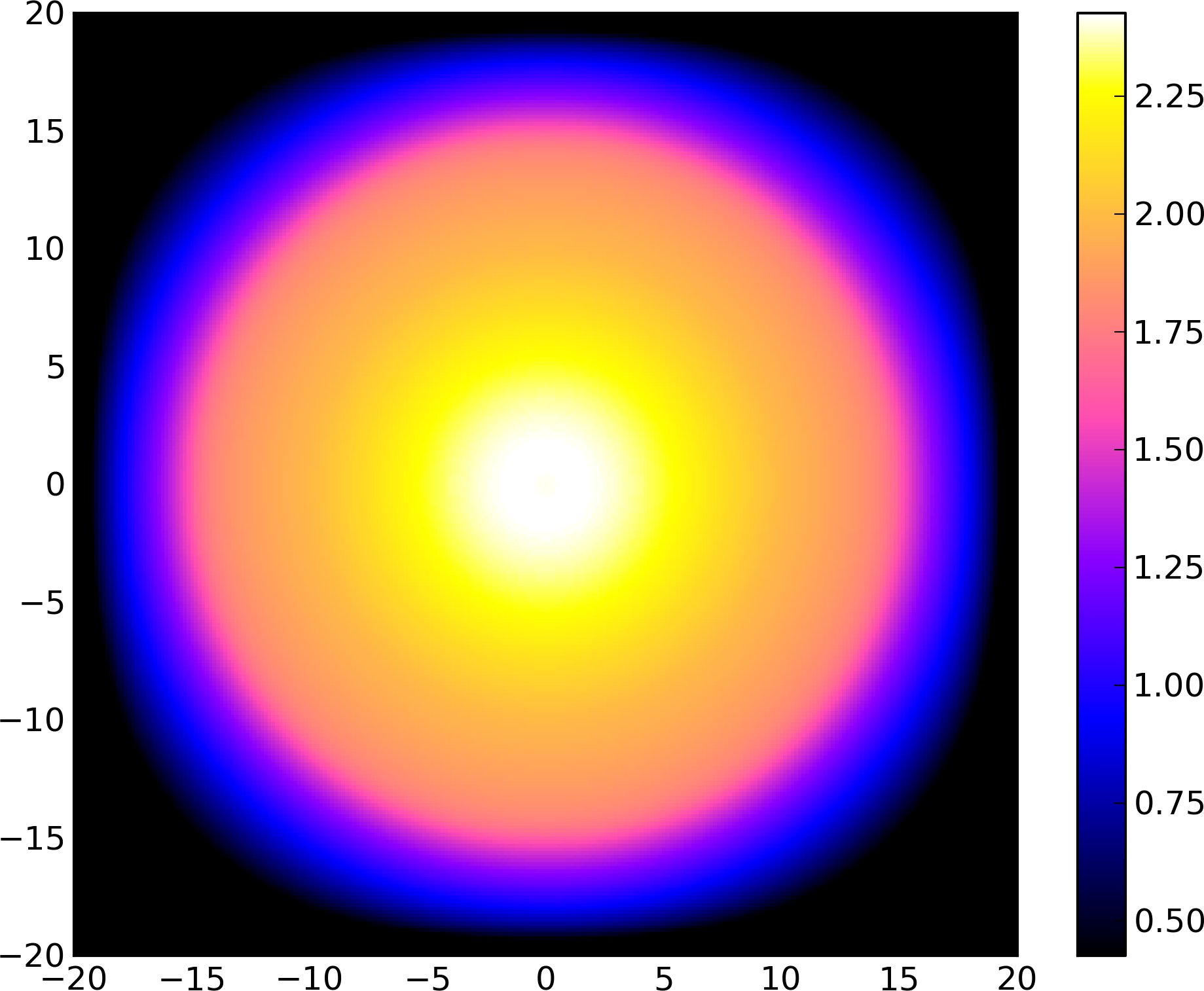}}
        \end{picture}
    \end{subfigure}
 	\quad
     \begin{subfigure}{500\unitlength}
        \begin{picture}(500,600)
        		\put(220,0){x}
        		\put(0,350){y}
        		\put(0,80){z}
        		\put(27,20){\includegraphics[trim=0cm 2cm 0cm 0cm, clip=true,width=410\unitlength]{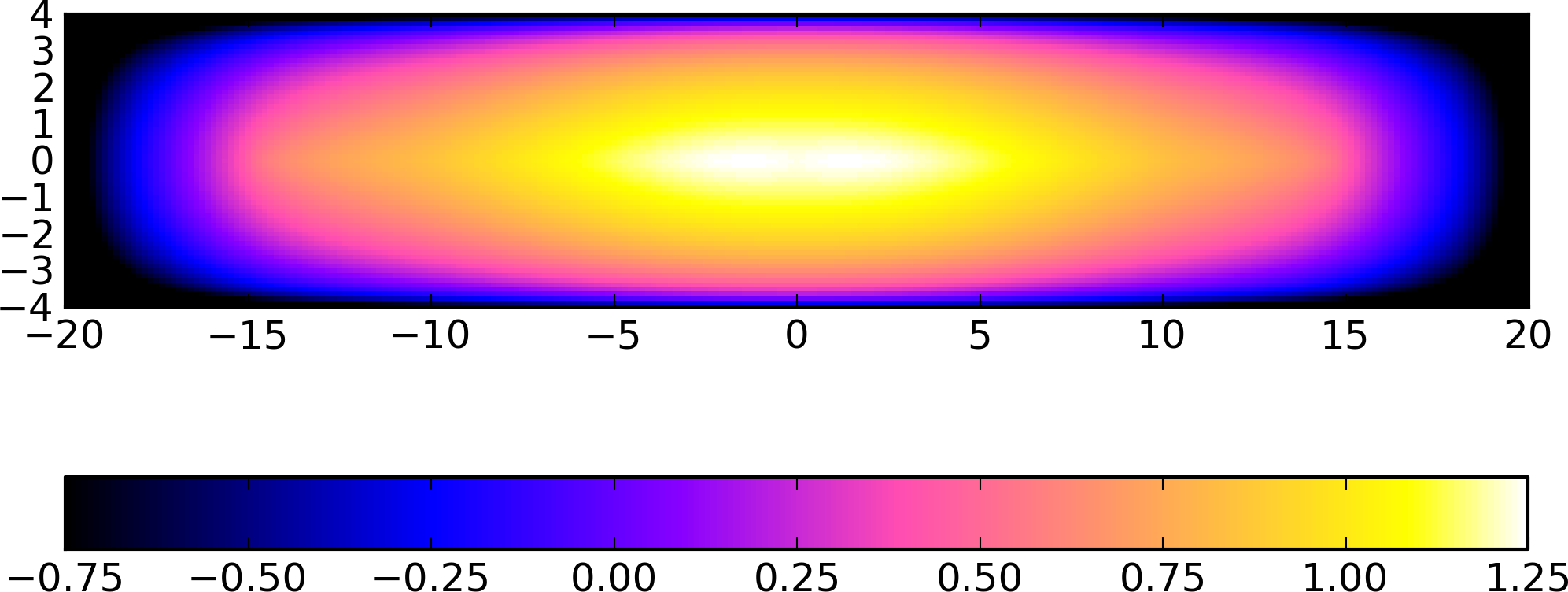}}
            \put(15,150){\includegraphics[width=500\unitlength]{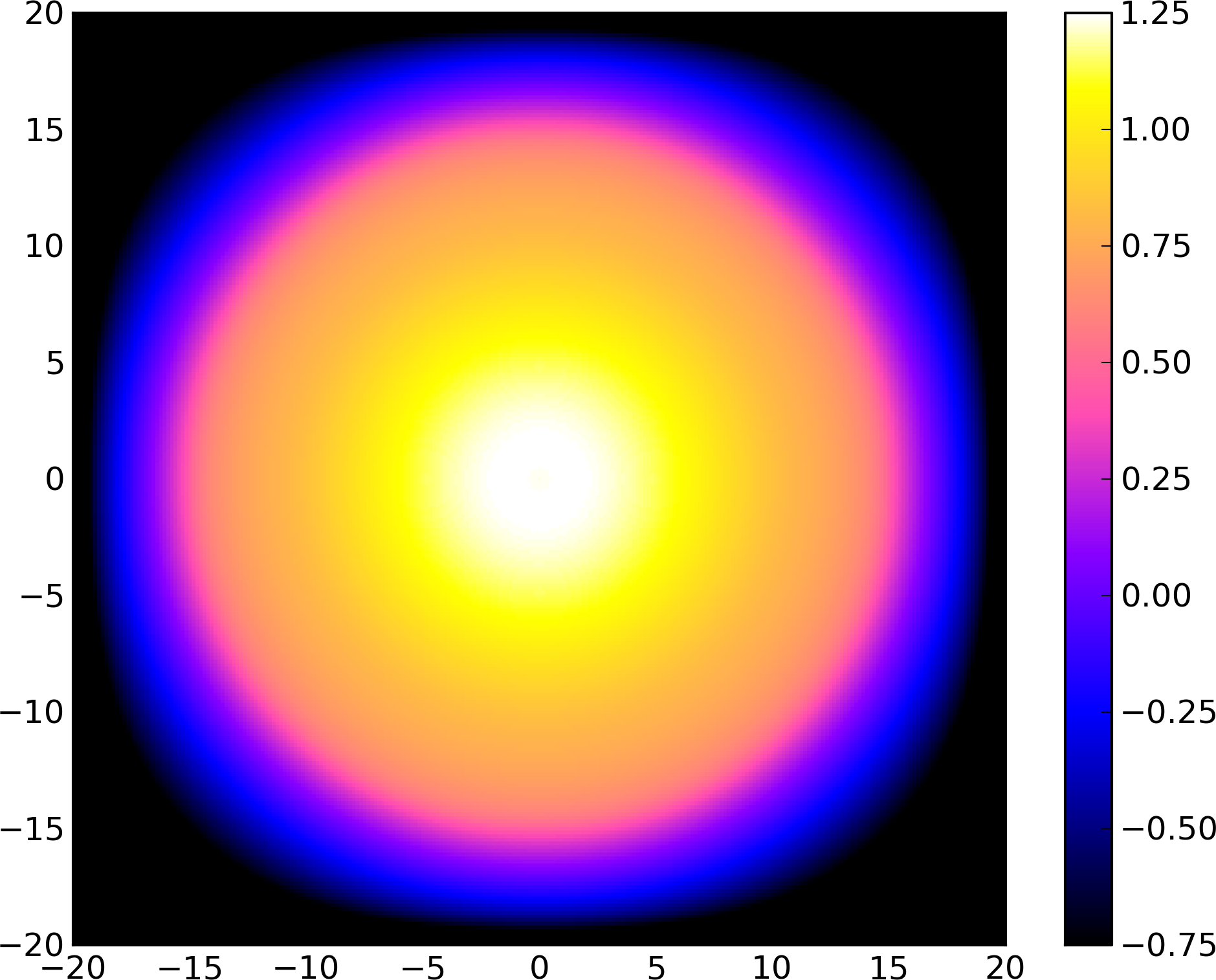}}
        \end{picture}
    \end{subfigure}\\
        \begin{subfigure}{500\unitlength}
        \begin{picture}(500,600)
            	\put(220,0){x}
        		\put(0,350){y}
        		\put(0,80){z}
        		\put(27,20){\includegraphics[trim=0cm 2cm 0cm 0cm, clip=true,width=410\unitlength]{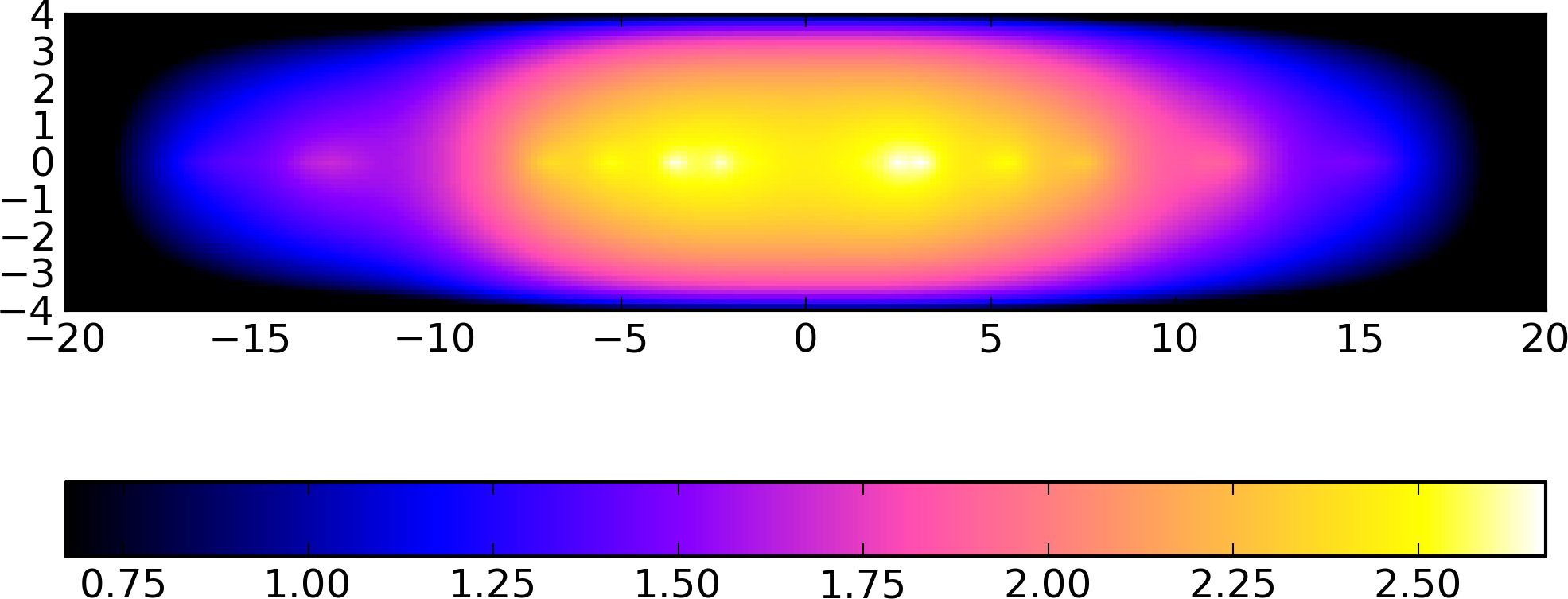}}
            \put(15,150){\includegraphics[width=500\unitlength]{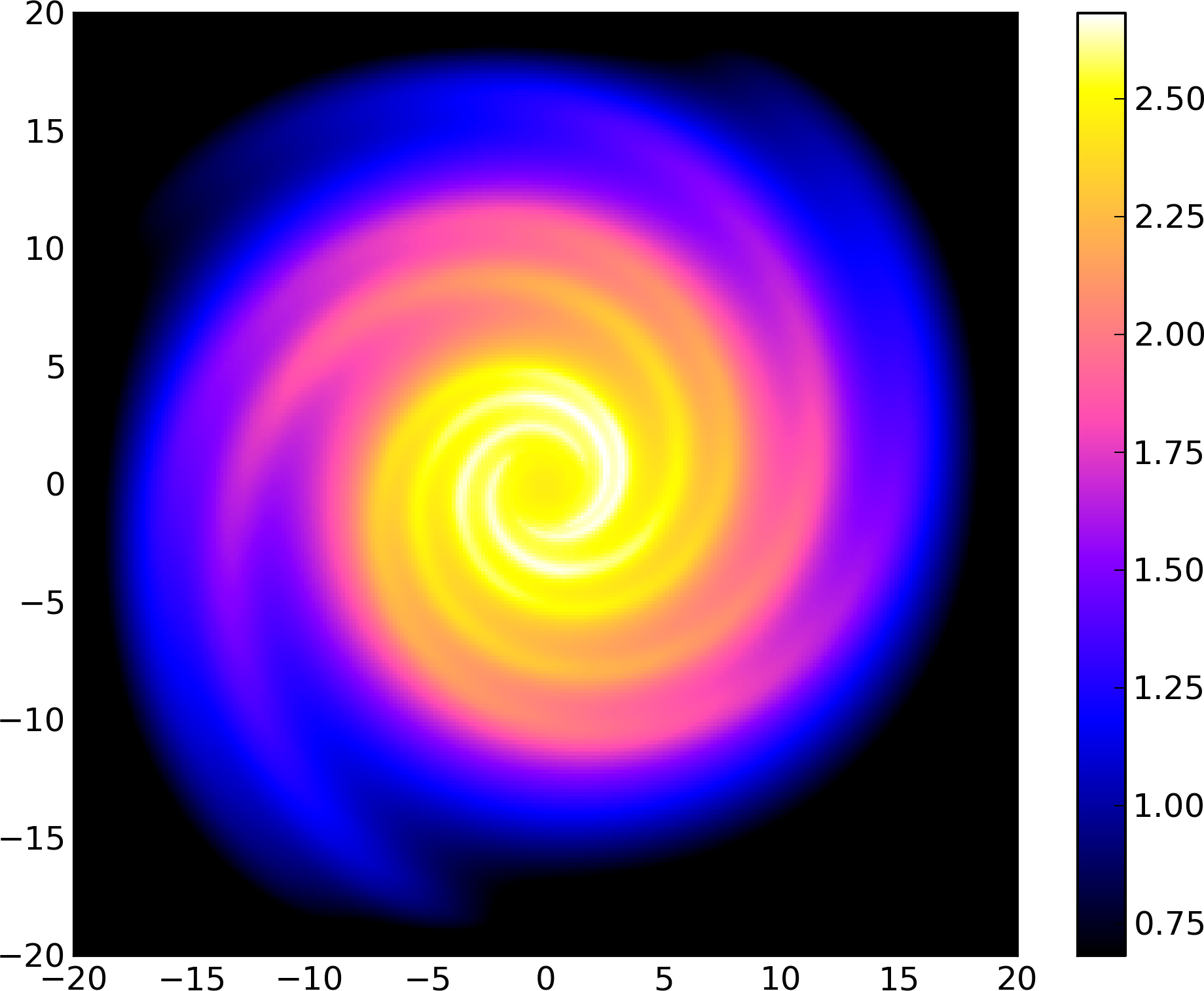}}
        \end{picture}
    \end{subfigure}
 	\quad
     \begin{subfigure}{500\unitlength}
        \begin{picture}(500,600)
        		 \put(220,0){x}
        		\put(0,350){y}
        		\put(0,80){z}
        		\put(27,20){\includegraphics[trim=0cm 2cm 0cm 0cm, clip=true,width=410\unitlength]{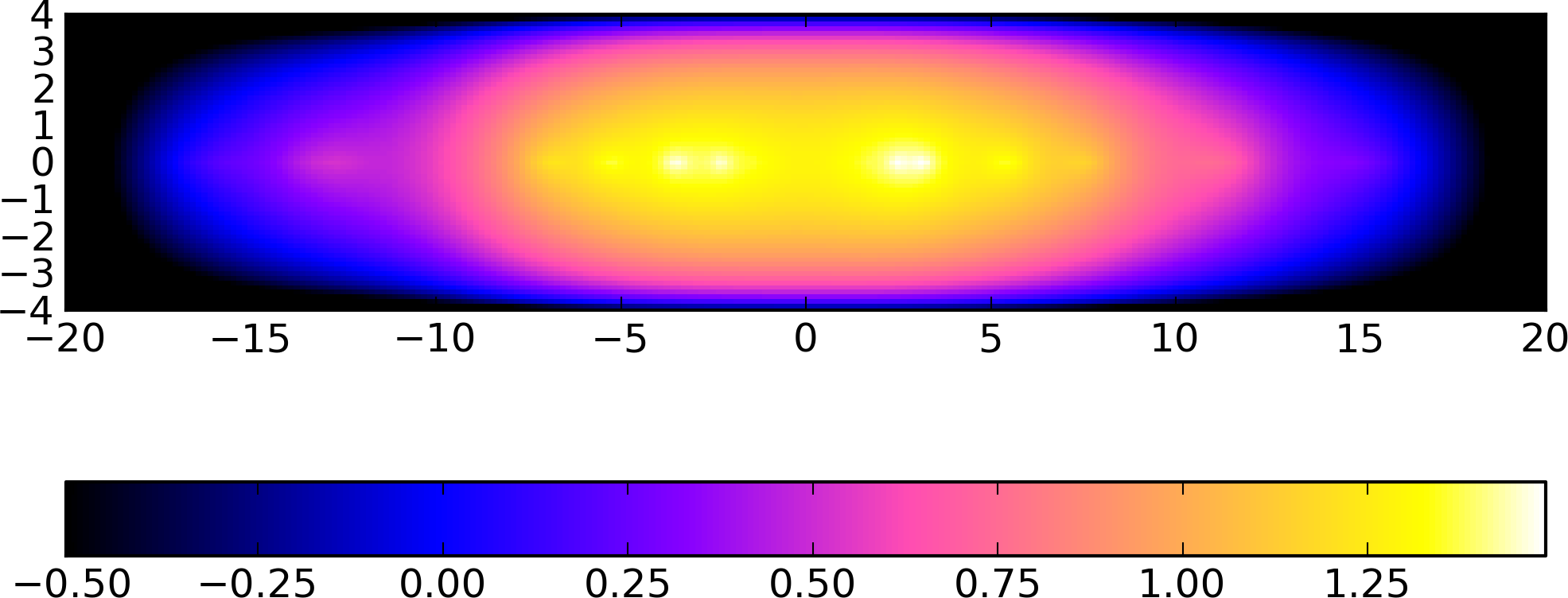}}
            \put(15,150){\includegraphics[width=500\unitlength]{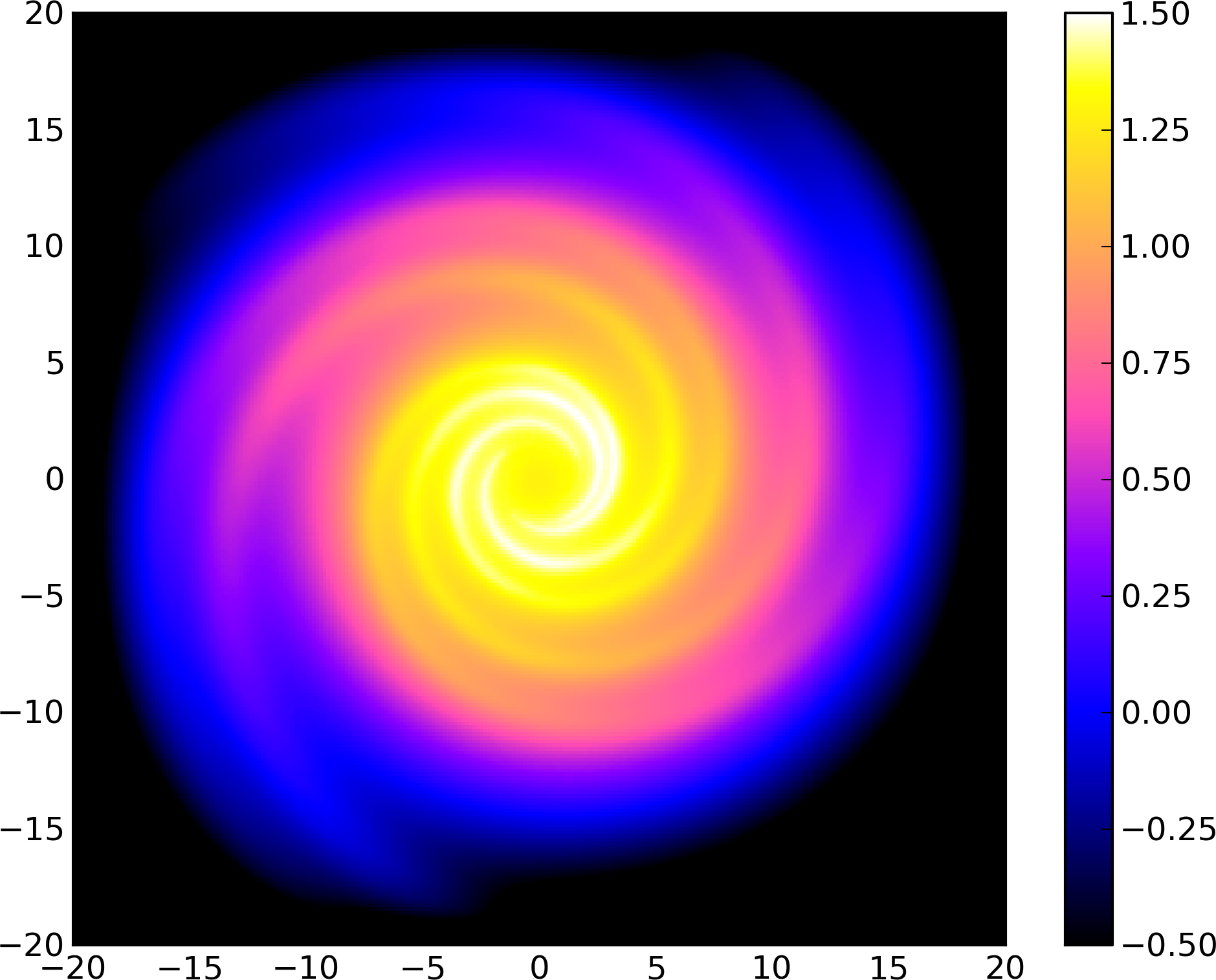}}
        \end{picture}
    \end{subfigure}
\caption{}
\end{figure}
\clearpage
\begin{figure}[h]
	\ContinuedFloat
    \setlength{\unitlength}{0.001\textwidth}
    \begin{subfigure}{500\unitlength}
        \begin{picture}(500,600)
        		\put(220,0){x}
        		\put(0,350){y}
        		\put(0,80){z}
        		\put(27,20){\includegraphics[trim=0cm 2cm 0cm 0cm, clip=true,width=410\unitlength]{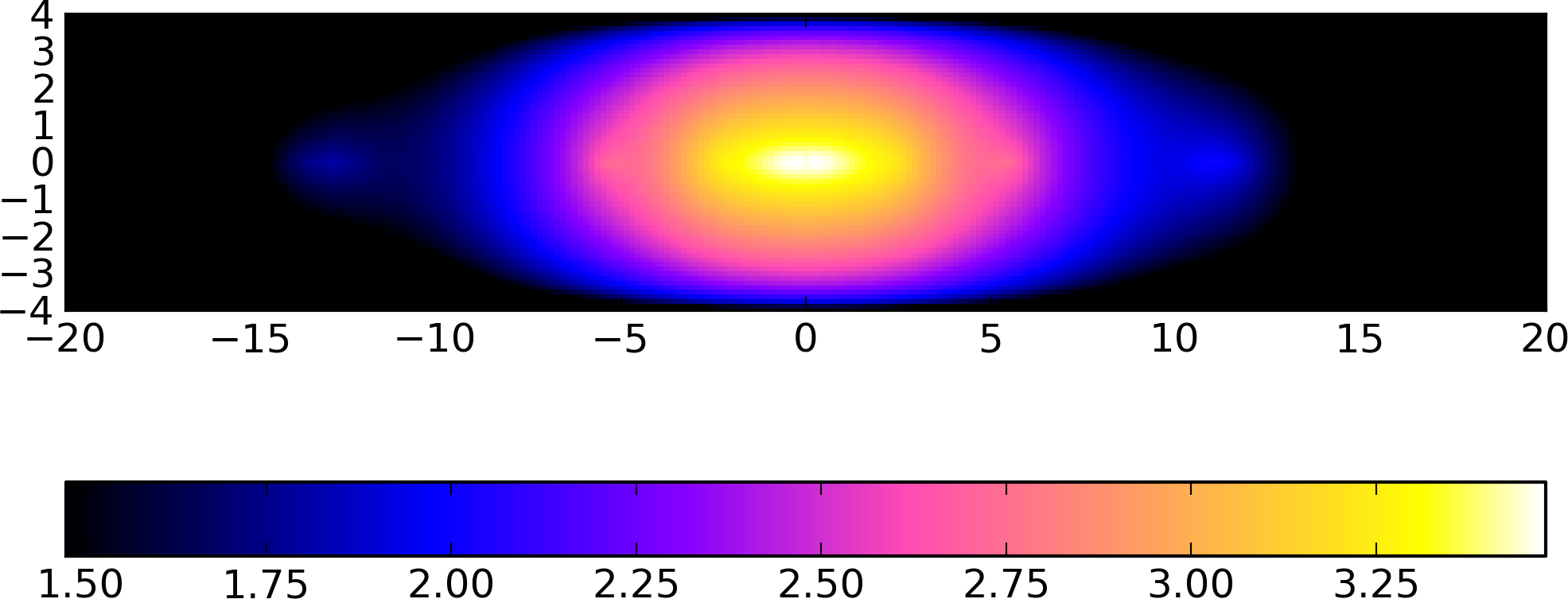}}
            \put(15,150){\includegraphics[width=500\unitlength]{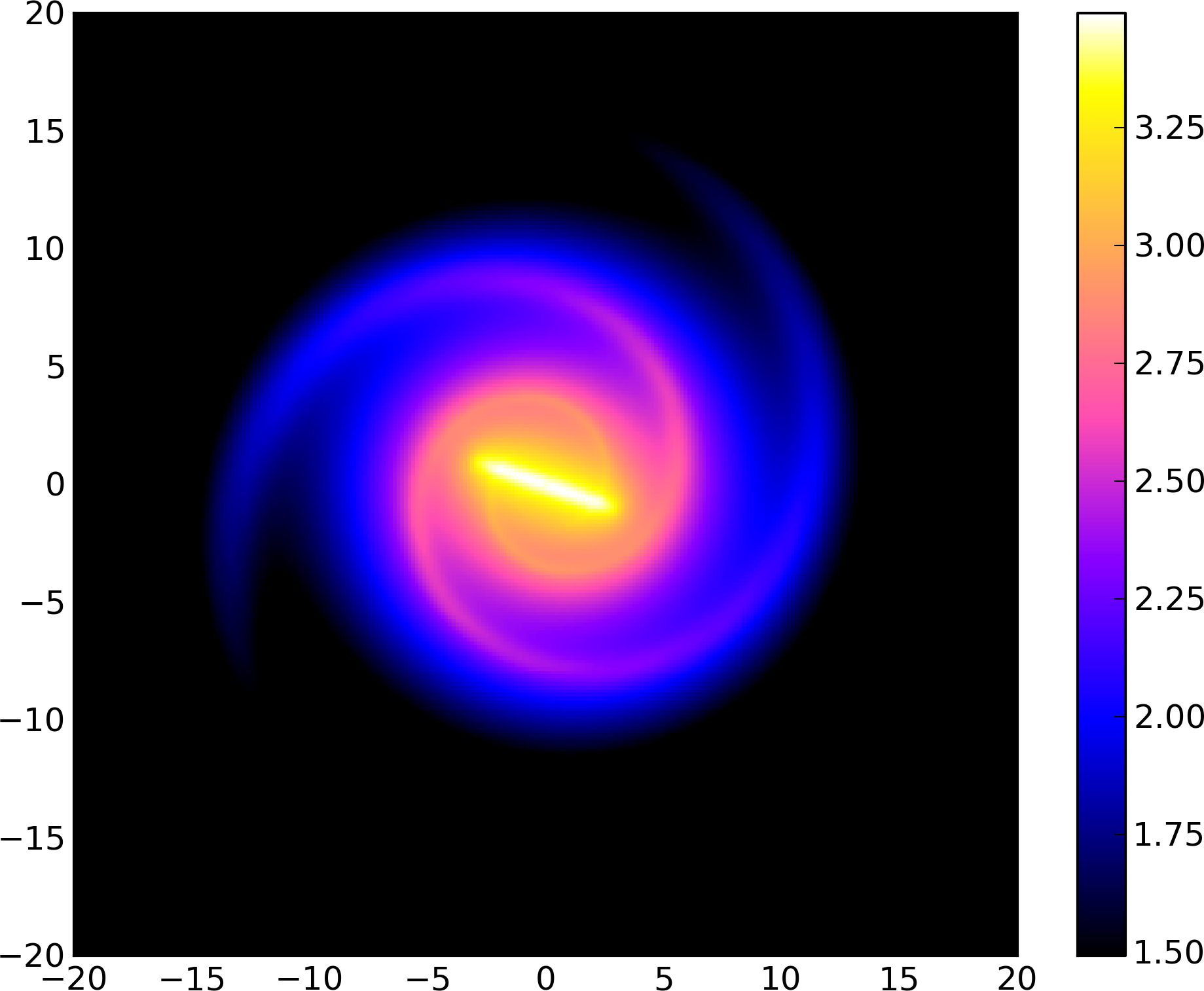}}
        \end{picture}
    \end{subfigure}
 	\quad
     \begin{subfigure}{500\unitlength}
        \begin{picture}(500,600)
        		\put(220,0){x}
        		\put(0,350){y}
        		\put(0,80){z}
        		\put(27,20){\includegraphics[trim=0cm 2cm 0cm 0cm, clip=true,width=410\unitlength]{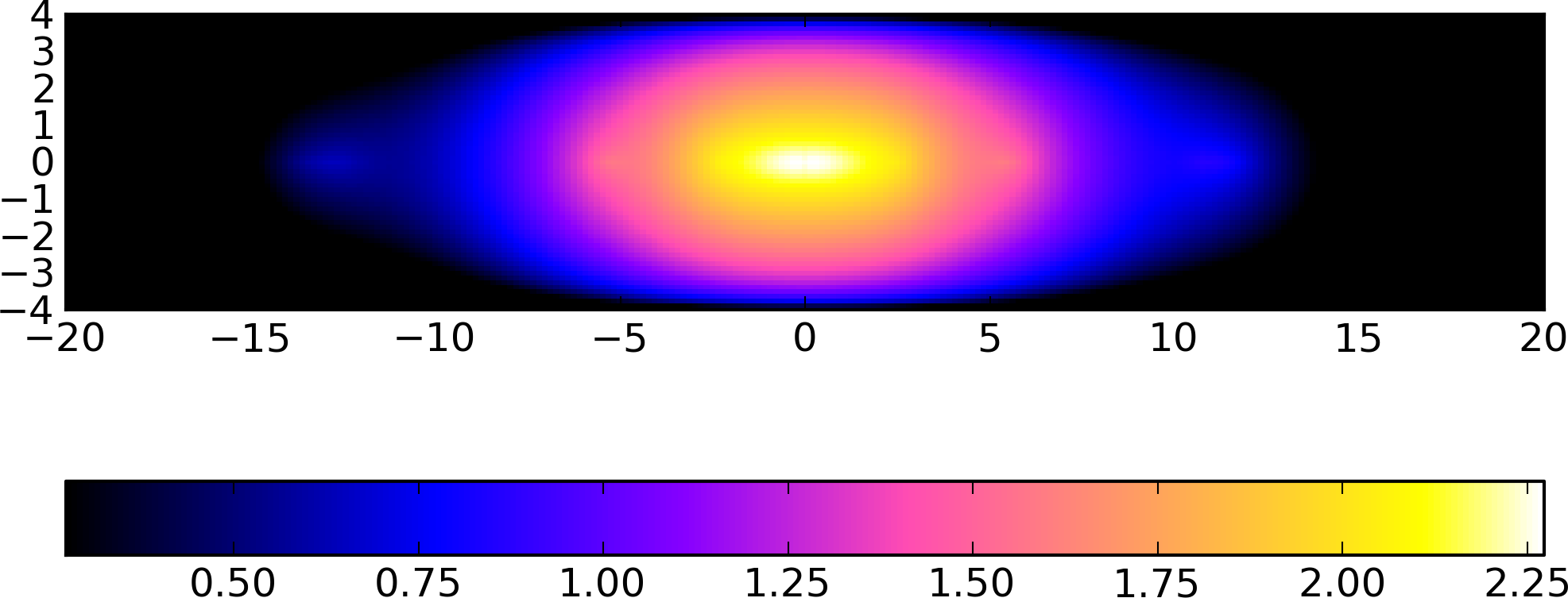}}
            \put(15,150){\includegraphics[width=500\unitlength]{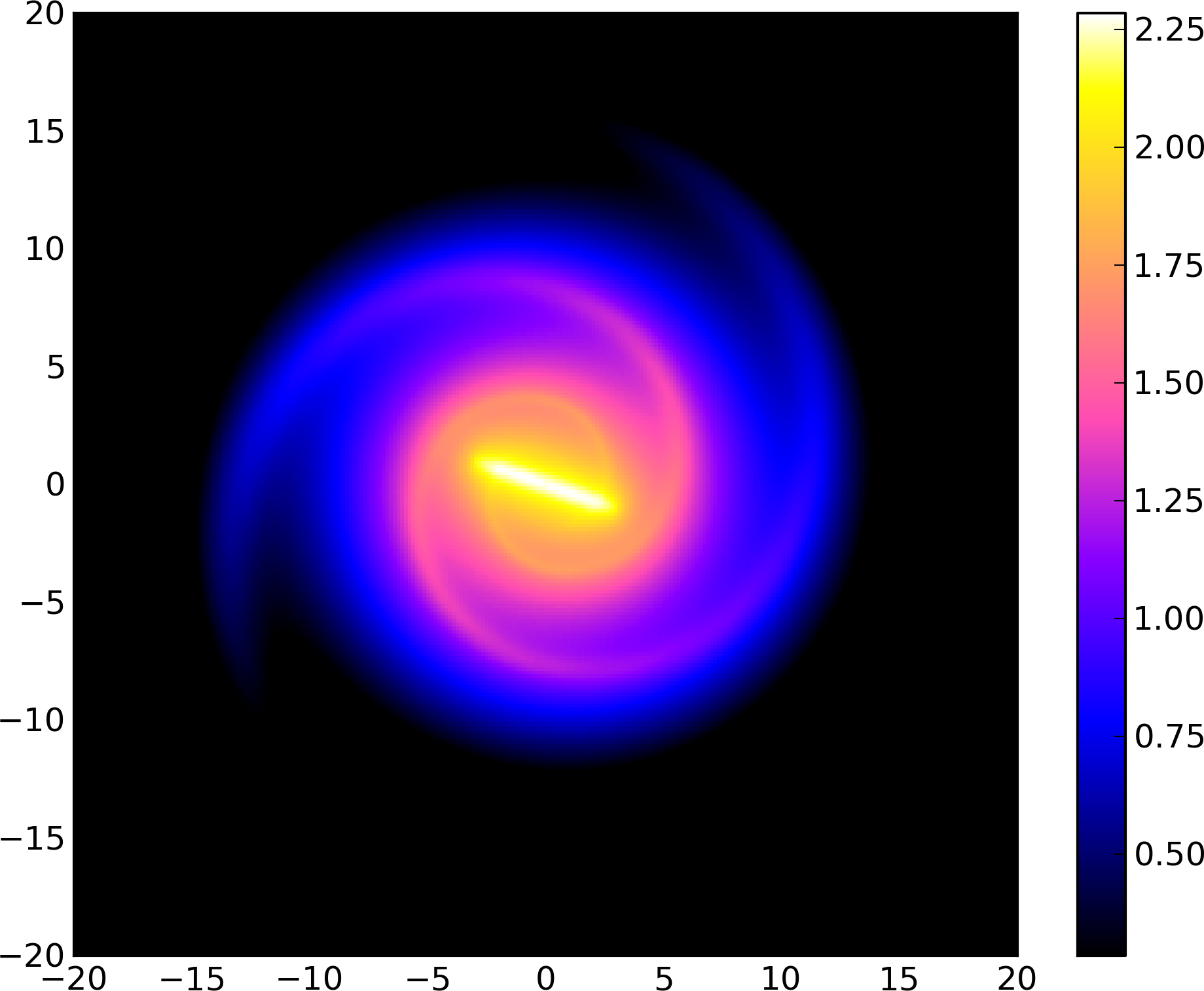}}
        \end{picture}
    \end{subfigure}\\
        \begin{subfigure}{500\unitlength}
        \begin{picture}(500,600)
        		 \put(220,0){x}
        		\put(0,350){y}
        		\put(0,80){z}
        		\put(27,20){\includegraphics[trim=0cm 2cm 0cm 0cm, clip=true,width=410\unitlength]{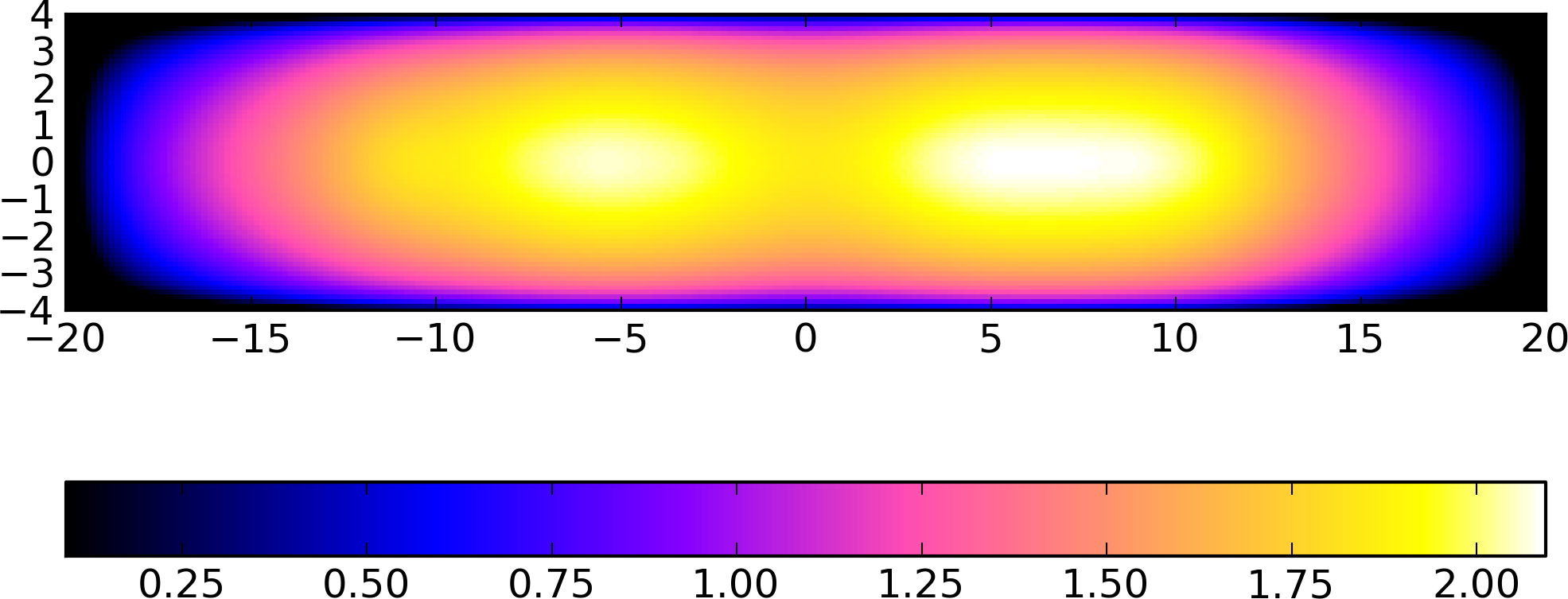}}
            \put(15,150){\includegraphics[width=500\unitlength]{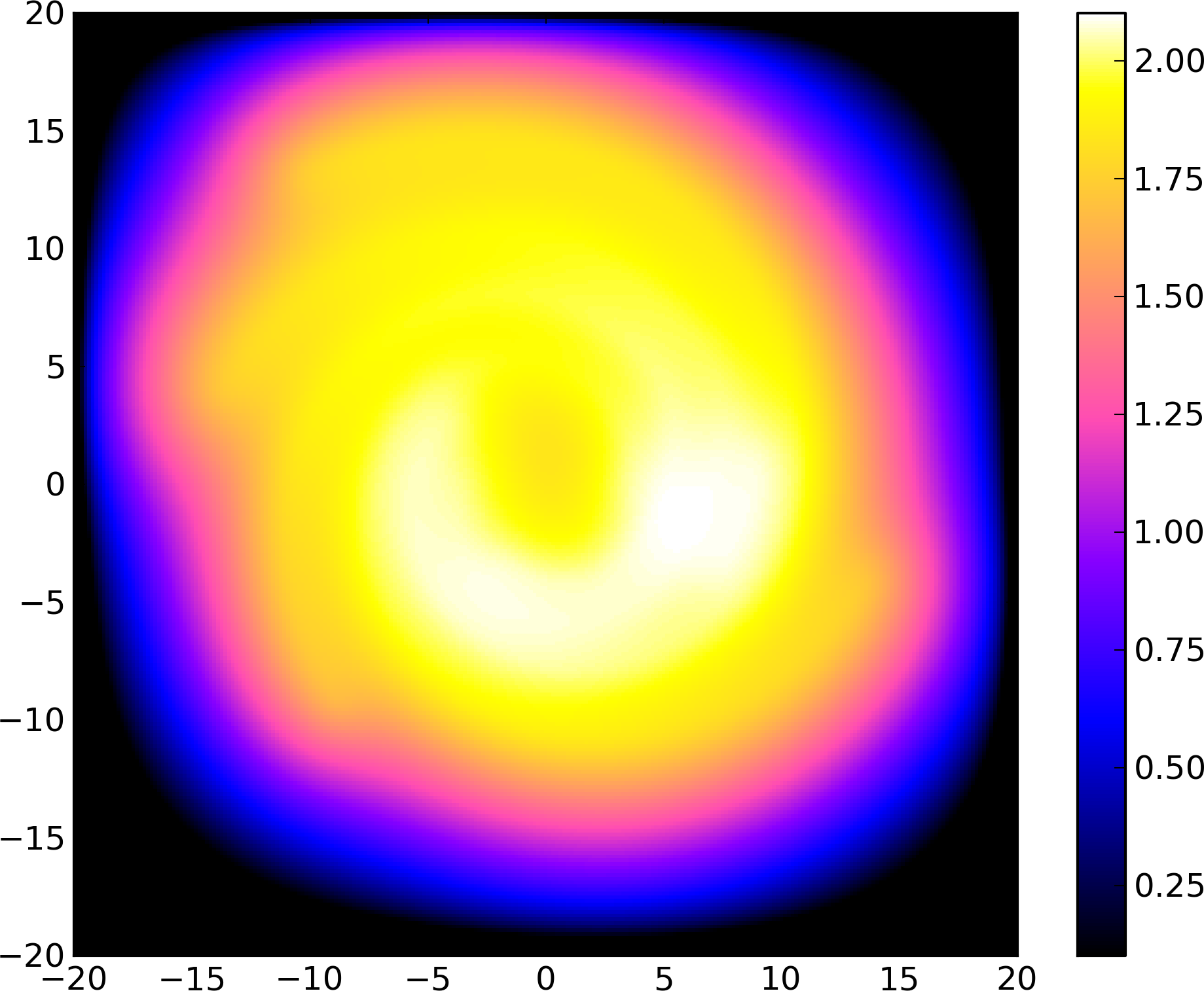}}
        \end{picture}
    \end{subfigure}
 	\quad
     \begin{subfigure}{500\unitlength}
        \begin{picture}(500,600)
        		\put(220,0){x}
        		\put(0,350){y}
        		\put(0,80){z}
        		\put(27,20){\includegraphics[trim=0cm 2cm 0cm 0cm, clip=true,width=410\unitlength]{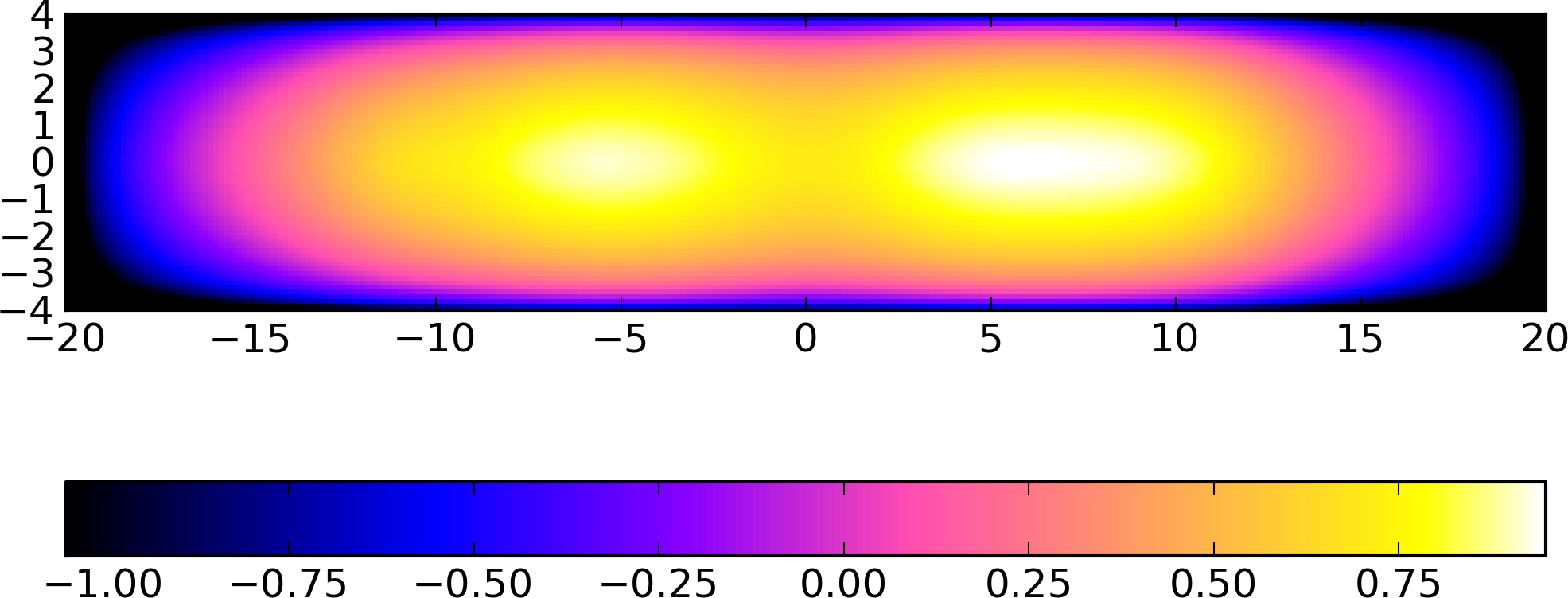}}
            \put(15,150){\includegraphics[width=500\unitlength]{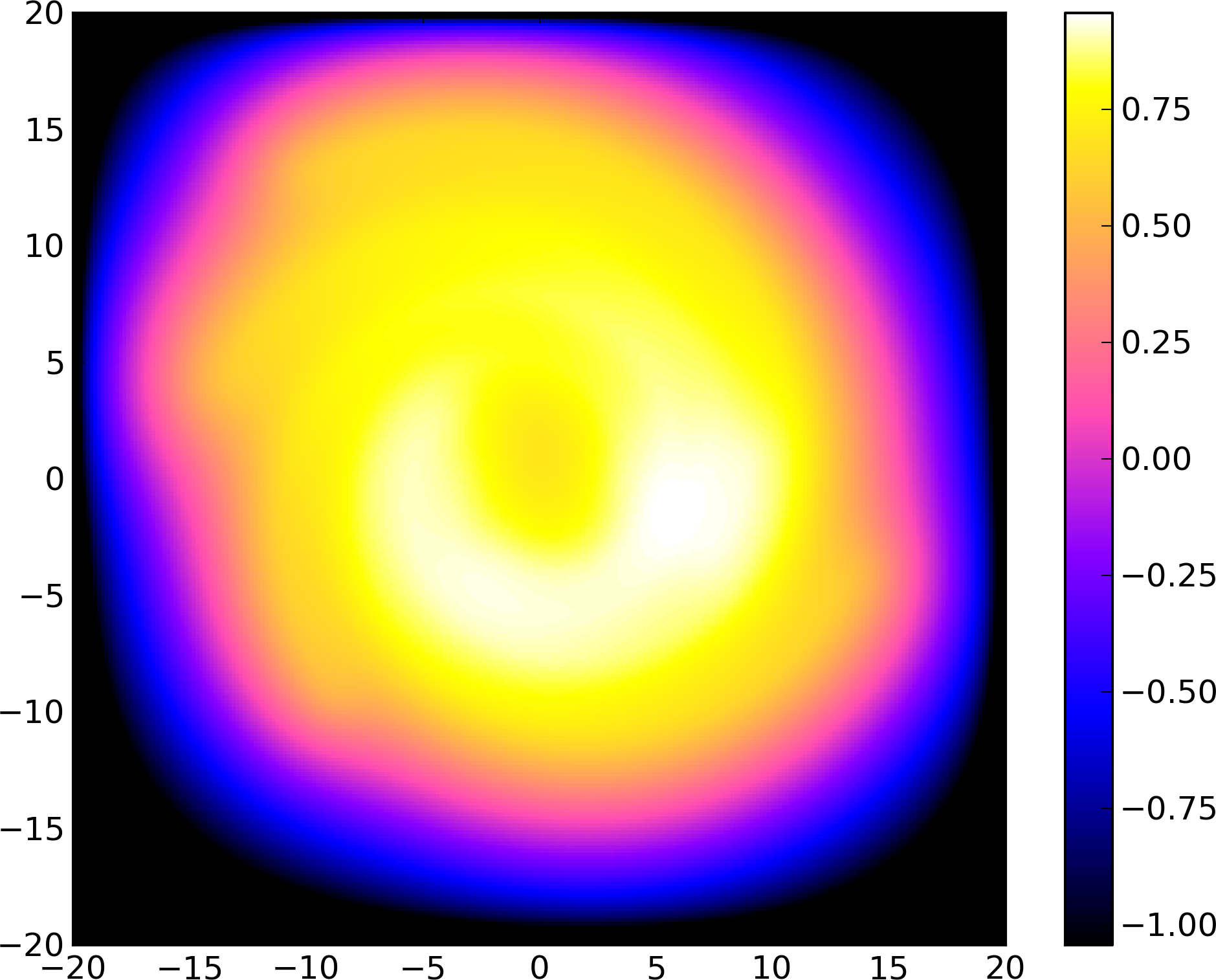}}
        \end{picture}
    \end{subfigure}
\caption{Depiction of the proton density distribution using the four models described in Section \ref{spirals} as CR source distributions. Each row contains the simulation results of the respective model. From top to bottom: \textit{Reference}-Model, \textit{Steiman}-Model, \textit{Dame-Model} and \textit{NE2001}-Model. The first column contains the $x$,$y$-slices at $z = 0$ and $x$,$z$-slices for $y = 0$ of the proton density distribution at a kinetic energy $E_{kin} = 1.1$ GeV. The second column contains the same but for a kinetic energy of $E_{kin} = 1.1$ TeV. X,y,z dimensions are in units of kpc. The logarithmic color scale represents the normalized proton density in arbitrary units. No smoothing is applied, pixels represent the actual computational grid.}
\label{Hdistributions}
\end{figure}

\begin{figure}[h]
    \setlength{\unitlength}{0.001\textwidth}
    \begin{subfigure}{500\unitlength}
        \begin{picture}(500,600)
        		\put(220,0){x}
        		\put(0,350){y}
        		\put(0,80){z}
        		\put(27,20){\includegraphics[trim=0cm 2cm 0cm 0cm, clip=true,width=410\unitlength]{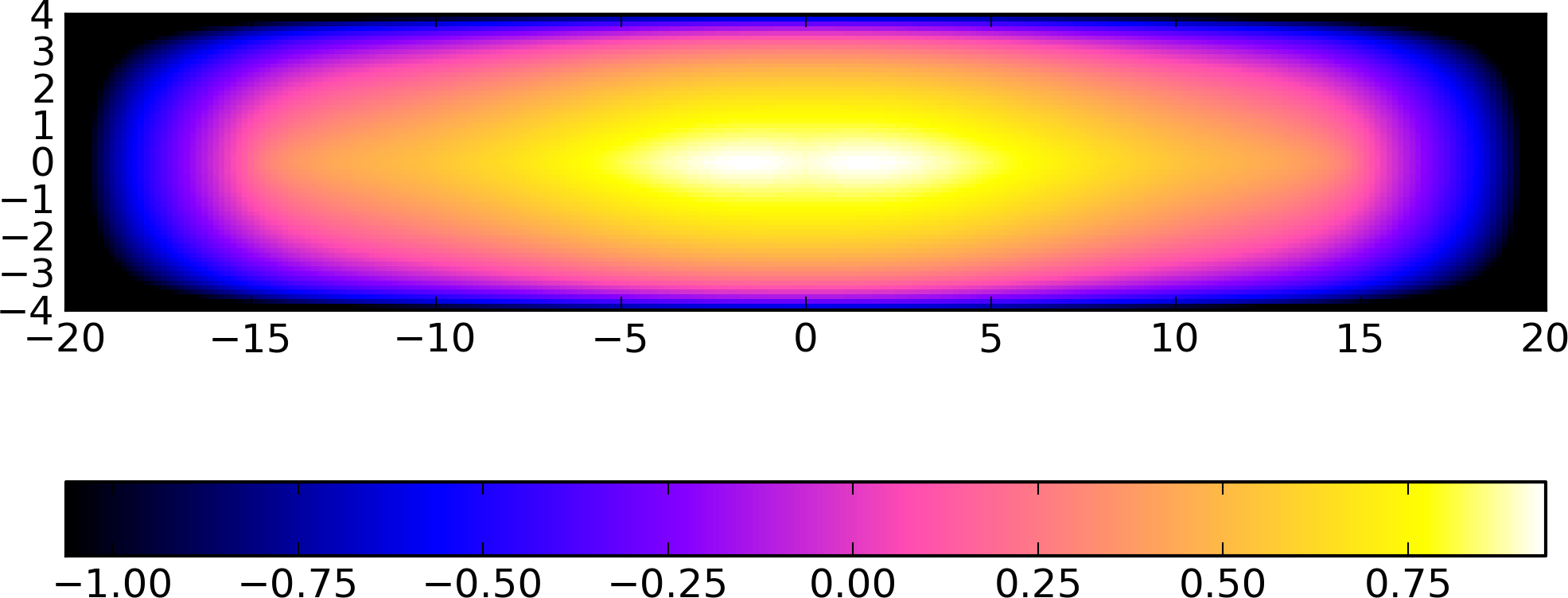}}
            \put(15,150){\includegraphics[width=500\unitlength]{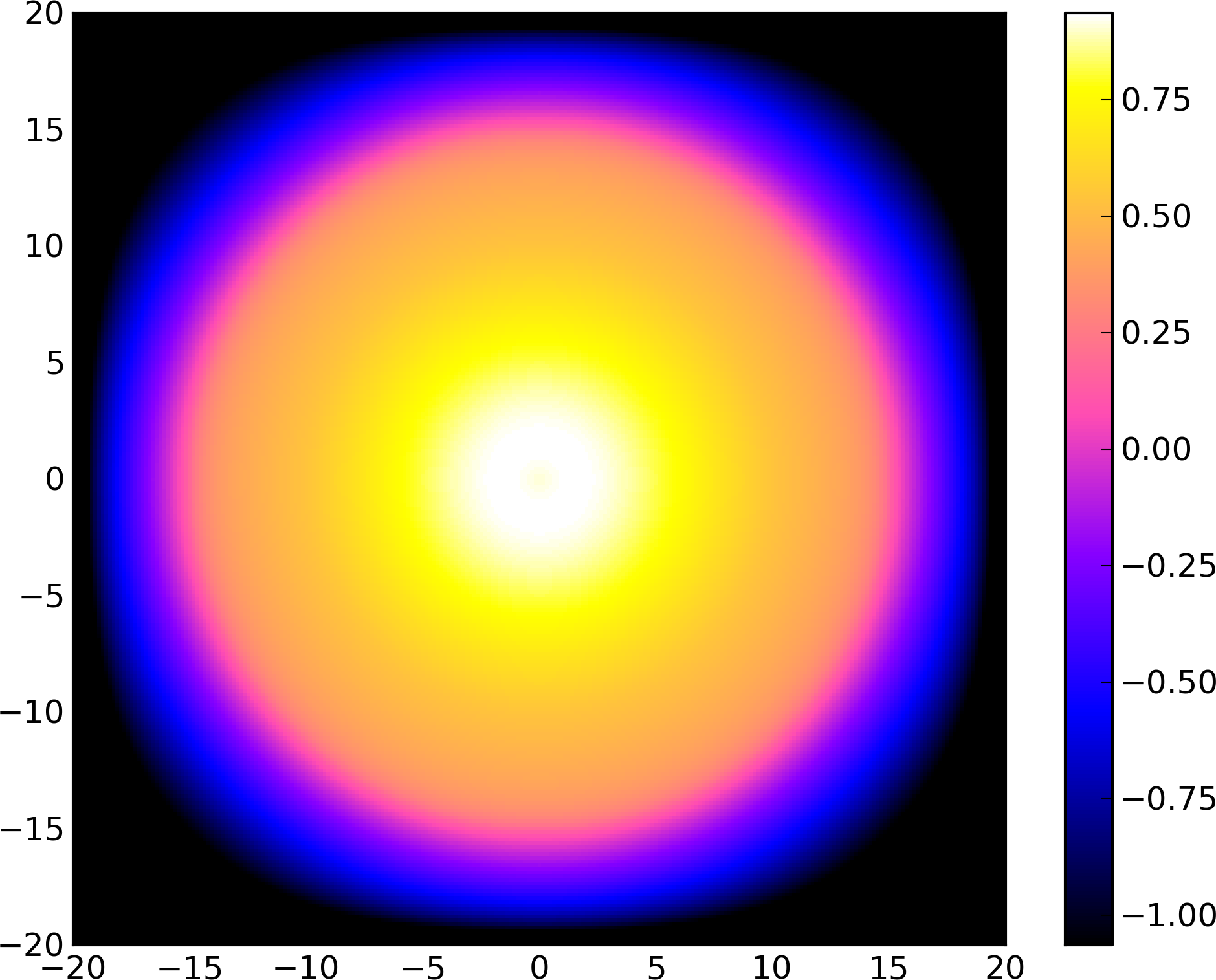}}
        \end{picture}
    \end{subfigure}
 	\quad
     \begin{subfigure}{500\unitlength}
        \begin{picture}(500,600)
        		\put(220,0){x}
        		\put(0,350){y}
        		\put(0,80){z}
        		\put(27,20){\includegraphics[trim=0cm 2cm 0cm 0cm, clip=true,width=410\unitlength]{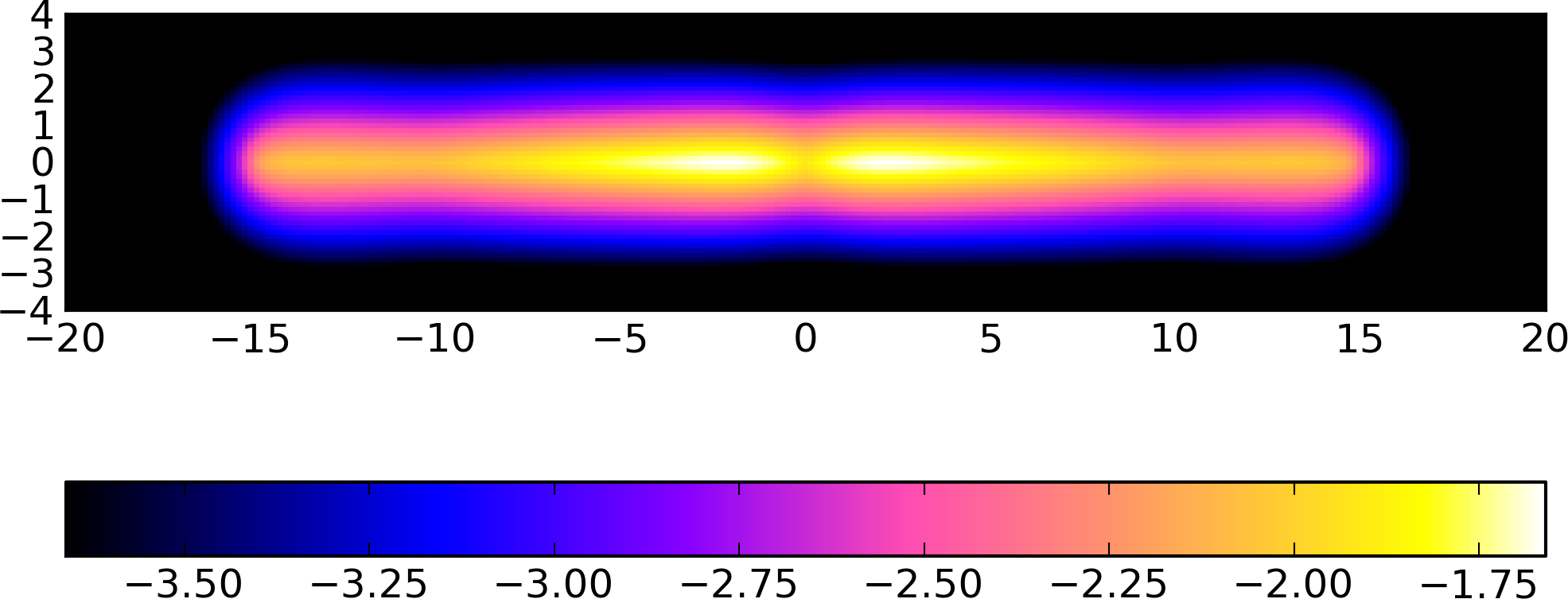}}
            \put(15,150){\includegraphics[width=500\unitlength]{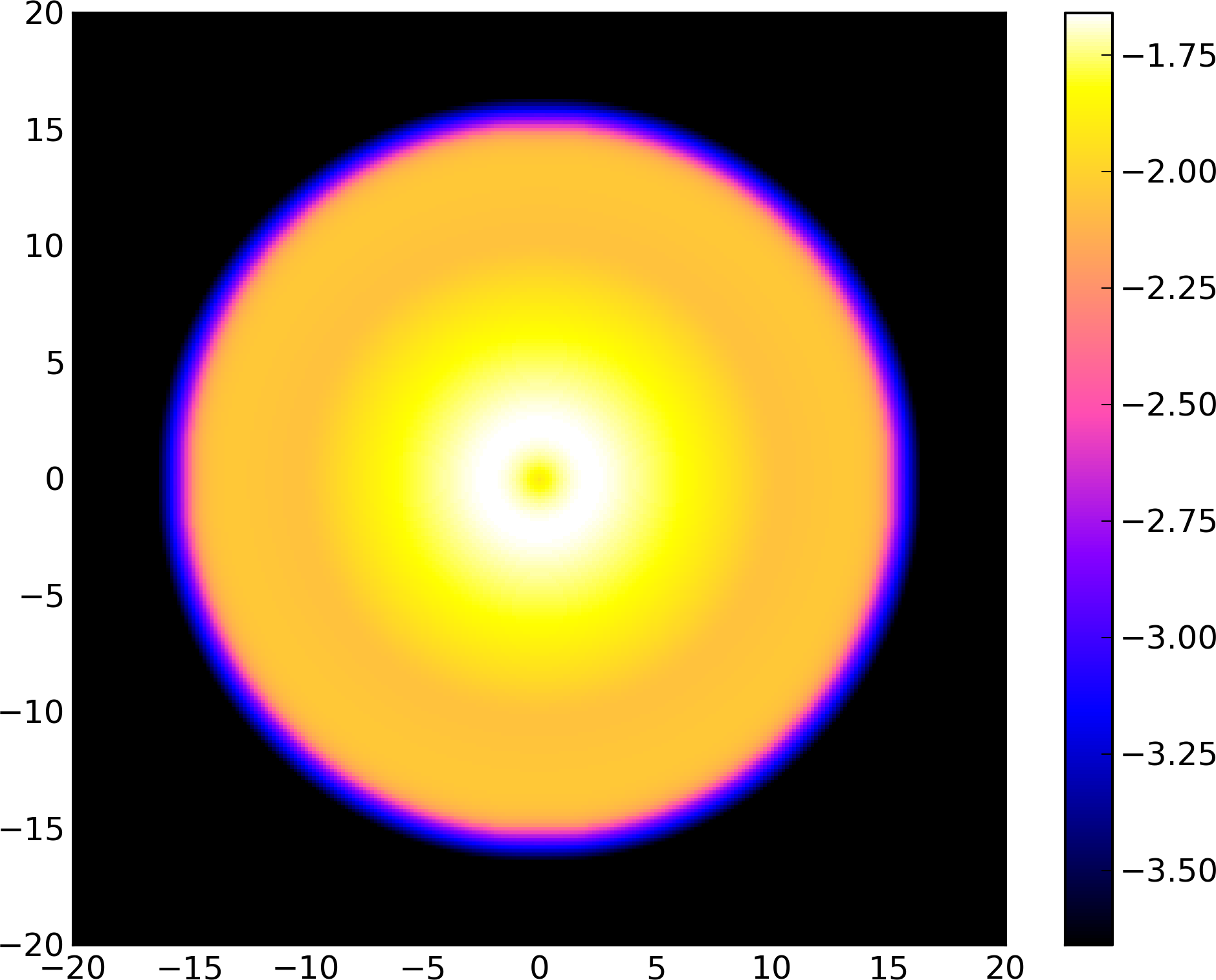}}
        \end{picture}
    \end{subfigure}\\
        \begin{subfigure}{500\unitlength}
        \begin{picture}(500,600)
        		\put(220,0){x}
        		\put(0,350){y}
        		\put(0,80){z}
        		\put(27,20){\includegraphics[trim=0cm 2cm 0cm 0cm, clip=true,width=410\unitlength]{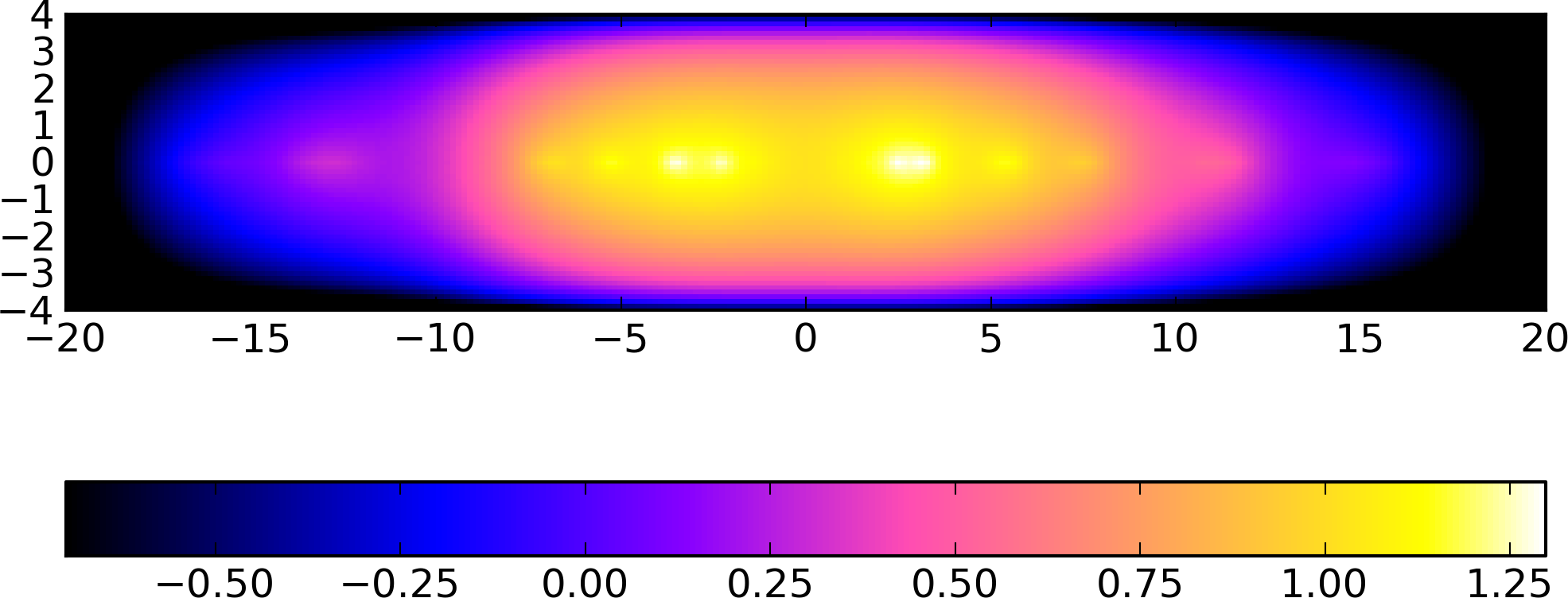}}
            \put(15,150){\includegraphics[width=500\unitlength]{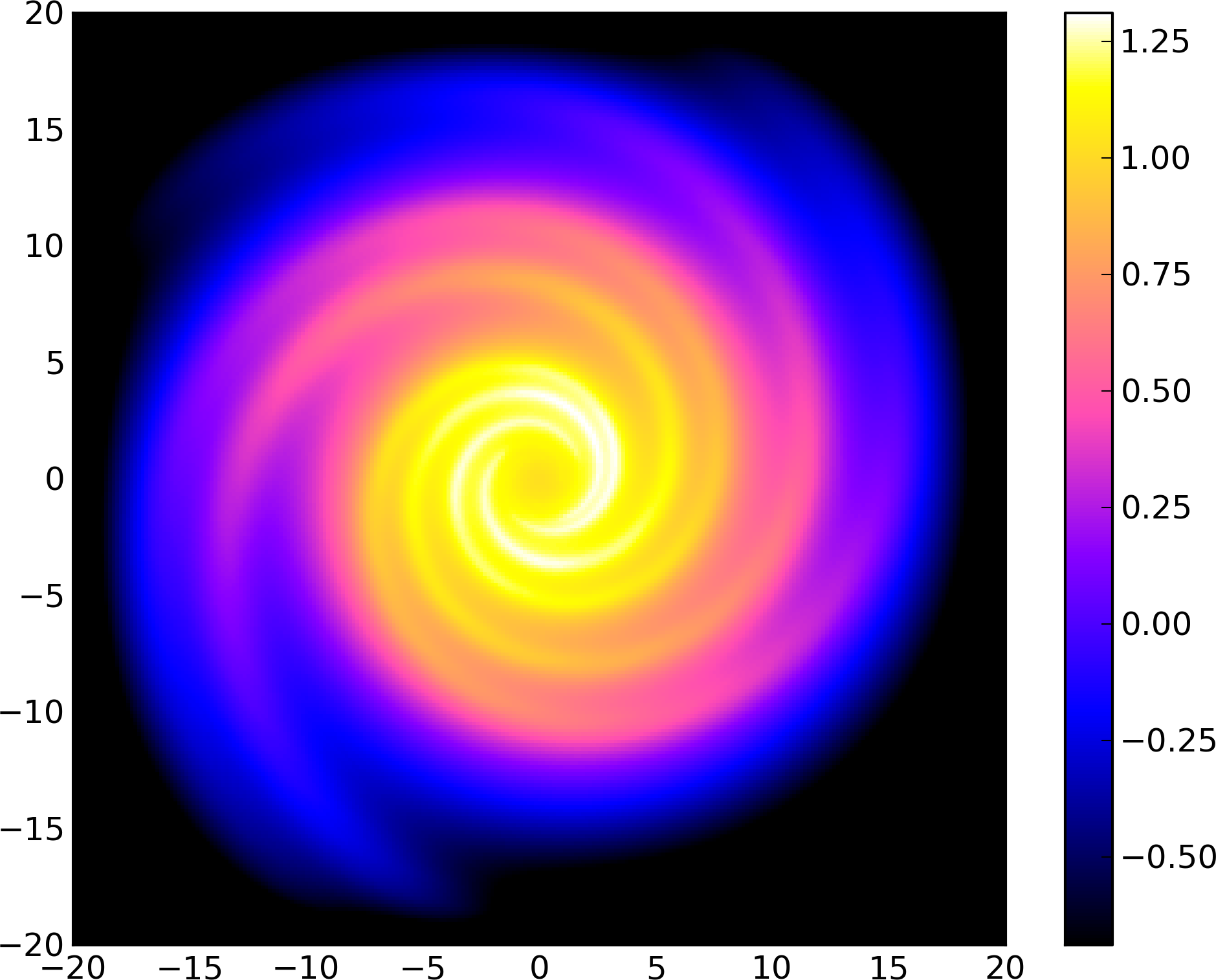}}
        \end{picture}
    \end{subfigure}
 	\quad
     \begin{subfigure}{500\unitlength}
        \begin{picture}(500,600)
        		\put(220,0){x}
        		\put(0,350){y}
        		\put(0,80){z}
        		\put(27,20){\includegraphics[trim=0cm 2cm 0cm 0cm, clip=true,width=410\unitlength]{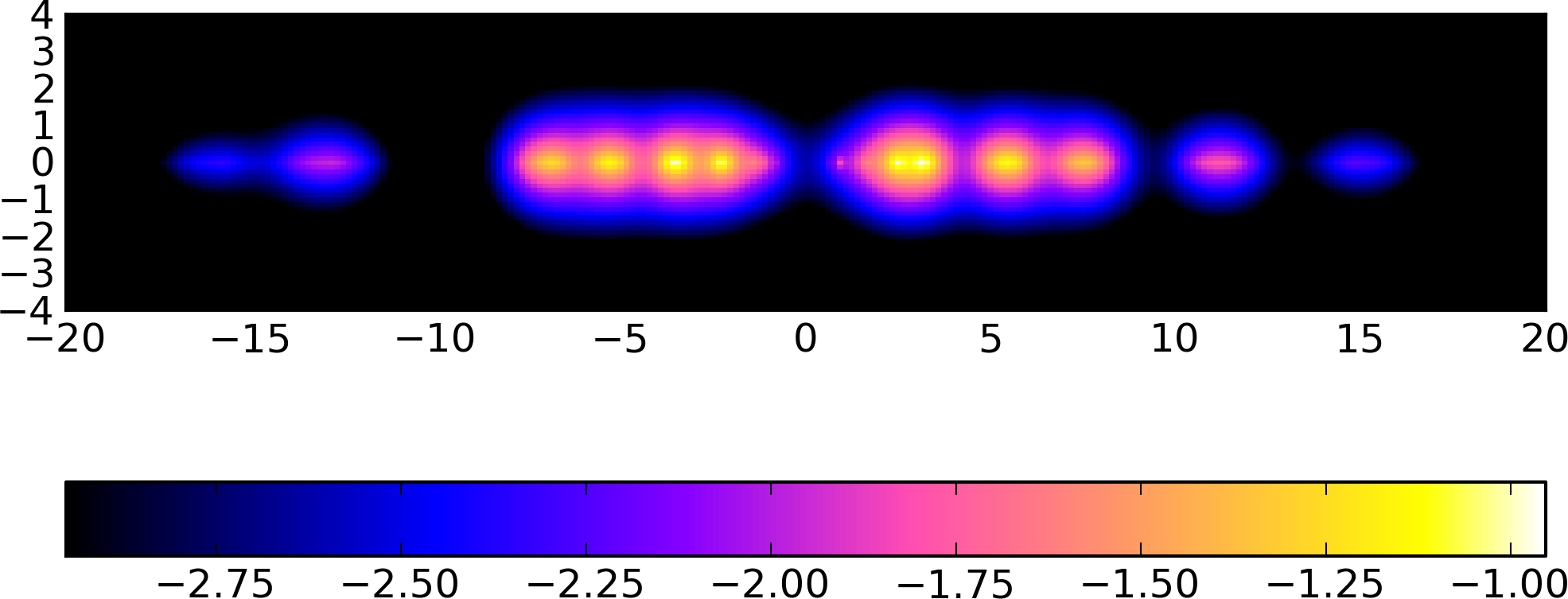}}
            \put(15,150){\includegraphics[width=500\unitlength]{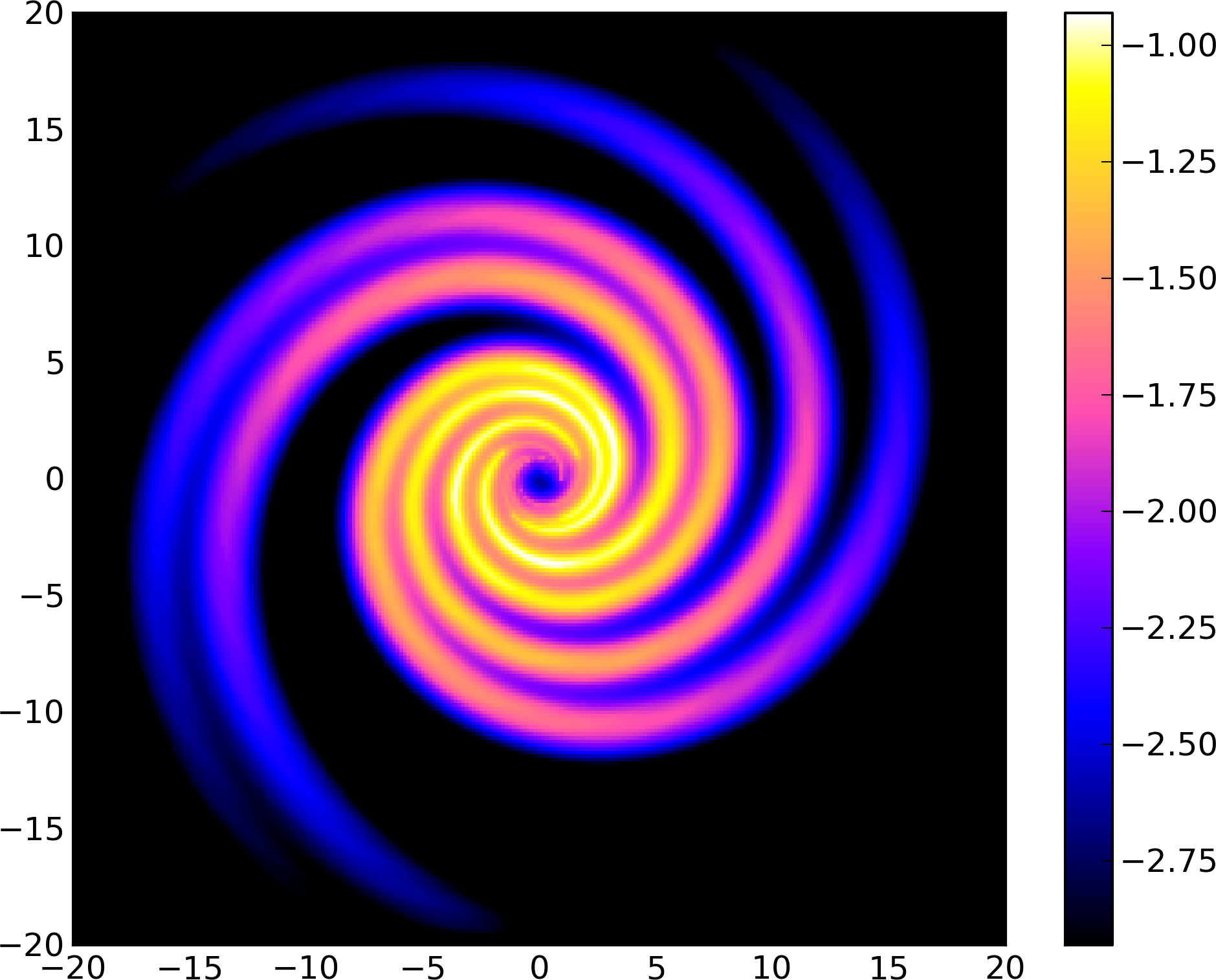}}
        \end{picture}
    \end{subfigure}
\caption{}
\end{figure}
\clearpage
\begin{figure}[h]
	\ContinuedFloat
    \setlength{\unitlength}{0.001\textwidth}
    \begin{subfigure}{500\unitlength}
        \begin{picture}(500,600)
        		\put(220,0){x}
        		\put(0,350){y}
        		\put(0,80){z}
        		\put(27,20){\includegraphics[trim=0cm 2cm 0cm 0cm, clip=true,width=410\unitlength]{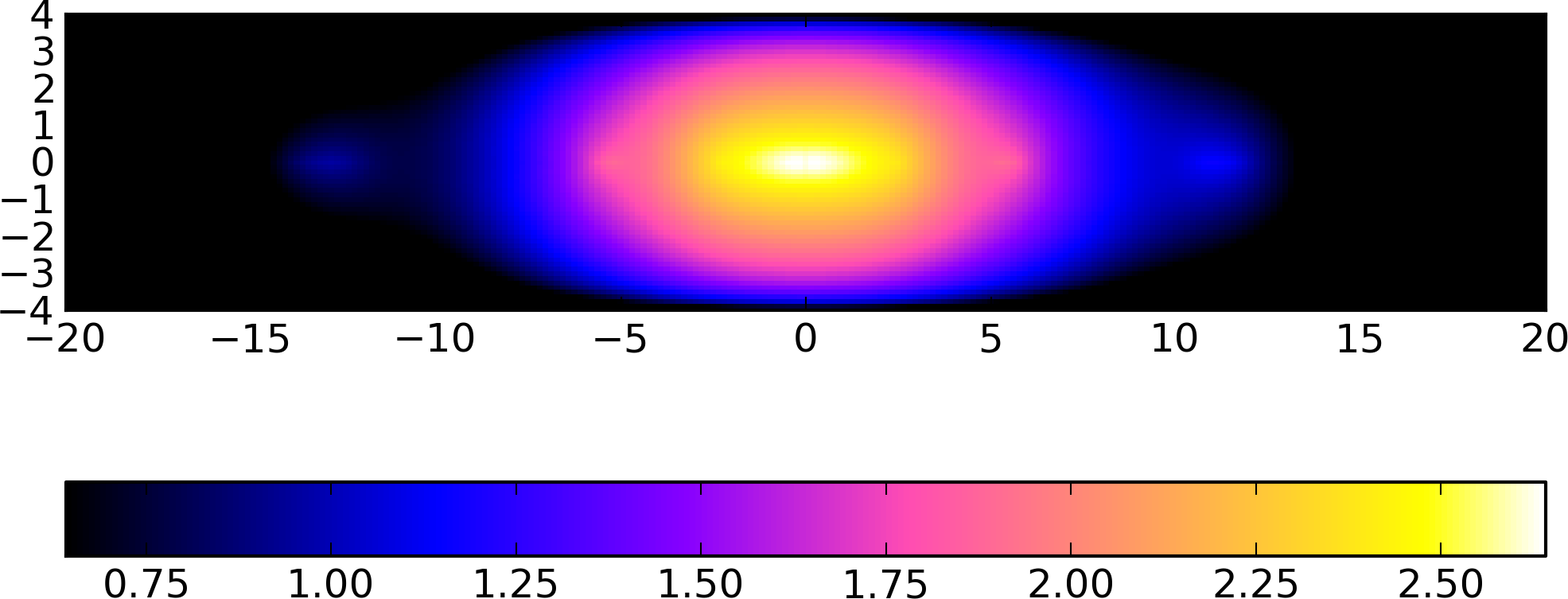}}
            \put(15,150){\includegraphics[width=500\unitlength]{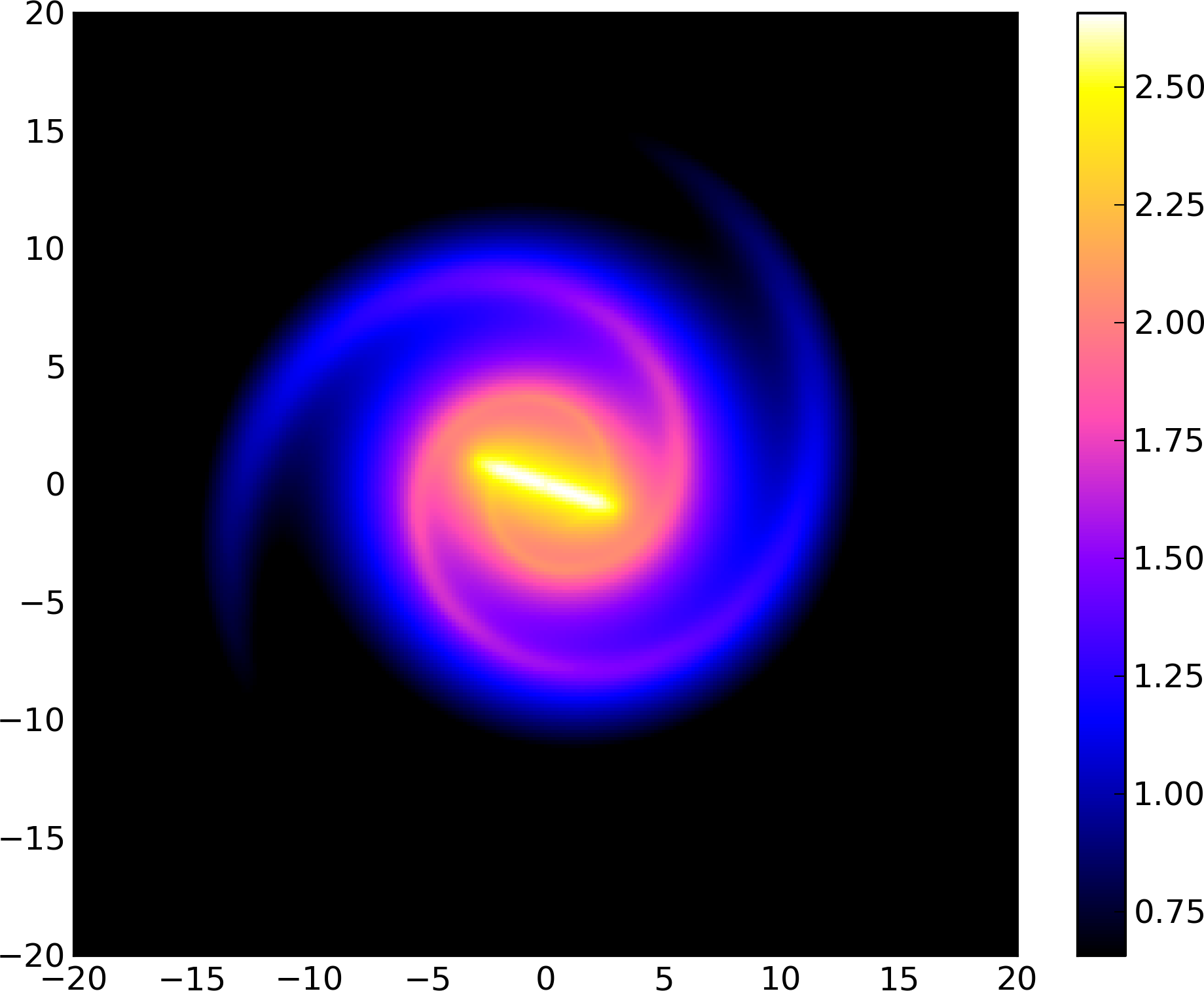}}
        \end{picture}
    \end{subfigure}
 	\quad
     \begin{subfigure}{500\unitlength}
        \begin{picture}(500,600)
        		\put(220,0){x}
        		\put(0,350){y}
        		\put(0,80){z}
        		\put(27,20){\includegraphics[trim=0cm 2cm 0cm 0cm, clip=true,width=410\unitlength]{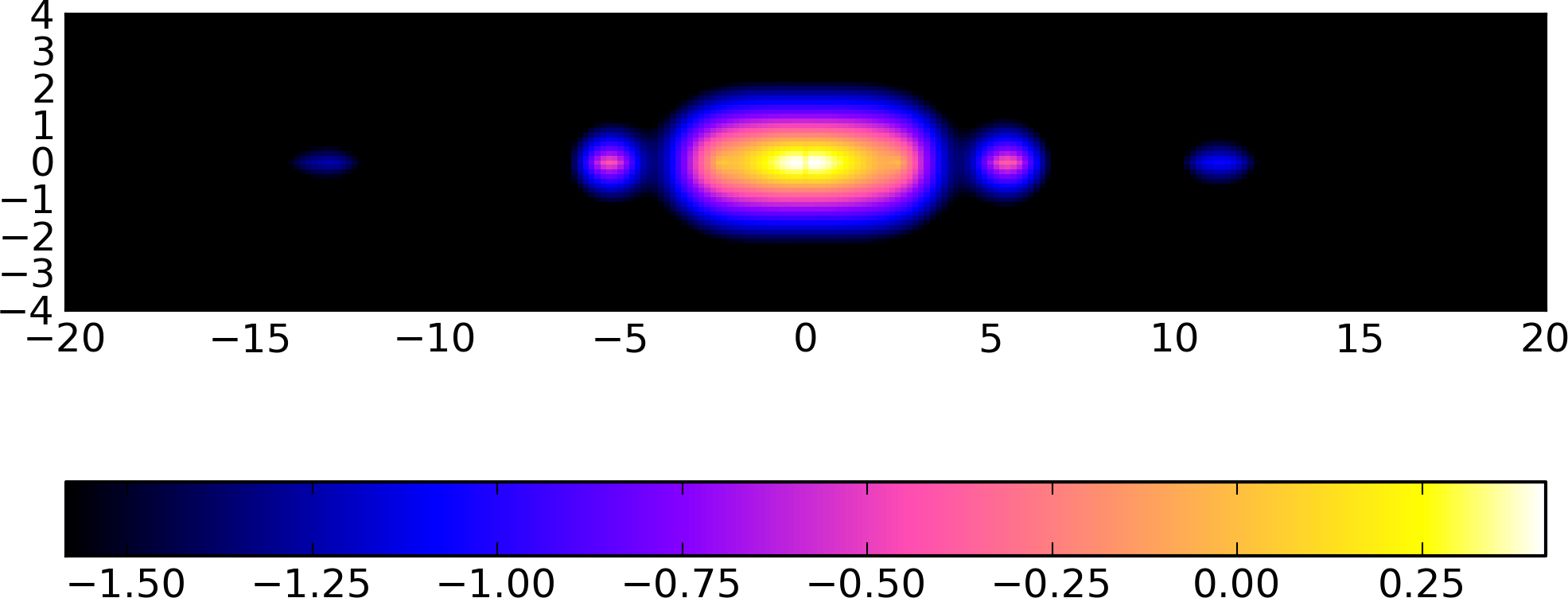}}
            \put(15,150){\includegraphics[width=500\unitlength]{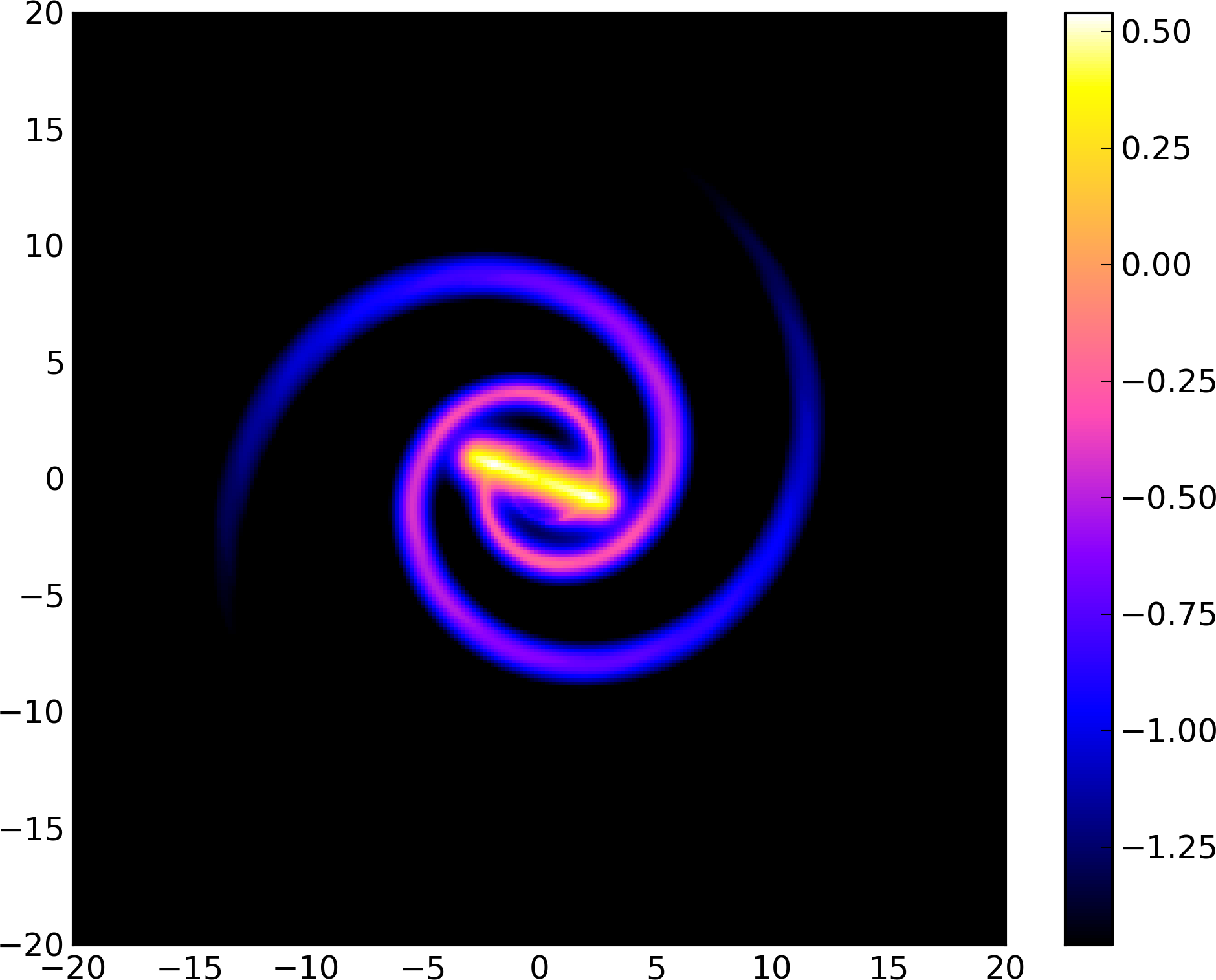}}
        \end{picture}
    \end{subfigure}\\
        \begin{subfigure}{500\unitlength}
        \begin{picture}(500,600)
        		\put(220,0){x}
        		\put(0,350){y}
        		\put(0,80){z}
        		\put(27,20){\includegraphics[trim=0cm 2cm 0cm 0cm, clip=true,width=410\unitlength]{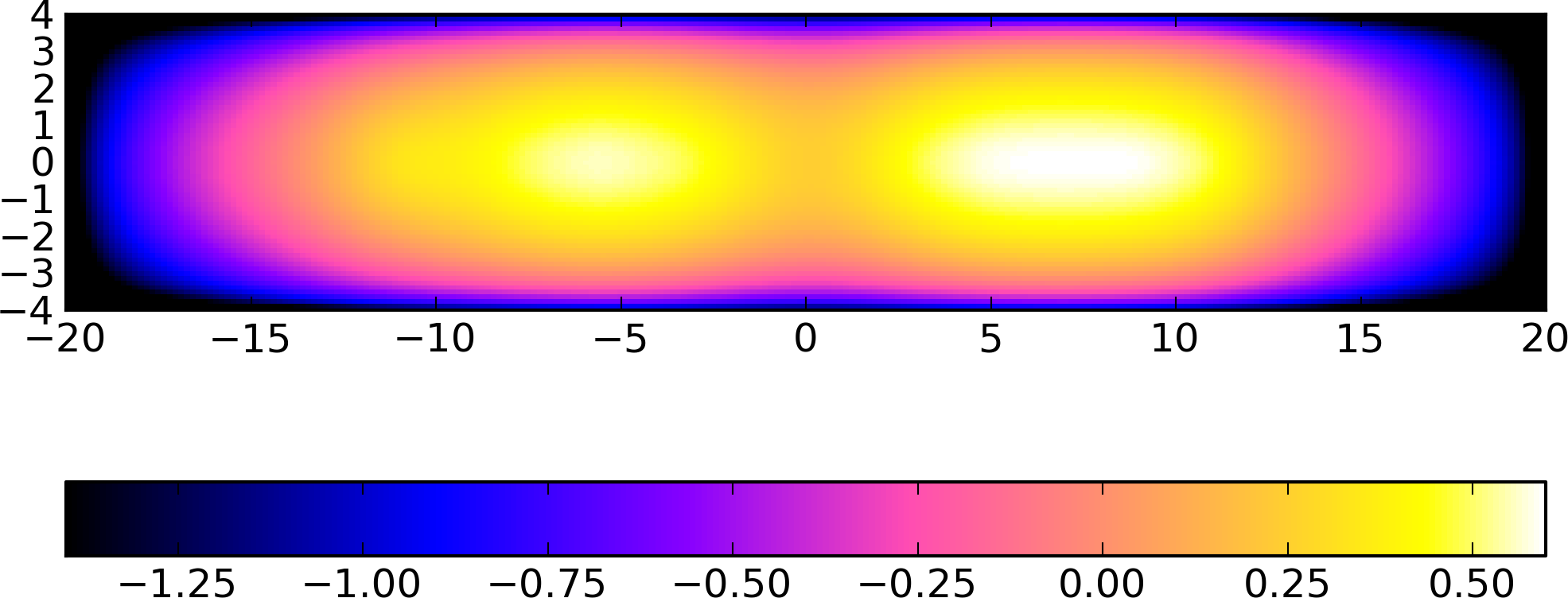}}
            \put(15,150){\includegraphics[width=500\unitlength]{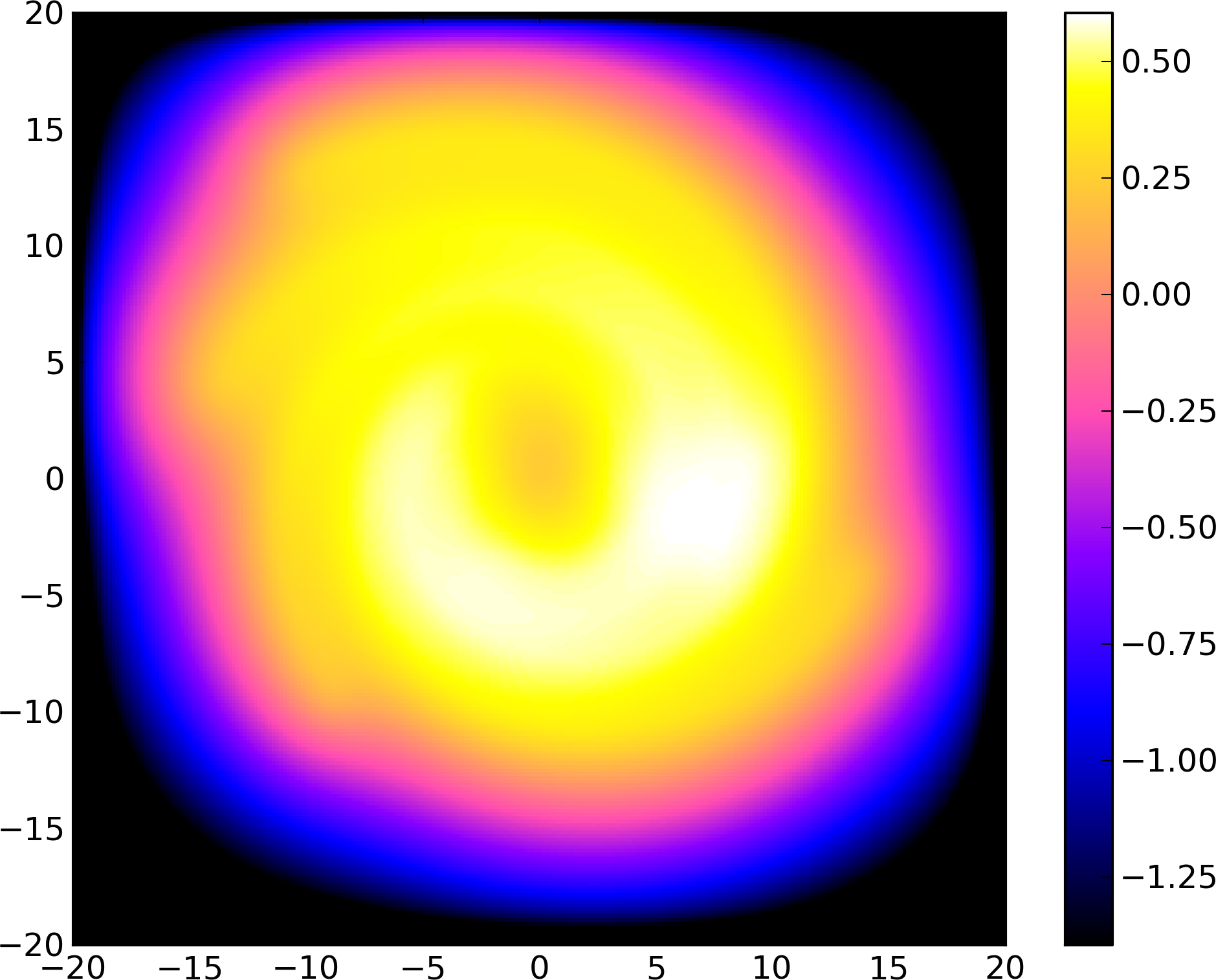}}
        \end{picture}
    \end{subfigure}
 	\quad
     \begin{subfigure}{500\unitlength}
        \begin{picture}(500,600)
        		\put(220,0){x}
        		\put(0,350){y}
        		\put(0,80){z}
        		\put(27,20){\includegraphics[trim=0cm 2cm 0cm 0cm, clip=true,width=410\unitlength]{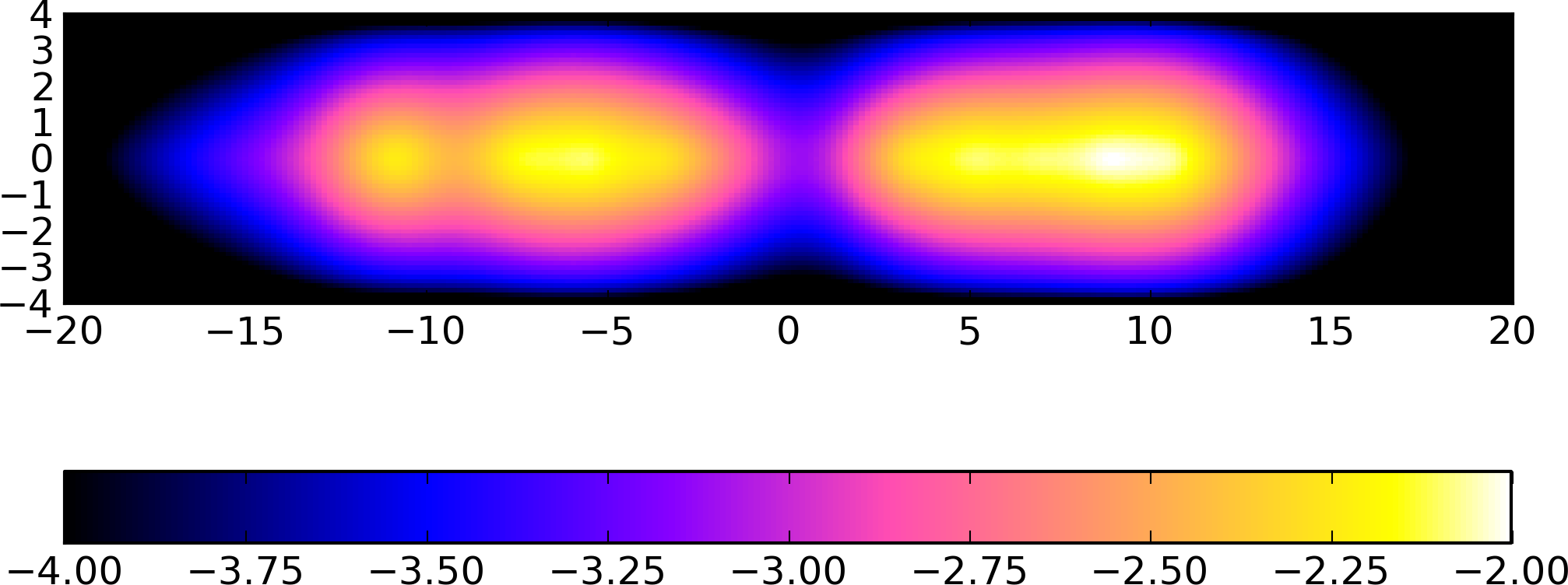}}
            \put(15,150){\includegraphics[width=500\unitlength]{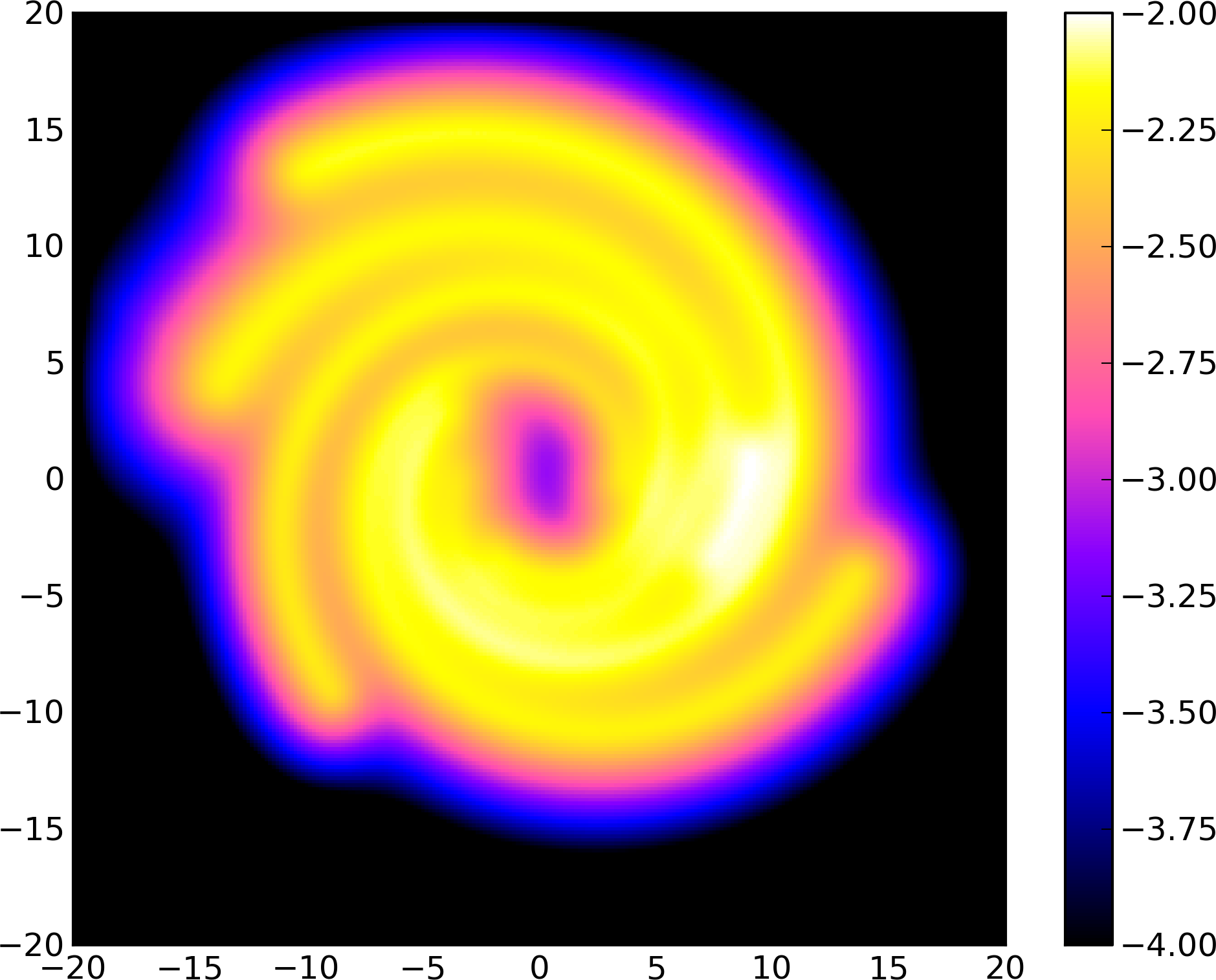}}
        \end{picture}
    \end{subfigure}
\caption{Same as Figure \ref{Hdistributions}, but for electrons.}
\label{Edistributions}
\end{figure}

\begin{figure}[h]
    \setlength{\unitlength}{0.001\textwidth}
    \begin{subfigure}{500\unitlength}
        \begin{picture}(500,500)
        		\put(0,200){\rotatebox{90}{$N^{p}$}}
            \put(10,10){\includegraphics[width=500\unitlength]{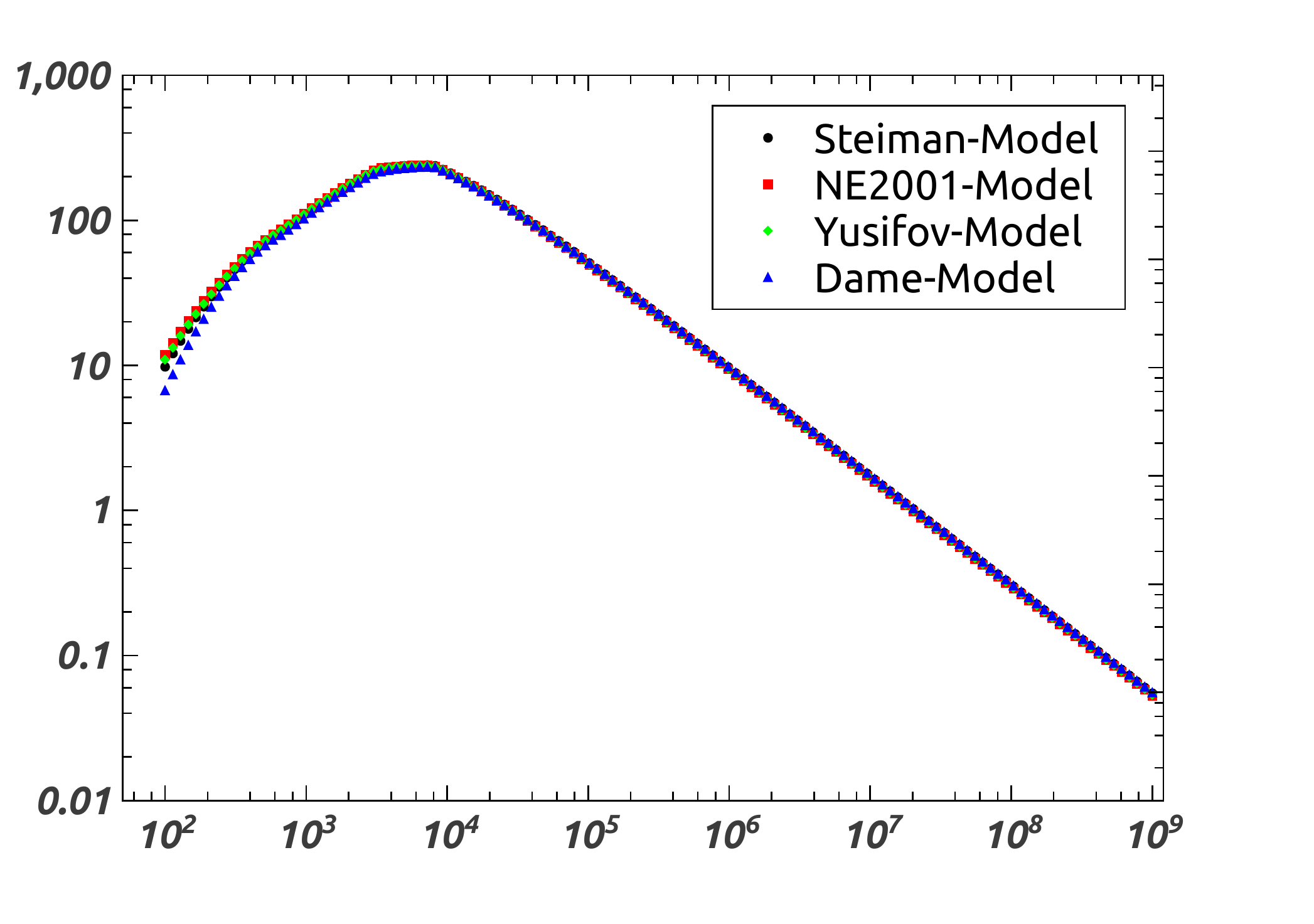}}
            \put(180,0){$E_{kin} \, \left[ \mbox{MeV} \right]$ } 
        \end{picture}
    \end{subfigure}
 	\quad
    \begin{subfigure}{500\unitlength}
        \begin{picture}(500,500)
        		\put(0,200){\rotatebox{90}{$N^{e}$}}
            \put(20,10){\includegraphics[width=500\unitlength]{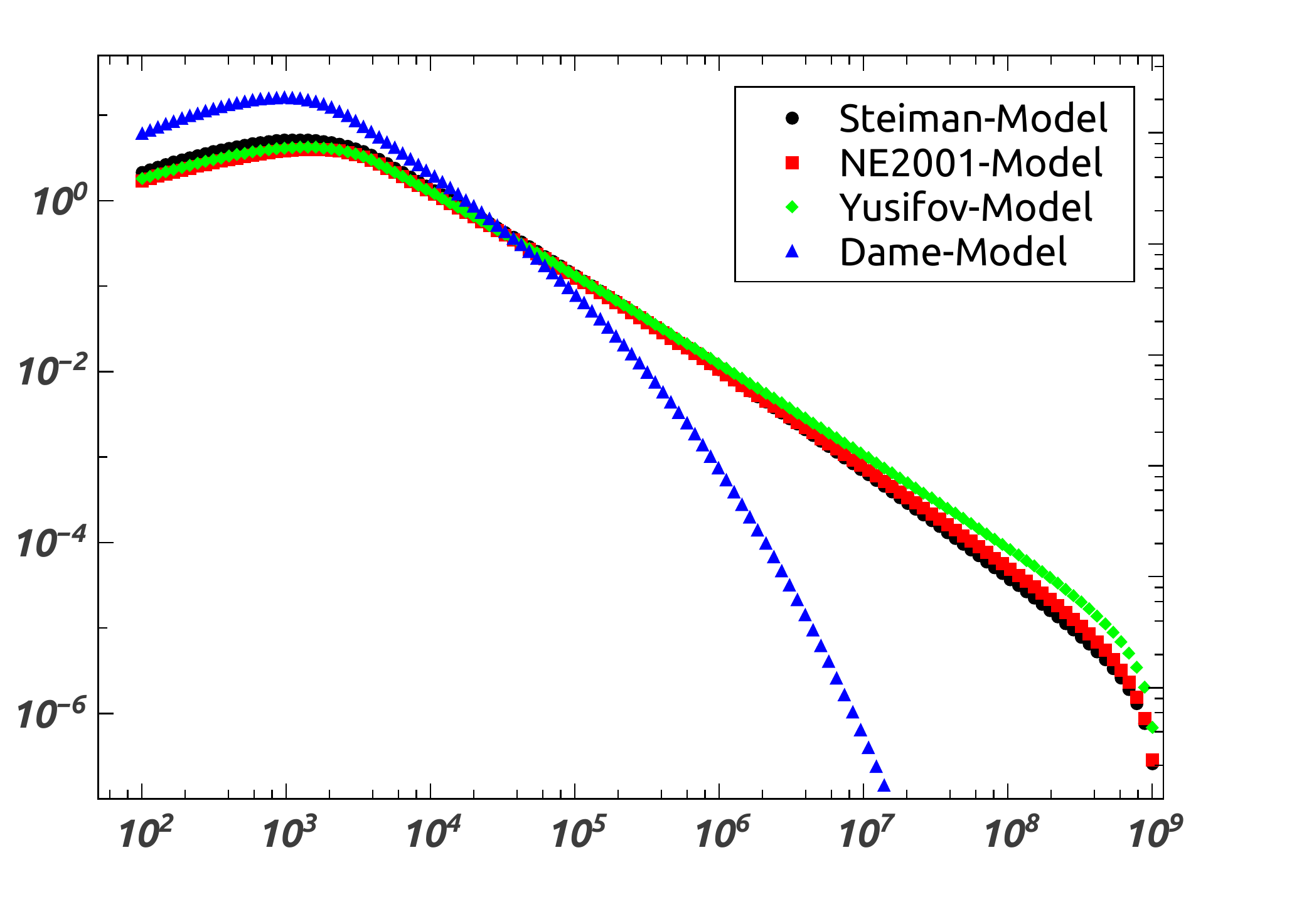}}
            \put(180,0){$E_{kin} \, \left[ \mbox{MeV} \right]$ }
        \end{picture}
    \end{subfigure}
\caption{\textit{Left:} Proton spectra at the Earth obtained using the four different source models (See Section \ref{setup}). Y-axis is in units of normalized flux $N^{p} = 5 \times 10^{-9}$ \flux. \\ \textit{Right:} Same as \textit{Right} but for electrons. Y-axis is in units of normalized flux $N^{e} = 3.2 \times 10^{-10}$ \flux.}
\label{HESpectra}
\end{figure}

\begin{figure}[h]
    \setlength{\unitlength}{0.001\textwidth}
    \begin{subfigure}{500\unitlength}
        \begin{picture}(500,500)
        		\put(0,180){\rotatebox{90}{{\large $\frac{S_{model}}{S_{ref}}$}}}
            \put(40,10){\includegraphics[width=500\unitlength]{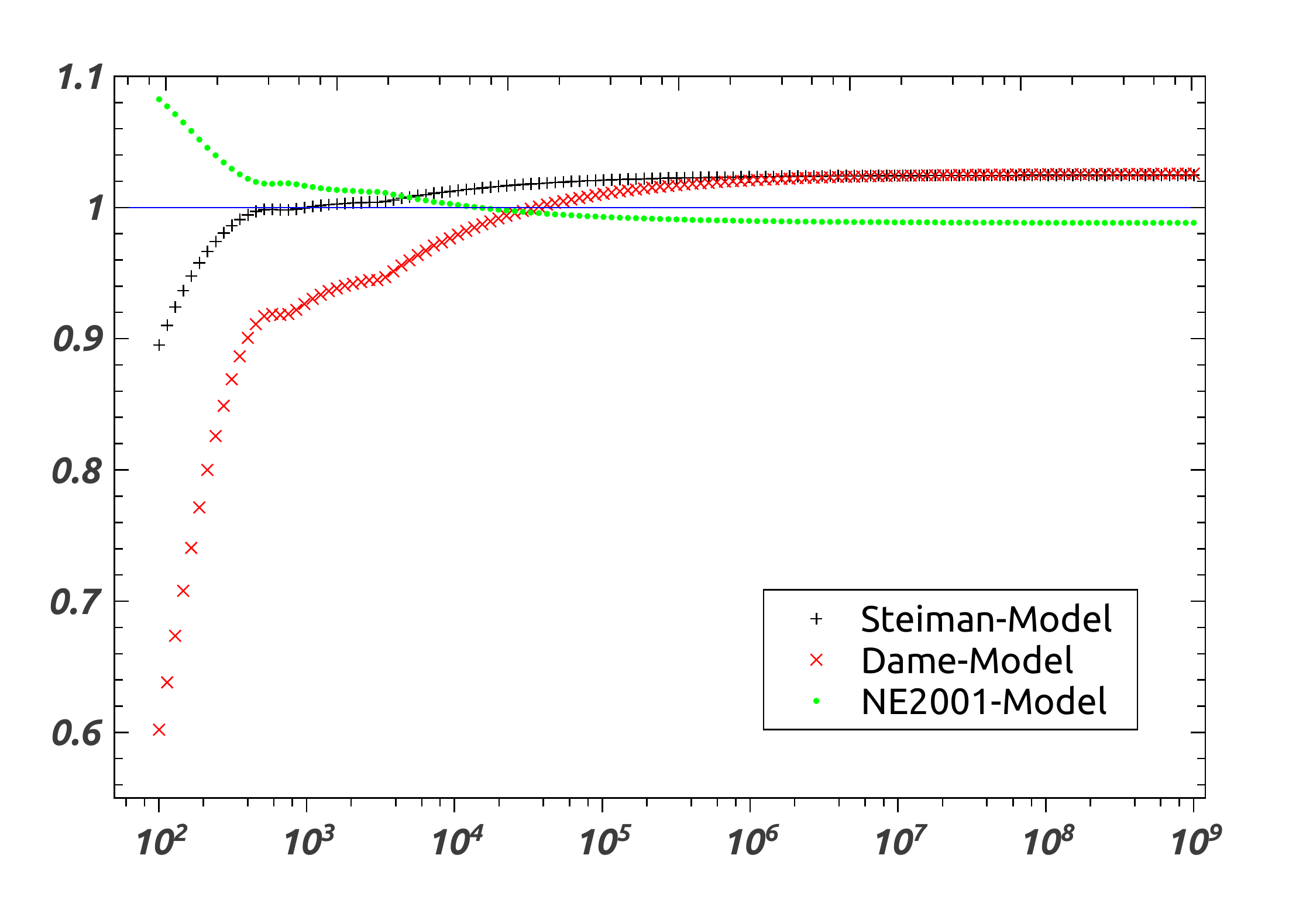}}
            \put(250,0){$E_{kin} \, \left[ \mbox{MeV} \right]$ } 
        \end{picture}
    \end{subfigure}
 	\quad
    \begin{subfigure}{500\unitlength}
        \begin{picture}(500,500)
        		\put(0,180){\rotatebox{90}{{\large $\frac{S_{model}}{S_{ref}}$}}}
            \put(40,10){\includegraphics[width=500\unitlength]{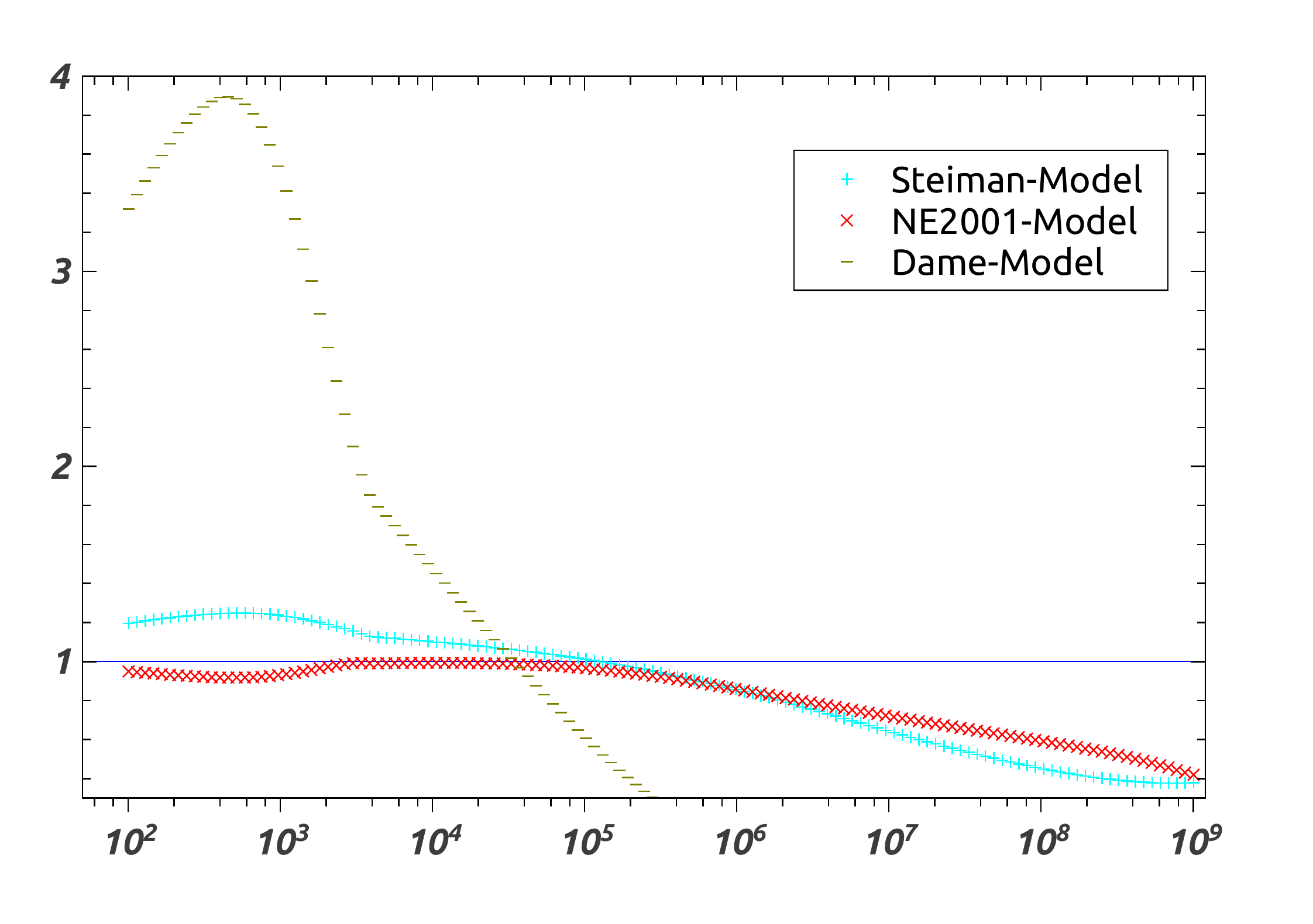}}
             \put(250,0){$E_{kin} \, \left[ \mbox{MeV} \right]$ } 
        \end{picture}
    \end{subfigure}
\caption{\textit{Left:} Deviation of proton spectra $S_{model}$ obtained with the corresponding spiral arm models (See Section \ref{spirals}) and the proton spectrum $S_{ref}$ obtained using the \textit{Reference}-Model which is our reference axisymmetric source distribution (See Section \ref{crProp}).\\ \textit{Right:} Same but for electrons.}
\label{relHESpectra}
\end{figure}

\begin{figure}[h]
    \setlength{\unitlength}{0.001\textwidth}
    \begin{subfigure}{500\unitlength}
        \begin{picture}(500,500)
        		\put(0,180){\rotatebox{90}{{\large $\frac{S_{model}}{S_{ref}}$}}}
            \put(40,20){\includegraphics[width=500\unitlength]{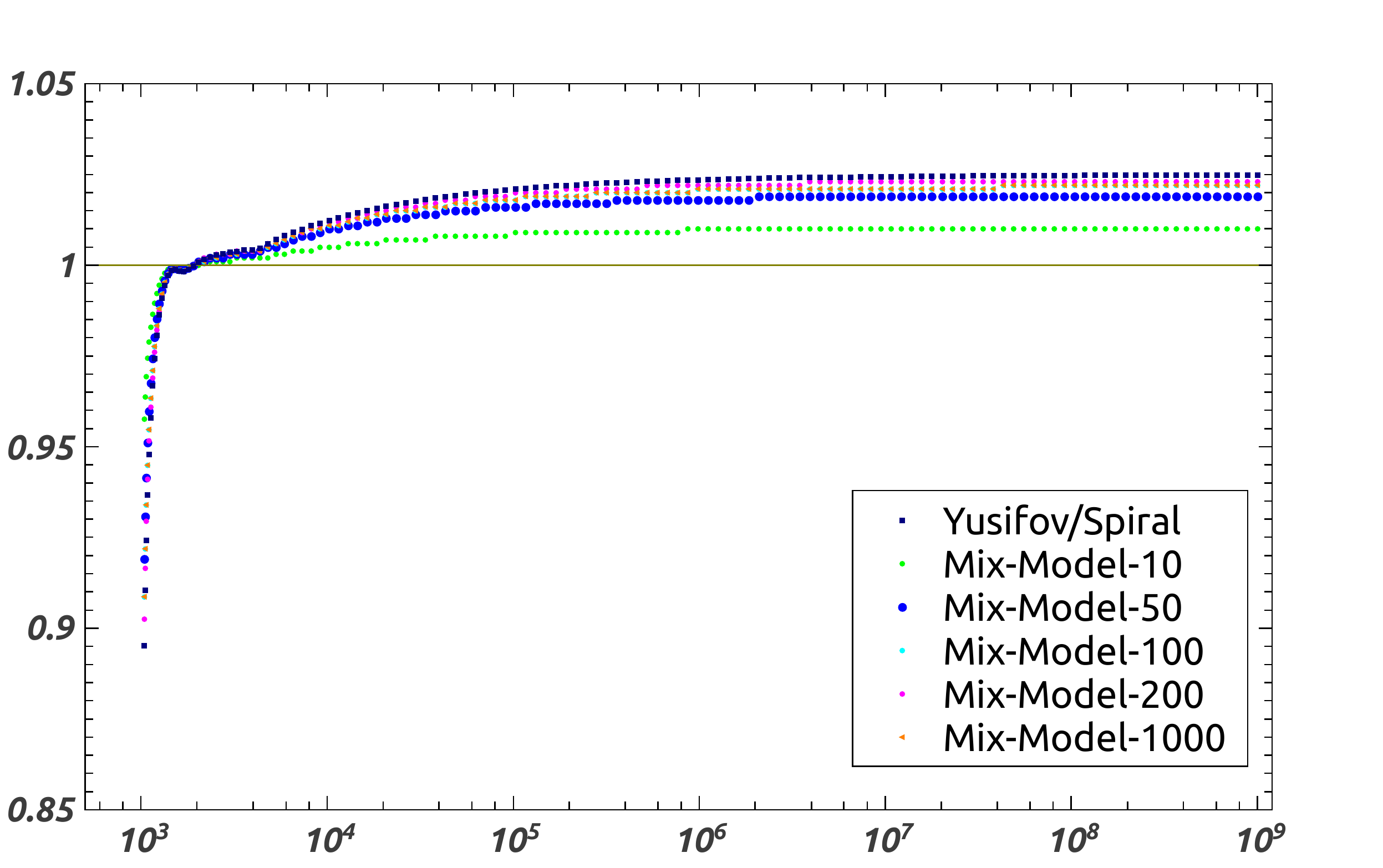}}
            \put(250,0){$E_{kin} \, \left[ \mbox{MeV} \right]$ } 
        \end{picture}
    \end{subfigure}
 	\quad
    \begin{subfigure}{500\unitlength}
        \begin{picture}(500,500)
        		\put(0,180){\rotatebox{90}{{\large $\frac{S_{model}}{S_{ref}}$}}}
            \put(40,20){\includegraphics[width=500\unitlength]{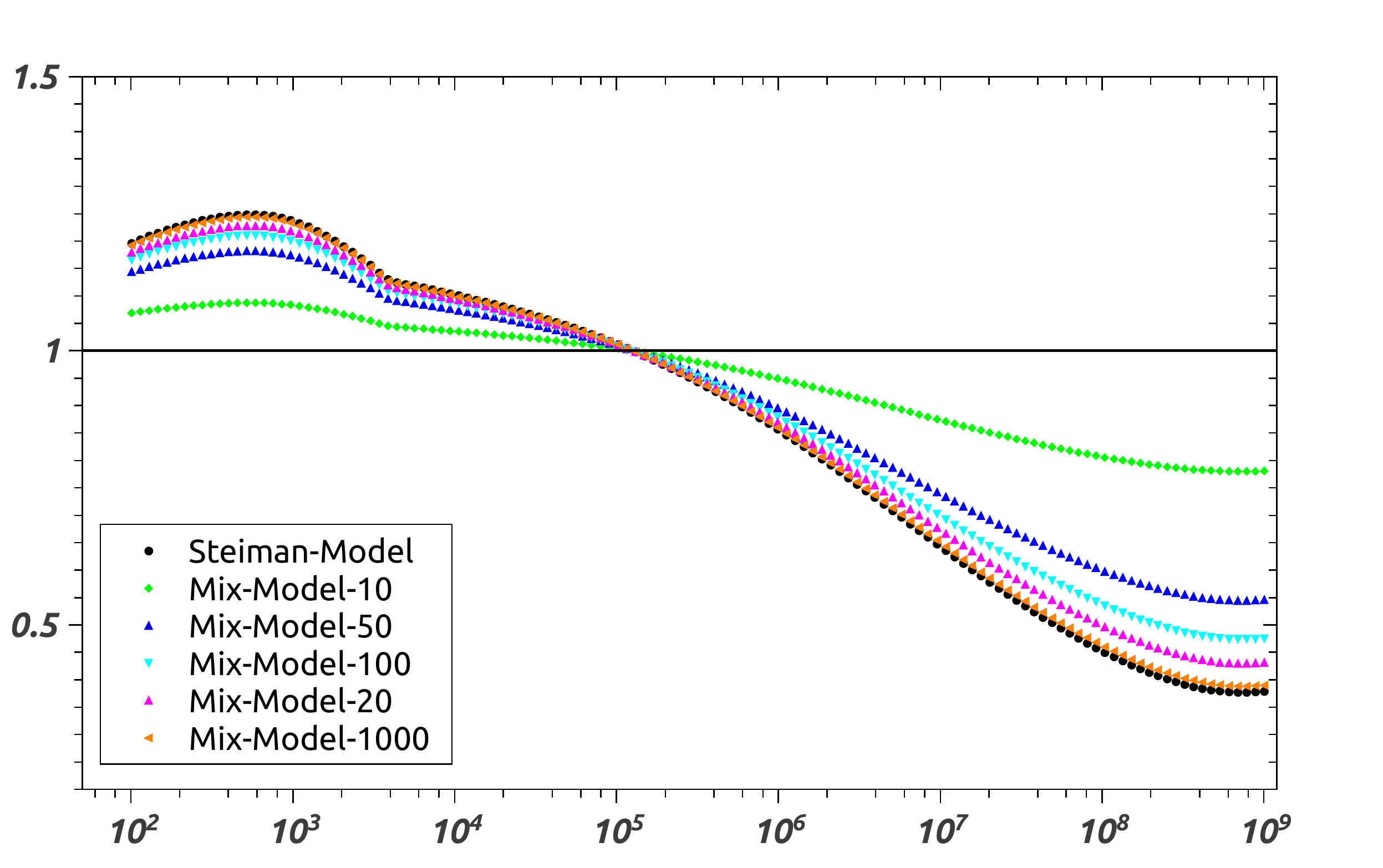}}
             \put(250,0){$E_{kin} \, \left[ \mbox{MeV} \right]$ } 
        \end{picture}
    \end{subfigure}
\caption{Deviation of proton spectra $S_{model}$ obtained with the corresponding \textit{Mix}-Models (See Section \ref{spirals}) and the proton spectrum $S_{ref}$ obtained using the \textit{Reference}-Model which is our reference axisymmetric source distribution (See Section \ref{crProp}).\\ \textit{Right:} Same but for electrons.}
\label{relYusifov}
\end{figure}

\begin{figure}[h]
    \setlength{\unitlength}{0.001\textwidth}
    \begin{subfigure}{500\unitlength}
        \begin{picture}(500,500)
        		\put(0,180){\rotatebox{90}{{\large $\frac{S_{bar}}{S_{noBar}}$}}}
            \put(35,10){\includegraphics[width=500\unitlength]{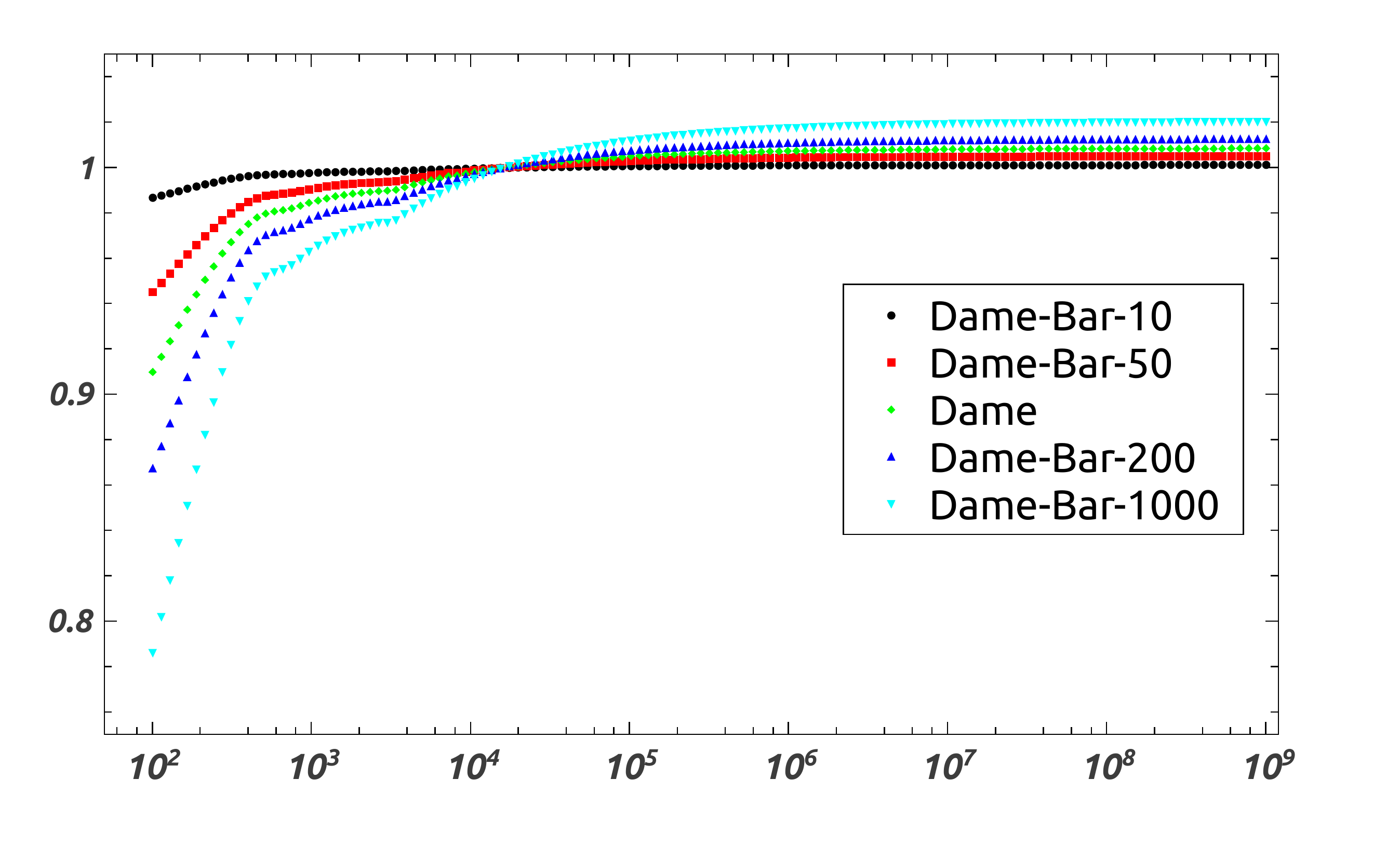}}
            \put(250,0){$E_{kin} \, \left[ \mbox{MeV} \right]$ } 
        \end{picture}
    \end{subfigure}
 	\quad
    \begin{subfigure}{500\unitlength}
        \begin{picture}(500,500)
        	\put(0,180){\rotatebox{90}{{\large $\frac{S_{bar}}{S_{noBar}}$}}}
            \put(35,10){\includegraphics[width=500\unitlength]{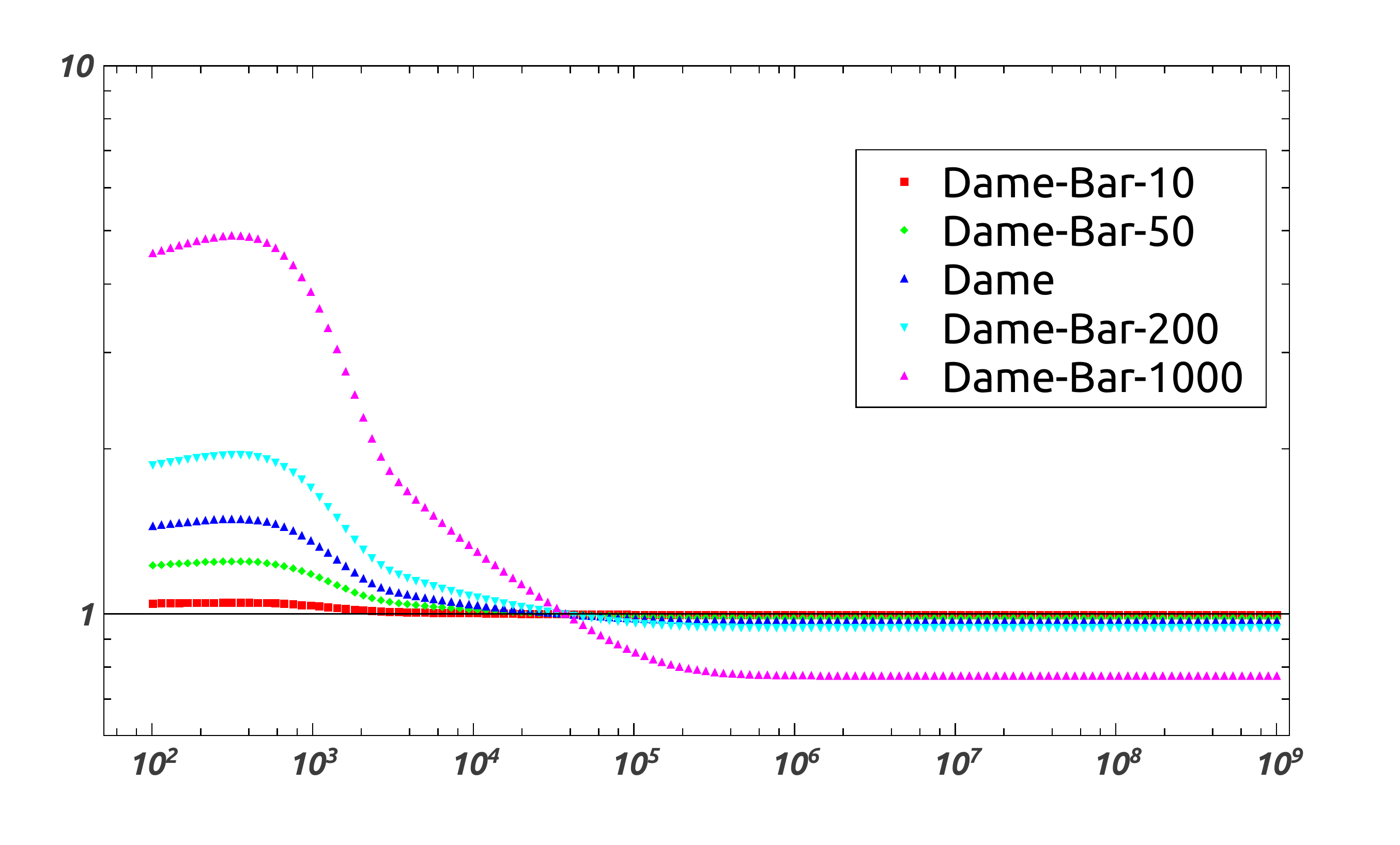}}
            \put(250,0){$E_{kin} \, \left[ \mbox{MeV} \right]$ }
        \end{picture}
    \end{subfigure}
\caption{Deviation of proton spectra $S_{bar}$ obtained with the corresponding \textit{Dame-Bar}-Models (See Section \ref{spirals}) and the proton spectrum $S_{noBar}$ obtained using the \textit{Dame-No-Bar}-Model.\\ \textit{Right:} Same but for electrons.}
\label{relBar}
\end{figure}

\begin{figure}[t]
\setlength{\unitlength}{0.001\textwidth}
	\centering
	\begin{picture}(800,600)
		\put(0,100){\rotatebox{90}{Proton flux [normalized to Earth]}}
		\put(5,10){\includegraphics[trim=0cm 0cm 0cm 0cm, clip=true,width=800\unitlength]{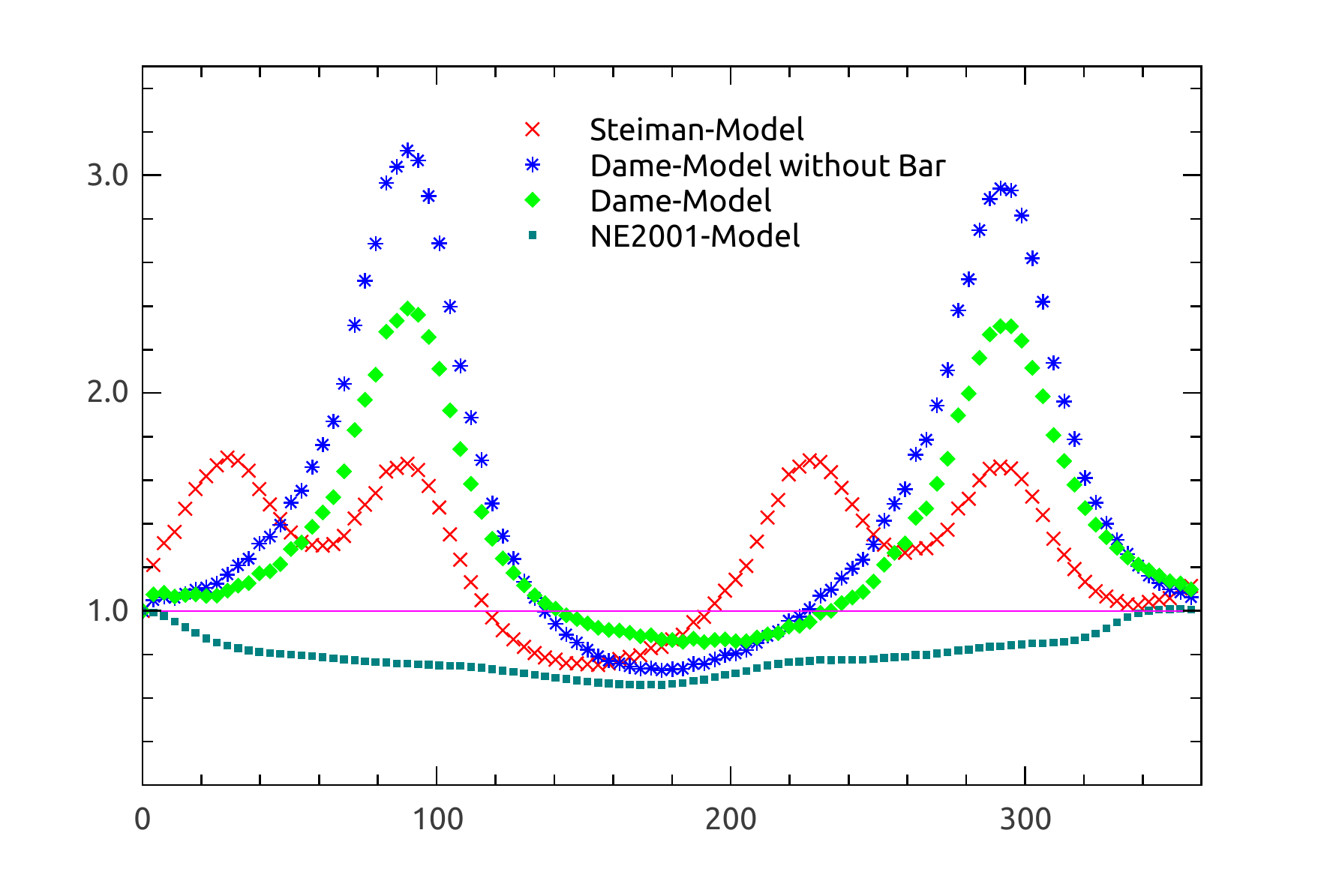}}
		\put(350,0){{\large $\phi \, \left[ ^{\circ} \right]$} }
	\end{picture}
\caption{Total Proton flux at the Earth obtained using the spiral CR source distributions (see Section \ref{spirals}) for kinetic energies $100 \mbox{MeV} < E_{kin} <  1 \mbox{PeV}$ at different positions of the Earth on a circle centred on the Galactic Center with radius $r = R_{\odot}$. Position are given by the angle $\phi$ between the line connecting the nominal position $x=8.5 \mbox{kpc, }  y = 0 \mbox{kpc, } z = 0 \mbox{kpc}$ of the Earth with the Galactic Center and the line connecting the Galactic Center with the test position on the circle. The proton flux is normalized to the flux at the nominal position of the Earth and is given in arbitrary units. \label{fig:orbitModels}}
\end{figure}

\begin{figure}[t]
\setlength{\unitlength}{0.001\textwidth}
	\centering
	\begin{picture}(800,600)
		\put(0,100){\rotatebox{90}{Proton flux [normalized to Earth]}}
		\put(5,10){\includegraphics[trim=0cm 0cm 0cm 0cm, clip=true,width=800\unitlength]{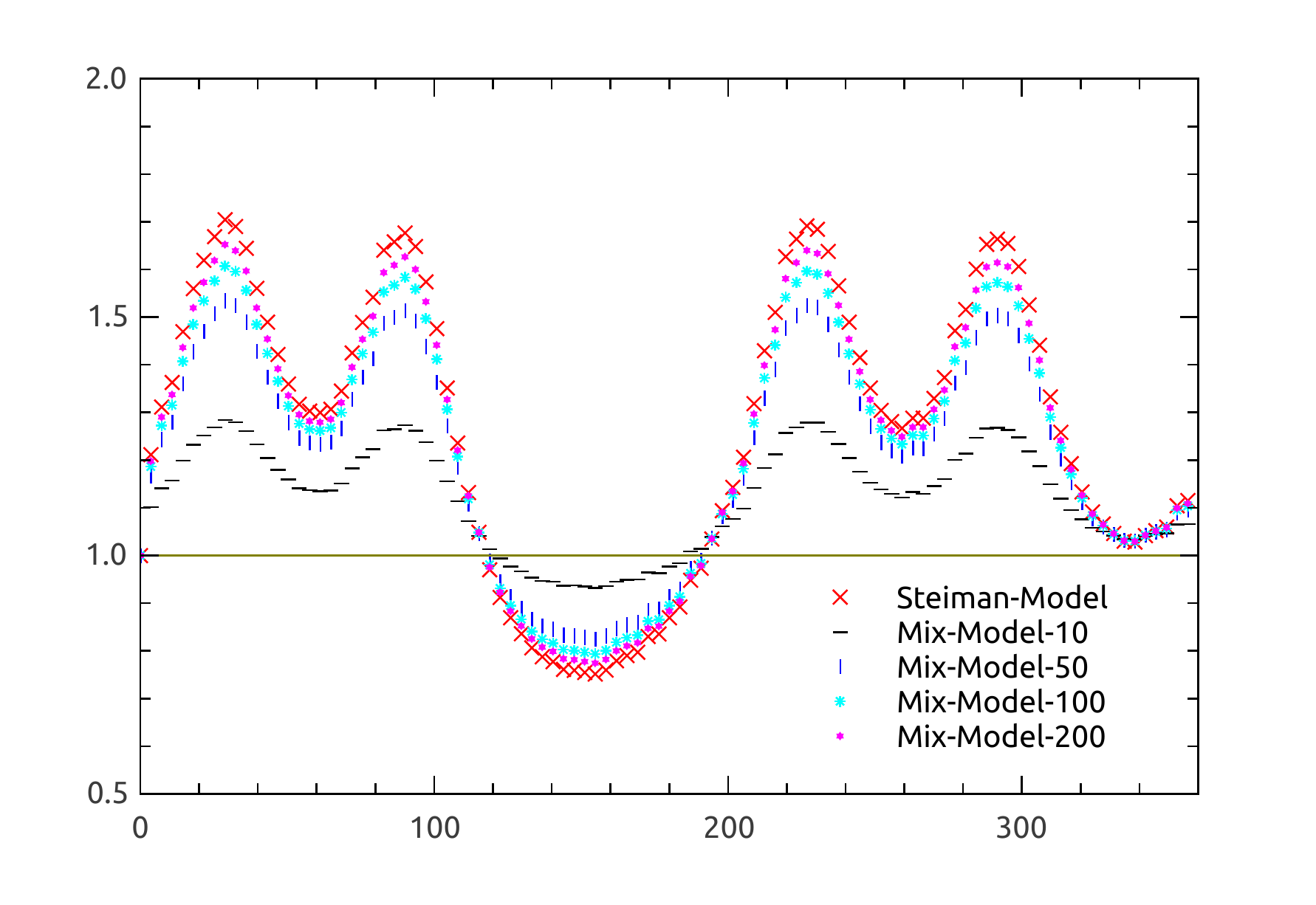}}
		\put(350,0){{\large $\phi \, \left[ ^{\circ} \right]$} }
	\end{picture}
\caption{Total Proton flux at the Earth obtained using the \textit{Mix}-Models (see Section \ref{spirals}) for kinetic energies $100 \mbox{MeV} < E_{kin} <  1 \mbox{PeV}$ at different positions of the Earth on a circle around the Galactic Center with radius $r = R_{\odot}$. Position are given by the angle $\phi$ between the line connecting the nominal position $x=8.5 \mbox{kpc, }  y = 0 \mbox{kpc, } z = 0 \mbox{kpc}$ of the Earth with the Galactic Center and the line connecting the Galactic Center with the test position on the circle. The proton flux is normalized to the flux at the nominal position of the Earth and is given in arbitrary units. \label{fig:orbitYusifov}}
\end{figure}

\begin{figure}[t]
    \setlength{\unitlength}{0.001\textwidth}
    \begin{subfigure}{500\unitlength}
        \begin{picture}(500,400)
        		\put(0,150){\rotatebox{90}{$E_{kin} \, \left[ \mbox{MeV} \right]$}}
        		\put(30,25){\includegraphics[trim=0cm 0cm 0cm 0cm, clip=true,width=410\unitlength]{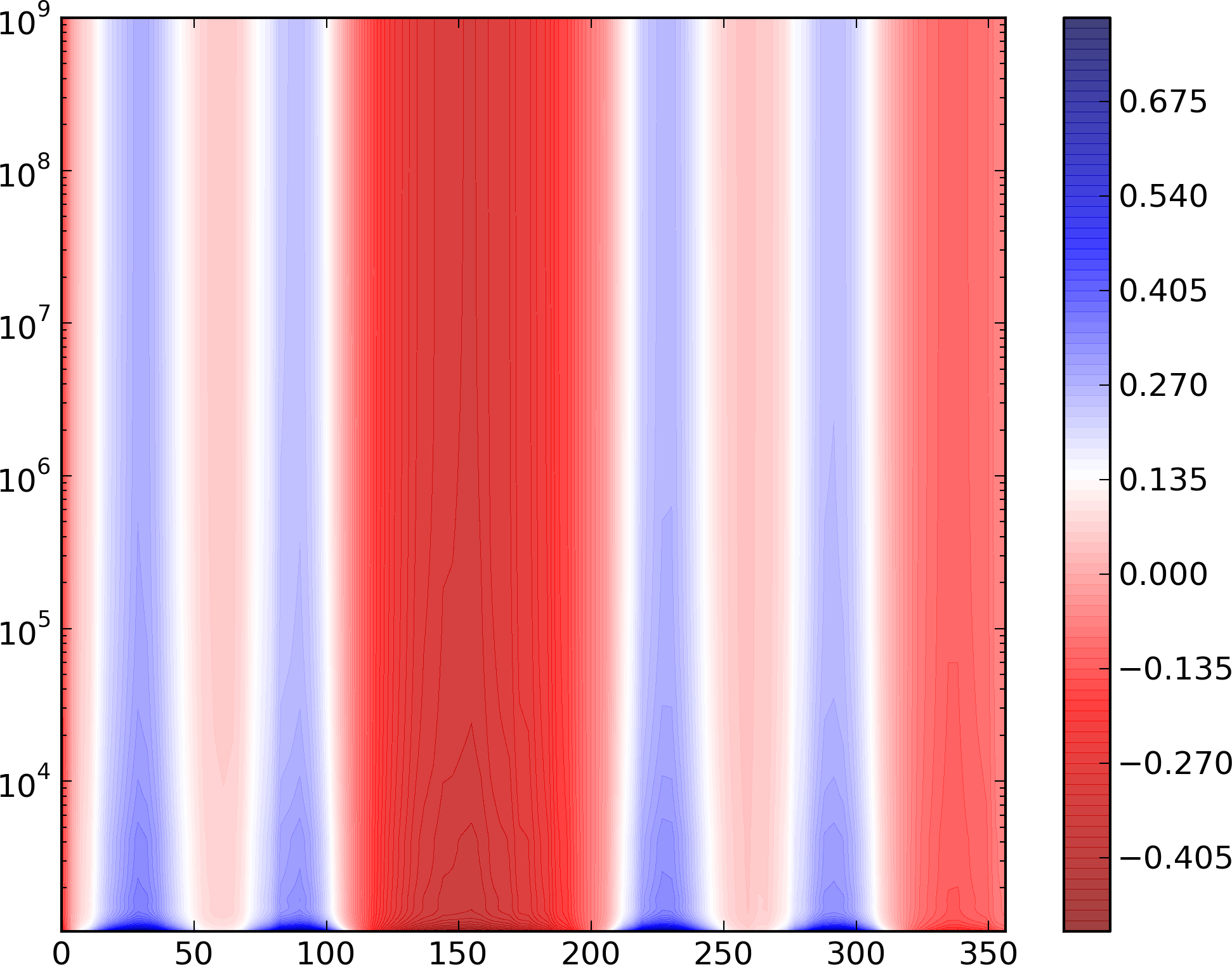}}
        		\put(450,300){\rotatebox{270}{$ \left(S_{\phi} - S_{earth} \right) / S_{earth}$}}
        		\put(200,0){ $\phi \, \left[ ^{\circ} \right]$} 
        \end{picture}
    \end{subfigure}
 	\quad
     \begin{subfigure}{500\unitlength}
        \begin{picture}(500,400)
	        \put(0,150){\rotatebox{90}{$E_{kin} \, \left[ \mbox{MeV} \right]$}}
        		\put(30,25){\includegraphics[trim=0cm 0cm 0cm 0cm, clip=true,width=410\unitlength]{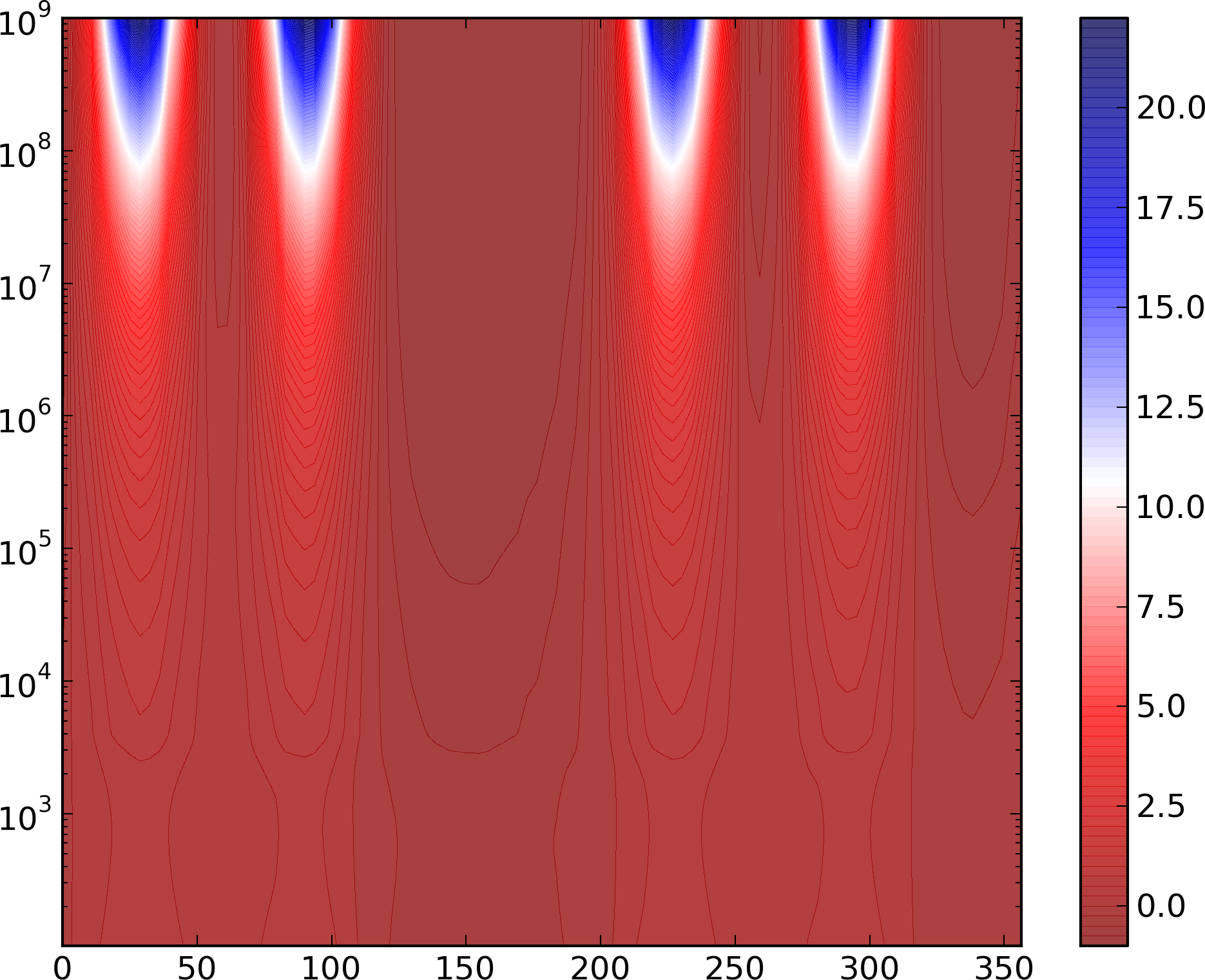}}
        		 \put(450,300){\rotatebox{270}{$ \left(S_{\phi} - S_{earth} \right) / S_{earth}$}}
        		\put(200,0){ $\phi \, \left[ ^{\circ} \right]$} 
        \end{picture}
    \end{subfigure}\\
    \begin{subfigure}{500\unitlength}
        \begin{picture}(500,400)
        		\put(0,150){\rotatebox{90}{$E_{kin} \, \left[ \mbox{MeV} \right]$}}
        		\put(30,25){\includegraphics[trim=0cm 0cm 0cm 0cm, clip=true,width=410\unitlength]{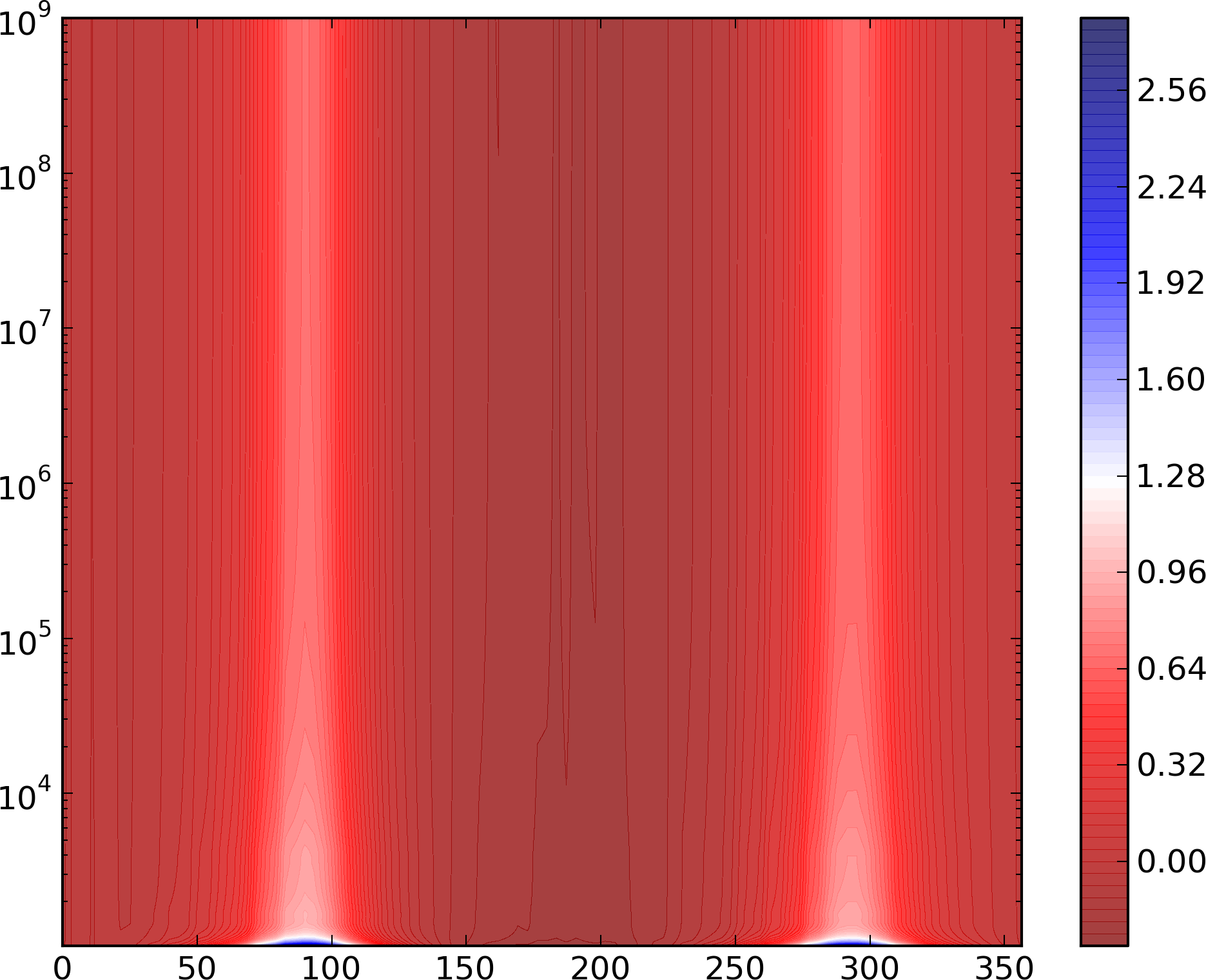}}
        		 \put(450,300){\rotatebox{270}{$ \left(S_{\phi} - S_{earth} \right) / S_{earth}$}}
        		\put(200,0){ $\phi \, \left[ ^{\circ} \right]$} 
        \end{picture}
    \end{subfigure}
 	\quad
     \begin{subfigure}{500\unitlength}
        \begin{picture}(500,400)
        		\put(0,150){\rotatebox{90}{$E_{kin} \, \left[ \mbox{MeV} \right]$}}
        		\put(30,25){\includegraphics[trim=0cm 0cm 0cm 0cm, clip=true,width=410\unitlength]{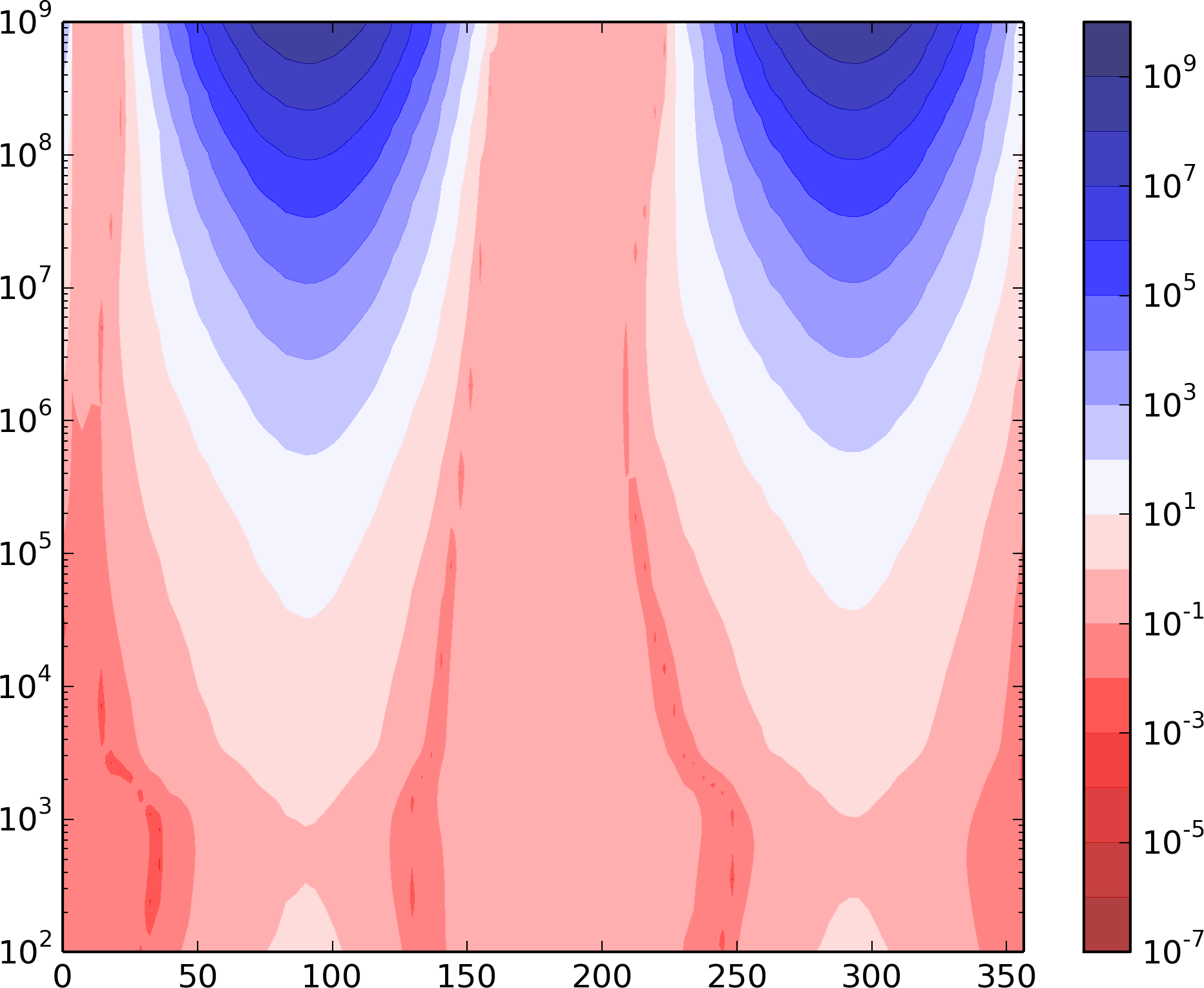}}
        		 \put(450,300){\rotatebox{270}{$\left(\vert S_{\phi} - S_{earth}\vert\right) /S_{earth} $}}
        		\put(200,0){ $\phi \, \left[ ^{\circ} \right]$} 
        \end{picture}
    \end{subfigure}
    \begin{subfigure}{500\unitlength}
        \begin{picture}(500,400)
	        \put(0,150){\rotatebox{90}{$E_{kin} \, \left[ \mbox{MeV} \right]$}}
        		\put(30,25){\includegraphics[trim=0cm 0cm 0cm 0cm, clip=true,width=410\unitlength]{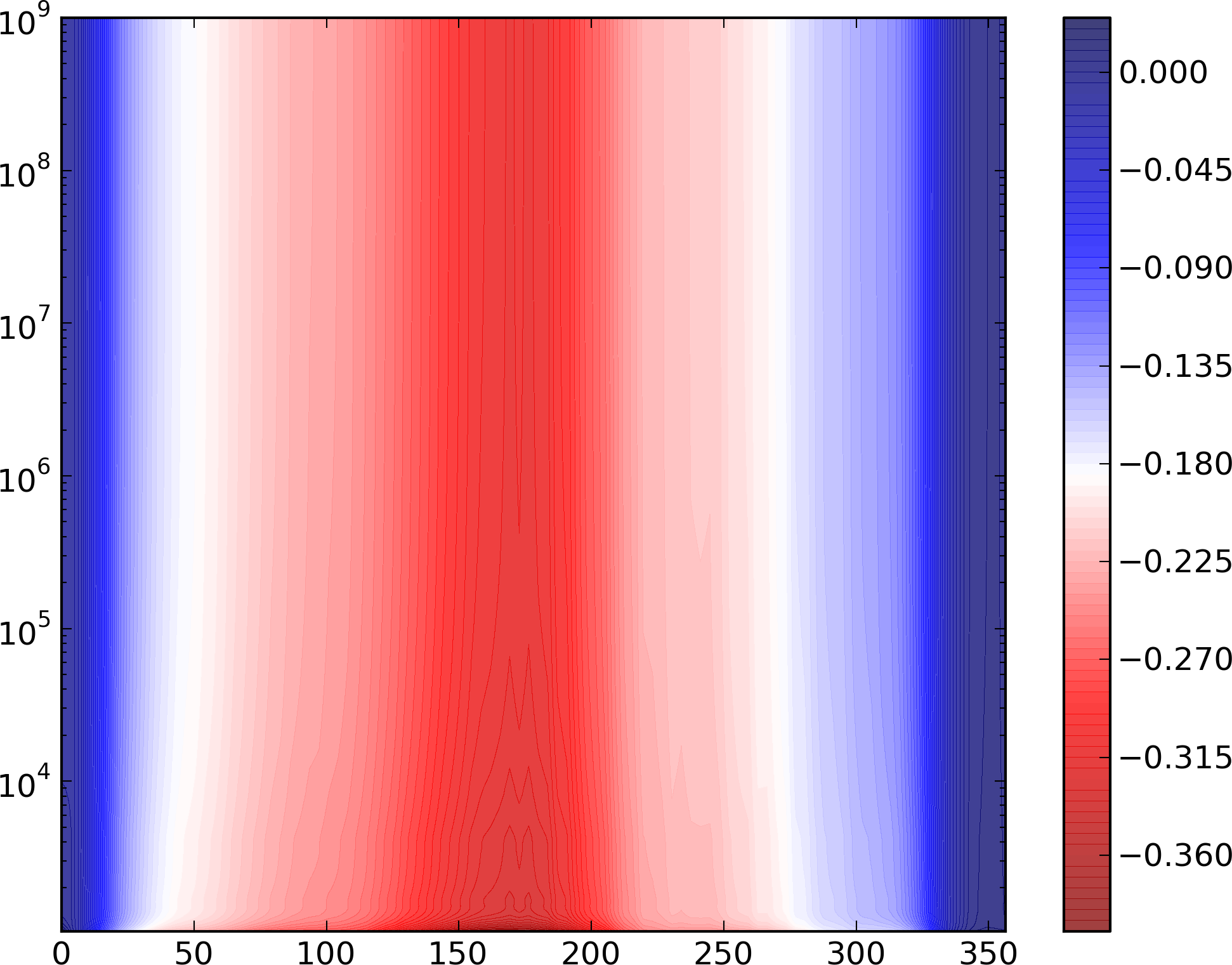}}
        		 \put(450,300){\rotatebox{270}{$ \left(S_{\phi} - S_{earth} \right) / S_{earth}$}}
        		\put(200,0){ $\phi \, \left[ ^{\circ} \right]$} 
        \end{picture}
    \end{subfigure}
 	\quad
     \begin{subfigure}{500\unitlength}
        \begin{picture}(500,400)
        		\put(0,150){\rotatebox{90}{$E_{kin} \, \left[ \mbox{MeV} \right]$}}
        		\put(30,25){\includegraphics[trim=0cm 0cm 0cm 0cm, clip=true,width=410\unitlength]{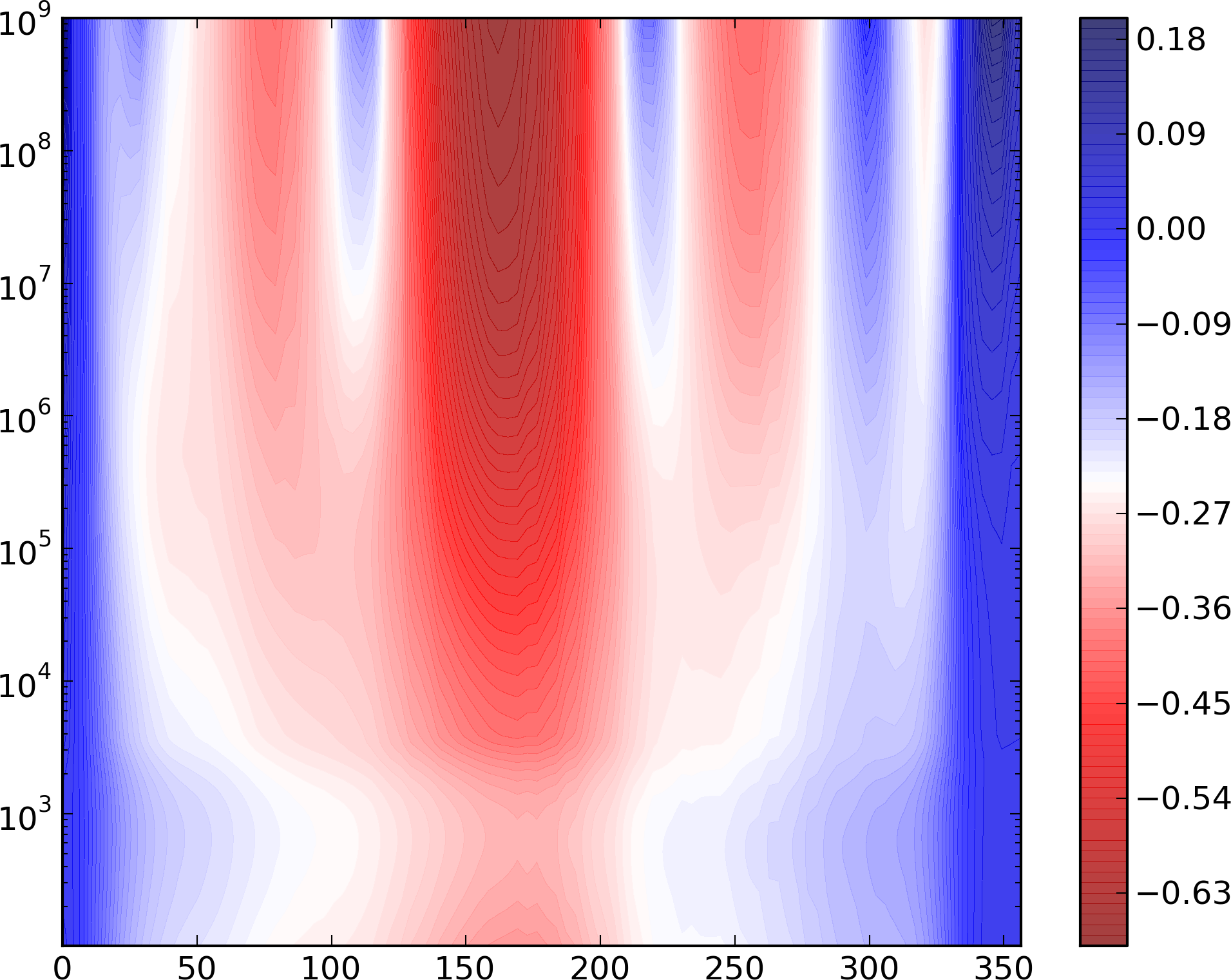}}
        		\put(450,300){\rotatebox{270}{$ \left(S_{\phi} - S_{earth} \right) / S_{earth}$}}
        		\put(200,0){ $\phi \, \left[ ^{\circ} \right]$} 
        \end{picture}
    \end{subfigure}
\caption{\textit{Left:} Spectrogram (see Section \ref{spectrogram} for details) of proton spectra for the different source models, \textit{Steiman}-Model (\textit{top-row}), \textit{Dame-Model} (\textit{middle-row}) and \textit{NE2001}-Model (\textit{bottom-row}) obtained at different positions on a circle around the Galactic Center with radius $r = R_{\odot}$. Position are given by the angle $\phi$ between the line connecting the nominal position $x=8.5 \mbox{kpc, }  y = 0 \mbox{kpc, } z = 0 \mbox{kpc}$ of the Earth with the Galactic Center and the line connecting the Galactic Center with the test position on the circle. Colour map indicates the ratio of the spectrum $S_{\phi}$ at a given angle $\phi$ and the spectrum at Earth nominal position $S_{earth}$ normalized by $S_{earth}$.\\
\textit{Right: } Same but for electrons.}
\label{Spectrograms}
\end{figure}
\clearpage

\appendix

\section{Comparison to Cosmic Ray Data}
\label{dataFit}
We discuss the implications of various spiral arm CR source distributions in relation to  - and only in relation to - the \textit{Reference}-Model (see Section \ref{setup}).
 
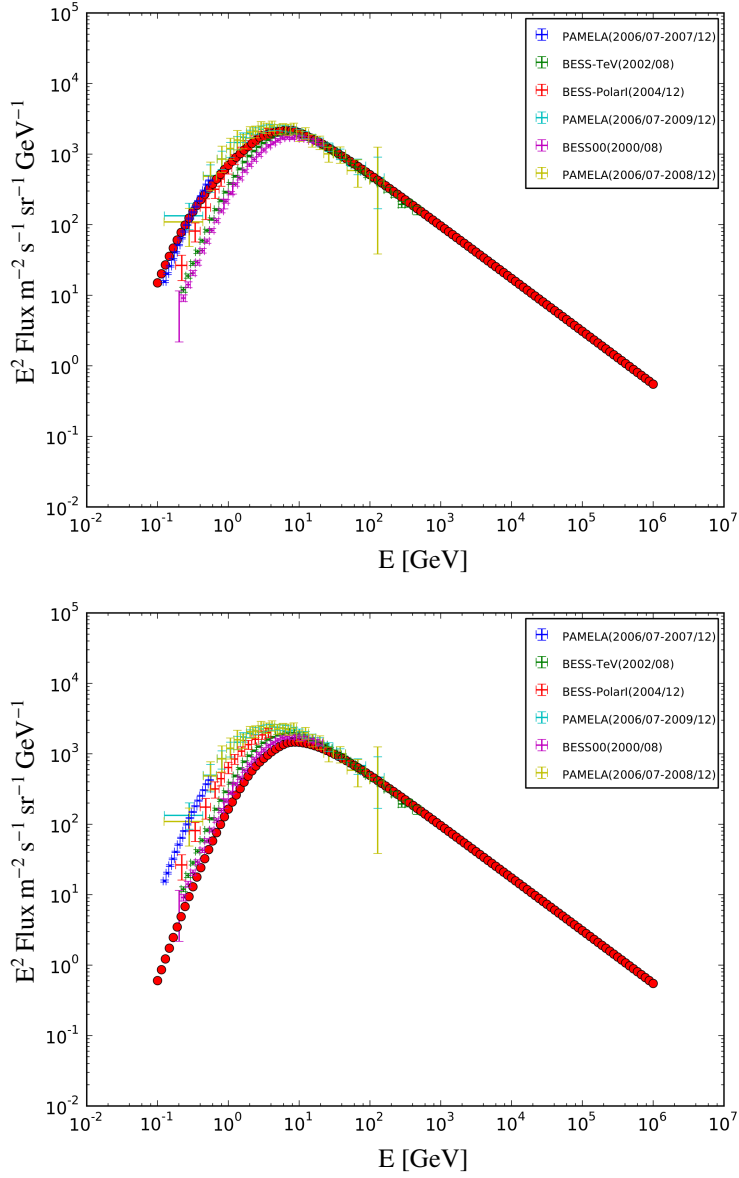
\begin{figure}[h!]
    \setlength{\unitlength}{0.001\textwidth}
    \begin{subfigure}{1000\unitlength}
        \begin{picture}(800,650)
        		\put(400,0){E [GeV]}
        		\put(0,200){\rotatebox{90}{$\mbox{E}^{2} \, \mbox{Flux} \, \mbox{m}^{-2} \, \mbox{s}^{-1} \, \mbox{sr}^{-1} \, \mbox{GeV}^{-1}$}}
            \put(40,40){\includegraphics[width=750\unitlength]{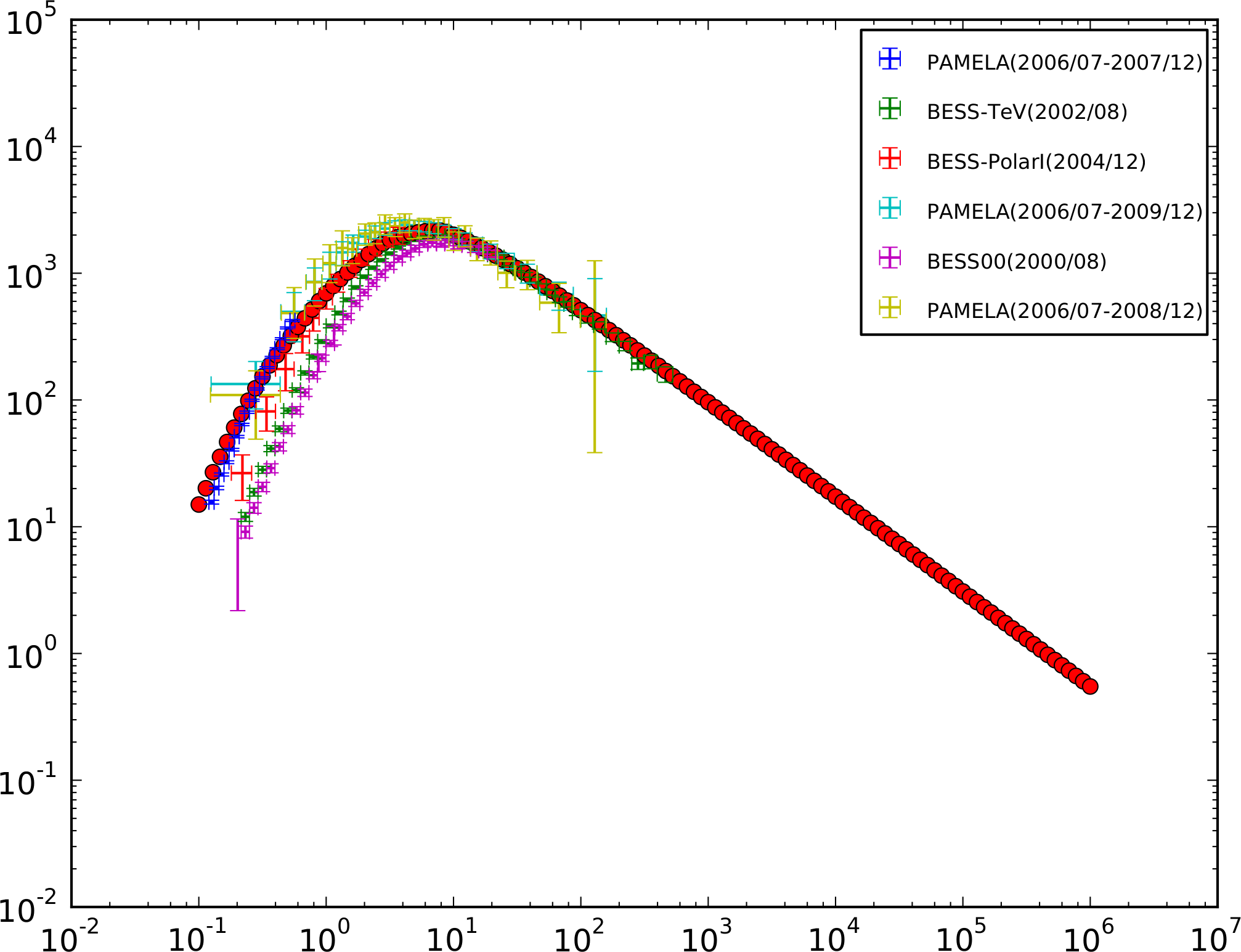}}
        \end{picture}
    \end{subfigure}\\
     \begin{subfigure}{1000\unitlength}
        \begin{picture}(800,650)
        		\put(400,0){E [GeV]}
        		\put(0,200){\rotatebox{90}{$\mbox{E}^{2} \, \mbox{Flux} \, \mbox{m}^{-2} \, \mbox{s}^{-1} \, \mbox{sr}^{-1} \, \mbox{GeV}^{-1}$}}
            \put(40,40){\includegraphics[width=750\unitlength]{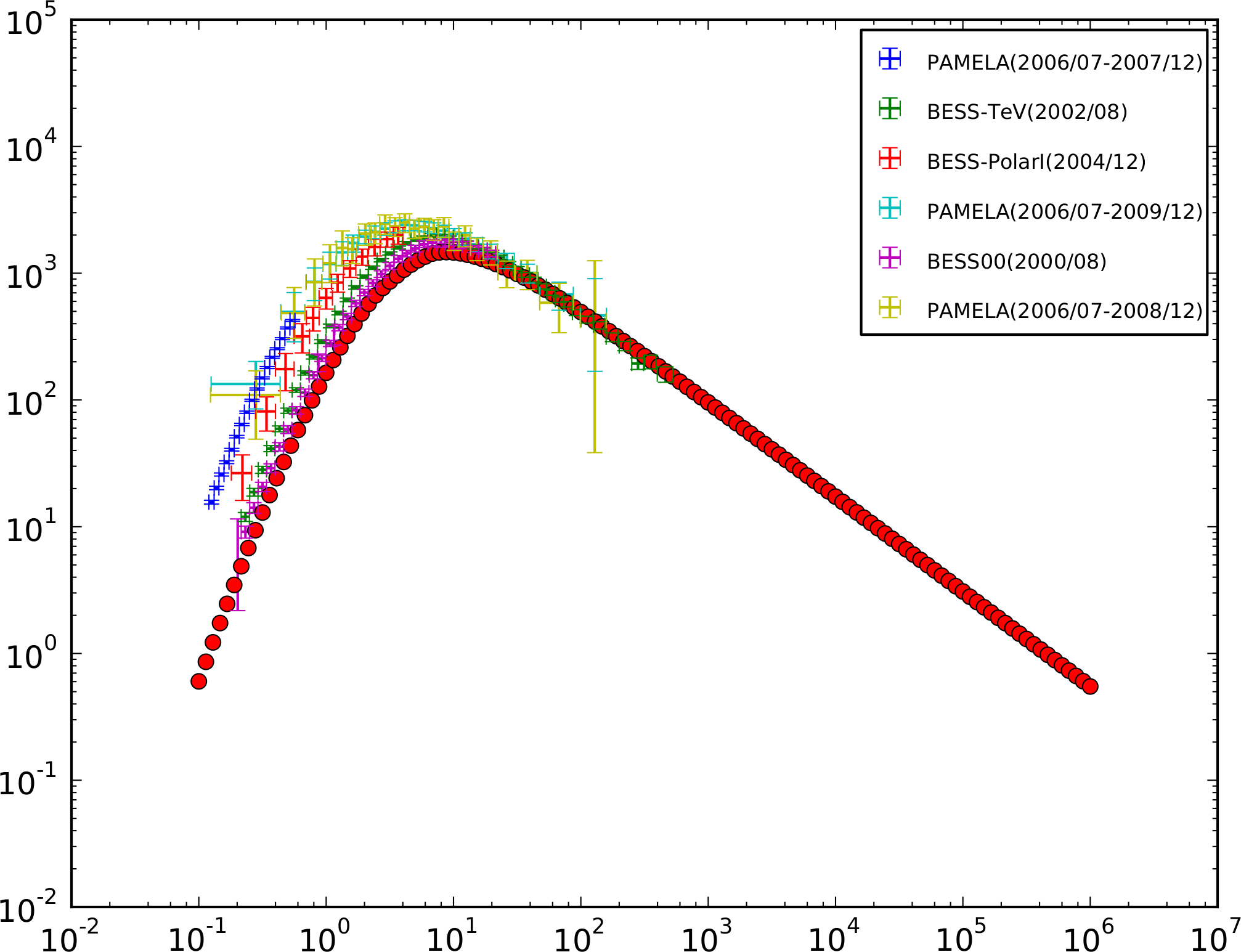}}
        \end{picture}
    \end{subfigure}
\caption{\textit{Top:} Proton spectra (\textit{red dots}) obtained using the four-arm \textit{Steiman}-Model compared to CR data taken from \cite{Abe2008,Shikaze2007,Adriani2010,Kim2013,Adriani2013,Adriani2013a}. 200 MV (\textit{top}) and 1000 MV (\textit{bottom}) force fields have been applied to the simulation data to take solar modulation \cite{Axford1968} into account.}
\label{pFitData}
\end{figure}

In Figure \ref{pFitData} we show that the four-arm \textit{Steiman}-Model is consistent with CR proton data. Above the spectral break (see Table \ref{table:PropParams}), our modelling results and data agree very well. Below that, effects, including, but not limited to, solar modulation, reacceleration and convection might have substantial influence on the agreement with CR data. We do not discuss the influence of these effects. Instead we focus on higher energies where only Galactic propagation effects are relevant. 

\begin{figure}[h!]
	\centering
    \setlength{\unitlength}{0.001\textwidth}
    \begin{subfigure}{1000\unitlength}
        \begin{picture}(800,650)
        		\put(400,0){E [MeV]}
        		\put(0,300){\rotatebox{90}{B/C-ratio}}
            \put(40,40){\includegraphics[width=750\unitlength]{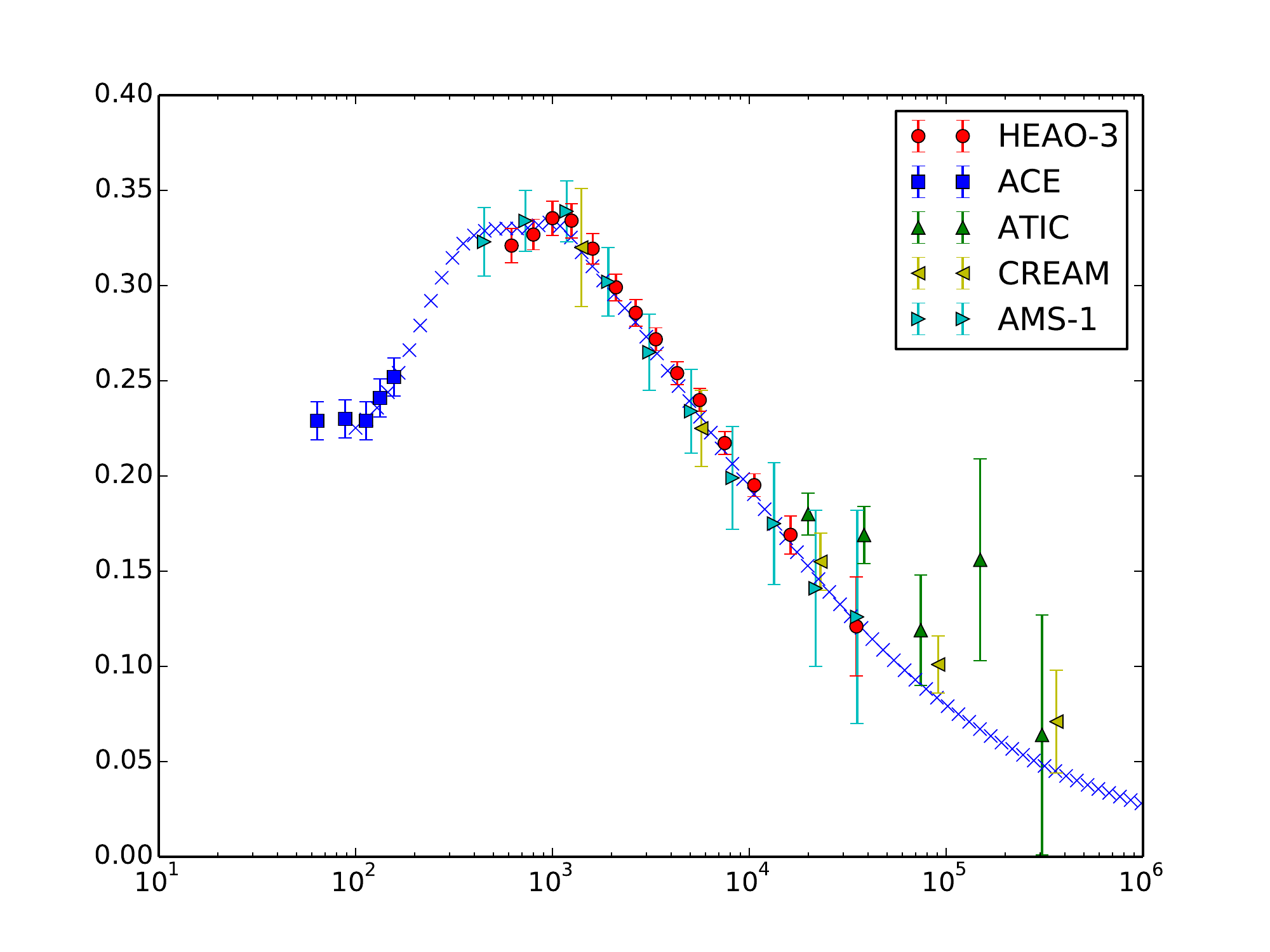}}
        \end{picture}
    \end{subfigure}\\
\caption{B/C-ratio (\textit{blue crosses}) obtained using the \textit{Reference}-Model compared to CR data taken from \cite{Engelmann1990,Aguilar2002,Davis2000,Nolfo2001,Panov2008,Ahn2008}. A 200 MV force field has been applied to the simulation data to take solar modulation \cite{Axford1968} into account.}
\label{BCfitdata}
\end{figure}

We remind the reader that the chosen propagation parameter set represents a plain diffusion model that was tuned to reproduce CR and $\gamma$-ray data (see \cite{Strong2010}). Consistency is confirmed with 3D \codename models using, e.g., the B/C-ratio shown in Figure \ref{BCfitdata}, which is a common test for propagation models.

\end{document}